\documentclass[aps,prx,twocolumn,amsmath,amssymb,nofootinbib,superscriptaddress,floatfix,reprint,longbibliography]{revtex4-2}
\usepackage{amsmath}
\usepackage{latexsym}
\usepackage{amssymb}
\usepackage{amsfonts}
\usepackage{stmaryrd}
\usepackage{graphicx}
\usepackage{esint}
\usepackage{subfigure}
\usepackage{xcolor}
\usepackage{mathtools,slashed}

\usepackage{ulem}

\makeatletter

\newcommand{\beq}{\begin{equation}}
\newcommand{\eeq}{\end{equation}}
\newcommand{\bea}{\begin{eqnarray}}
\newcommand{\eea}{\end{eqnarray}}
\newcommand{\bwt}{\begin{widetext}}
\newcommand{\ewt}{\end{widetext}}
\@ifundefined{textcolor}{}
{%
 \definecolor{BLACK}{gray}{0}
 \definecolor{WHITE}{gray}{1}
 \definecolor{RED}{rgb}{1,0,0}
 \definecolor{GREEN}{rgb}{0,1,0}
 \definecolor{BLUE}{rgb}{0,0,1}
 \definecolor{CYAN}{cmyk}{1,0,0,0}
 \definecolor{MAGENTA}{cmyk}{0,1,0,0}
 \definecolor{YELLOW}{cmyk}{0,0,1,0}
}

\newcommand{\bk}{\mathbf{k}}

\newcommand{\bK}{\mathbf{K}}

\newcommand{\bG}{\mathbf{G}}
\newcommand{\br}{\mathbf{r}}

\newcommand{\mtr}{\mathrm{Tr}}

\newcommand{\mcC}{\mathcal{C}}

\newcommand{\mcL}{\mathcal{L}}

\newcommand{\fvec}[1]{\boldsymbol{#1}}

\newcommand{\half}{\frac{1}{2}}
\newcommand{\pd}{\partial}
\newcommand{\rmd}{{\rm d}}

\newcommand{\bra}{{\langle}}
\newcommand{\ket}{{\rangle}}

\newcommand{\oRe}{{\operatorname{Re}}}
\newcommand{\oIm}{{\operatorname{Im}}}

\newcommand{\sn}{{\slashed{n}}}
\newcommand{\sa}{{\slashed{a}}}
\newcommand{\bpd}{{\bar{\partial}}}
\newcommand{\wpd}{{\widetilde{\partial}}}

\def\@buildmath#1{
    \expandafter\def\csname bb#1\endcsname{\mathbb{#1}}
    \expandafter\def\csname bs#1\endcsname{\boldsymbol{#1}}
    \expandafter\def\csname cal#1\endcsname{\mathcal{#1}}
    \expandafter\def\csname frak#1\endcsname{\mathfrak{#1}}
    \expandafter\def\csname rm#1\endcsname{\mathrm{#1}}
    \expandafter\def\csname bf#1\endcsname{\mathbf{#1}}
    \expandafter\def\csname sf#1\endcsname{\mathsf{#1}}
}
\def\@buildmathletters#1{
    \ifx#1\relax\else
        \@buildmath{#1}
        \expandafter\@buildmathletters
    \fi
}
\@buildmathletters ABCDEFGHIJKLMNOPQRSTUVWXYZabcdefghijklmnopqrstuvwxyz\relax

\makeatother

\usepackage{babel}

\begin{document}

\title{Field Response of the Orbital Moment Probed by Integrated Magnetic Circular Dichroism}

\author{Qian-Sheng Shi}
\affiliation{School of Physical Science and Technology, ShanghaiTech University, Shanghai 201210, China}

\author{Ziqiang Wang}
\email{ziqiang.wang@bc.edu}
\affiliation{Department of Physics, Boston College, Chestnut Hill, MA 02467, USA}

\author{Jian Kang}
\email{kangjian@shanghaitech.edu.cn}
\affiliation{School of Physical Science and Technology, ShanghaiTech University, Shanghai 201210, China}

\author{Yi Zhang}
\email{zhangyi821@shu.edu.cn}
\affiliation{Department of Physics and Institute for Quantum Science and Technology, Shanghai University, Shanghai 200444, China}
\affiliation{Shanghai Key Laboratory of High Temperature Superconductors, Shanghai University, Shanghai 200444, China}


\begin{abstract}
Magnetic circular dichroism (MCD) is an optical probe of orbital time reversal symmetry breaking. Although frequency integrated MCD is related to the self-rotation contribution to orbital magnetization at zero field, its magnetic response generally differs from the thermodynamic orbital susceptibility. Here, we define the MCD susceptibility and derive general expressions for multiband systems using an electromagnetic $U(1)$ gauge invariant approach. For insulators, the response contains contributions from the non-Abelian quantum metric, Berry curvature, and field induced interband mixing. For metals, a finite temperature Green function formulation separates the smooth response into Fermi sea and Fermi surface contributions.  An independent Landau level calculation confirms the expressions for the zero-field MCD moment and its linear magnetic response. In a continuum model for twisted bilayer graphene, the MCD susceptibility vanishes within the charge neutrality gap, while the thermodynamic orbital susceptibility remains markedly diamagnetic. Near the gap edges, the orbital moment term dominates the response. Features near the van Hove energies are associated with density of states enhancement. These results clarify how the two susceptibilities reflect different orbital processes even for the same electronic structure and thus establish the MCD susceptibility as a distinct probe of orbital magnetic response.
\end{abstract}

\maketitle

\section{Introduction}
Orbital time reversal symmetry breaking (TRSB) can generate circulating electronic currents and orbital magnetization, thereby linking the magnetic order to the topology of electronic state~\cite{NiuRMP2010, DaiNRP2021}. A paradigmatic example is the Haldane model, in which complex hoppings break time reversal symmetry and produce a Chern insulating phase without a net magnetic flux through the unit cell~\cite{HaldanePRL1988}. Moir\'e materials provide a versatile experimental platform for exploring orbital TRSB~\cite{PabloNature2018_1, PabloNature2018_2, DavidGGScience2019, WangNature2020, YazdaniNature2020, YankowitzNC2021, YoungNature2020, KinFaiMakNature2023, YoungNature2024, Efretov2019, Xiaodong21, JuLongChiralSC, FeldmanHofstadter, wus_2021, das_2021,  ste_2021, xie2021, FolkNC2022}, while theoretical studies have identified interaction driven valley polarization as an important route to anomalous Hall and Chern insulating states~\cite{KangPRL2019, DaiPRX2019, WuDasSarmaPRL2020, ChenBBNC2020, SenthilPRL2020, KTLawPRL2022, LiangFuPRX2022, WangPRX2024, SenthilPRL2024, MinxuanPRB2024,zhang_2020}. Graphene and transition metal dichalcogenide (TMD) based moire systems exhibit ferromagnetism, anomalous Hall responses, integer and fractional Chern insulating states, electrically switchable magnetic order, and local orbital magnetism~\cite{DavidGGScience2019, YoungScience2020, WangNature2020, YazdaniNature2020, YankowitzNC2021, KinFaiMakNature2021, JuLongNature2024, YankowitzNP2022, YoungNature2020, DavidGGNL2021, YoungScience2021, XiaodongCaiNature2023, KinFaiMakNature2023, XiaodongNature2023, YoungNature2024}. 
Beyond moir\'e systems, Kerr and circular dichroism signals have also been reported in the charge ordered phase of the kagome metals $A\mathrm{V}_3\mathrm{Sb}_5$ ($A=\mathrm{K},\mathrm{Rb},\mathrm{Cs}$) and interpreted as signatures of TRSB~\cite{xu_2022}.
These findings motivate a quantitative understanding of orbital magnetization and its response to an applied magnetic field, both to characterize orbital TRSB and to connect magnetic measurements with electronic structure~\cite{RestaPRL2005, RestaPRB2006, JunrenShiPRL2007, RaouxPRB2015, MontambauxPRL2016}. Here we focus exclusively on the orbital response, for which the magnetic field enters through the electromagnetic vector potential, with the Zeeman correction discussed separately in the Supplemental Material (SM).

Although orbital TRSB can produce a finite orbital magnetization, its direct detection is often challenging, particularly in low dimensional systems. The associated stray magnetic fields depend on sample geometry, magnetic domains, and domain walls, so extracting the underlying magnetization generally requires a corresponding magnetostatic analysis~\cite{YoungScience2021, YoungNature2024}. Magnetic circular dichroism (MCD) provides a complementary optical probe~\cite{KinFaiMakNature2020, XiaodongNature2022, XiaodongNature2023, KinFaiMakNature2023, KinFaiMakNC2025, KinFaiMakNature2026, JiangPRX2026}. Dichroic sum rules connect the frequency integrated optical signal to orbital circulation~\cite{SouzaVanderbiltPRB2008, RestaPRR2020,BrunoArXiv2507}. Integrated MCD therefore offers a natural optical probe of orbital TRSB when direct measurements of static orbital magnetization are difficult.

While dichroic sum rules characterize integrated MCD at zero field~\cite{RestaPRR2020}, the orbital processes governing its magnetic response and their relation to the thermodynamic orbital susceptibility remain to be clarified. This question is motivated by field dependent circular dichroism measurements in moir\'e materials~\cite{XiaodongCaiNature2023, KinFaiMakNature2023, KinFaiMakNC2025}, although these measurements generally probe selected optical resonances rather than the full spectral integral. Here, we derive general expressions for the smooth weak field MCD susceptibility in multiband systems entirely in terms of zero field electronic structure. In the continuum model for twisted bilayer graphene (TBG), the MCD susceptibility vanishes within the charge neutrality gap, while the thermodynamic orbital susceptibility remains diamagnetic. Outside the gap, $\chi_{\rm MCD}$ and $\chi_{\rm orb}$ can have either the same or opposite signs, depending on the chemical potential. The band edge peaks are dominated by the orbital moment Fermi surface contribution, whereas the van Hove features highlight density-of-states enhancement. These results show that the magnetic response of integrated MCD weights orbital processes differently from the thermodynamic orbital susceptibility, providing a quantitative basis for comparing optical and thermodynamic probes of orbital magnetism.

\section{Integrated MCD and its magnetic response}
In this work, we study the magnetic field response of integrated MCD using an electromagnetic (EM) $U(1)$ gauge-invariant formalism. Circularly polarized light carries opposite angular momenta for the two helicities and probes the corresponding circular components of electronic motion. Orbital TRSB can therefore lead to unequal absorption for the two polarizations. While the dichroism at a given frequency depends on specific optical resonances, its integral over positive frequencies obeys a dichroic sum rule. At zero field and $T=0$, within the electric dipole approximation, this sum rule relates the integrated spectral weight to the gauge invariant self-rotation contribution to the orbital magnetization~\cite{SouzaVanderbiltPRB2008, RestaPRR2020, sm}. This contribution characterizes the internal orbital circulation probed by integrated MCD.

Throughout this work, we consider the interband optical response, with the intraband magneto-optical contribution subtracted. At finite magnetic field, intraband absorption includes cyclotron resonances whose frequencies approach zero in the weak field limit. We denote the remaining regular response by $\sigma_\pm^{\rm reg}(\omega) = \sigma_\pm(\omega) -\sigma_\pm^{\rm intra}(\omega)$, where $\sigma_\pm$ is the total optical conductivity for left- and right-handed circularly polarized light. When the intraband and interband absorption are well separated, an appropriate experimental lower frequency cutoff can exclude the former without removing appreciable spectral weight from the latter. We define
\begin{align}
    M_{\rm MCD}(\mu, B) \equiv \frac{\hbar}{2\pi e} \int_{0^+}^\infty \rmd\omega\, \oRe\left( \sigma_+^{\rm reg}(\omega) - \sigma_-^{\rm reg}(\omega) \right) \ , \label{Eqn:MMCDDef}
\end{align}
with the intraband contribution subtracted before the frequency integration and magnetic field differentiation. The lower limit $0^+$ specifies integration over positive frequencies. The prefactor $\hbar/(2\pi e)$ converts the integrated conductivity difference into an orbital magnetic moment according to the dichroic sum rule. Under appropriate conditions, this quantity can also be related to the orbital angular momentum of Bloch states~\cite{LaanPRL1992, BurgosAP2024}.

To characterize how this optically probed orbital moment responds to a weak magnetic field, we define its susceptibility at fixed chemical potential as
\begin{align}
    \chi_{\rm MCD} \equiv \left( \pd M_{\rm MCD} / \pd B \right)_{\mu, B = 0} \ . \label{Eqn:ChiMCDDef}
\end{align} 
For metals, the derivative in Eq.~\eqref{Eqn:ChiMCDDef} denotes the smooth response obtained by taking $B\rightarrow0$ at finite temperature before $T\rightarrow0$, as specified in Eq.~\eqref{Eqn:ChiMCDMetalDefinition}.  Our central questions are which microscopic processes govern this response and how their weighting differs from that entering the thermodynamic orbital susceptibility. 

The thermodynamic orbital susceptibility describes the response of the full equilibrium orbital magnetization, including contributions beyond self-rotation. It is defined by $\chi_{\rm orb} = -\left. \pd^2 \Omega / \pd B^2 \right|_{\mu, B = 0}$, where $\Omega$ is the equilibrium grand potential. Consequently, $\chi_{\rm MCD}$ need not coincide with $\chi_{\rm orb}$. Closed-form expressions for the latter have been derived for non-interacting multiband systems~\cite{StauberPRL2011, GaoPRB2015, RaouxPRB2015}, with recent extensions to interacting systems within the Hartree-Fock (HF) approximation~\cite{KangPRL2026, XiaoArXiv2026}.

We first reformulate the integrated MCD in terms of band projectors and Green's functions, providing a convenient starting point for incorporating magnetic field effects. We then construct their EM $U(1)$ gauge invariant counterparts in the presence of an external magnetic field and expand them systematically to linear order in $B$. This procedure yields a closed form expression for $\chi_{\rm MCD}$ entirely in terms of zero field quantities and exposes its connection to the quantum geometry of the Bloch states. More generally, the same EM gauge invariant expansion provides a systematic route to the magnetic field response of other electronic observables.

\section{Electromagnetic Gauge invariant formulation}
We first consider an insulator at $T = 0$ and assume that its gap remains open in a weak magnetic field. Since there are no partially occupied bands, the intraband optical contribution vanishes. We can therefore express Eq.~\eqref{Eqn:MMCDDef} in terms of occupied and empty projectors, providing a convenient starting point for its field derivative. Orbital coupling in a tight binding Hamiltonian is implemented by the Peierls substitution, $t_{ij}\rightarrow t_{ij} \exp\left(i\frac{e}{\hbar} \int_{\fvec r_i}^{\fvec r_j} \fvec A \cdot \rmd \fvec s \right)$. At $T = 0$, the electric dipole approximation gives~\cite{SouzaVanderbiltPRB2008, RestaPRR2020, sm}:
\begin{align}
    M_{\rm MCD} & = \frac{e}{2\hbar } \sum_{\fvec k} \epsilon_{\mu\nu} \oIm\mtr\left( \big( \pd_{\mu} \hat H(\fvec k) \big) \big( \pd_{\nu} \hat P(\fvec k) \big) \hat Q(\fvec k)  \right) \nonumber \\
    & = - \frac{e}{2 \hbar } \epsilon_{\mu\nu}  \int \left( \prod_{i = 1}^3\rmd \fvec r_i \right) \oIm\mtr\left\{ \big[\hat r_{\mu},  \hat H\big]_{\fvec r_1 \fvec r_2}    \right. \nonumber \\
    & \qquad \times \left. \big[ \hat r_{\nu},  \hat P\big]_{\fvec r_2 \fvec r_3} \hat Q(\fvec r_3, \fvec r_1) \right\} \label{Eqn:MMCDZeroT}
\end{align}
Here, $\hat P$ and $\hat Q$ denote the projectors onto the occupied and empty states, respectively. In contrast to individual Bloch wavefunctions, these projectors are invariant under unitary transformations within the occupied and empty subspaces, and therefore remain smooth even in the presence of degeneracies within either subspace~\cite{ZhangArXiv2026}. The variable $\fvec k$ is the Bloch momentum, and $\pd_\mu \equiv \pd/\pd k_\mu$. The second line gives the equivalent real space representation of the first one and will be particularly useful for analyzing the magnetic field dependence below.

To expand Eq.~\eqref{Eqn:MMCDZeroT} in $B$, we use its real space form, which remains well defined when a uniform field breaks ordinary translation symmetry. Separating the gauge-dependent Peierls phase from a gauge-invariant kernel isolates the physical field response and allows us to express it using zero-field electronic structure.

To separate the physical field dependence from the choice of vector potential, for any two point fermion bilinear $O(\fvec r_1,\fvec r_2)$ we introduce its EM gauge invariant counterpart $\widetilde O(\fvec r_1,\fvec r_2)$ through~\cite{Gorkov62, LeePRB2011, RaouxPRB2015, KangPRL2026, YeArXiv2026} 
\begin{align}
    O(\fvec r_1,\fvec r_2) & = e^{i\varphi(\fvec r_1,\fvec r_2)} \widetilde O(\fvec r_1,\fvec r_2) \ , 
\end{align}
where $\varphi(\fvec r_1, \fvec r_2) = \frac{e}{\hbar} \int_{\fvec r_1}^{\fvec r_2}\fvec A\cdot\rmd\fvec s$, with the integral taken along the straight line connecting the two points. By construction, $\widetilde O$ is invariant under an EM gauge transformation and depends only on the magnetic field $B$, rather than on the particular choice of $\fvec A$.

For an insulator whose gap remains open, these gauge invariant kernels rewrite the real space expression in Eq.~\eqref{Eqn:MMCDZeroT} as
\begin{align}
    M_{\rm MCD} & = -\frac{e}{2\hbar } \epsilon_{\mu\nu} \int \prod_{i = 1}^3 \rmd \fvec r_i\, \oIm\mtr\left\{ e^{i\Phi_{123}} [ \hat r_\mu, \widetilde H]_{\fvec r_1, \fvec r_2} \right. \nonumber \\
    & \qquad \left. \times [\hat r_\nu, \widetilde P]_{\fvec r_2,\fvec r_3} \widetilde Q_{\fvec r_3,\fvec r_1} \right\} \ , \label{Eqn:MMCDGaugeInv}
\end{align}
where the phase accumulated around the triangular path is
\begin{align}
    \Phi_{123} & \equiv \varphi_{12} + \varphi_{23} + \varphi_{31} = \frac{e B}{2\hbar} \epsilon_{\mu\nu} (\fvec r_1 - \fvec r_3)_\mu (\fvec r_2 - \fvec r_1)_\nu . \label{Eqn:Phi123}
\end{align}
Here, $\varphi_{ij} \equiv \varphi(\fvec r_i, \fvec r_j)$. The phase $\Phi_{123}$ is proportional to the magnetic flux through the triangle formed by $\fvec r_1$, $\fvec r_2$, and $\fvec r_3$, and is itself EM gauge invariant. It can be viewed as the Aharonov-Bohm (AB) phase associated with this real space loop. 

The linear magnetic response of the integrated MCD thus has two distinct origins. The first comes from the explicit field dependence of the gauge invariant loop phase $e^{i\Phi_{123}}$, which represents the magnetic flux enclosed by the corresponding real space loop. The second comes from the field induced changes of the occupied and empty subspaces through $\widetilde P(B)$ and $\widetilde Q(B)$. Both contributions should be included in the MCD susceptibility.

The loop phase and the gauge invariant kernels preserve the original lattice periodicity under simultaneous translation of their coordinates~\cite{sm}. We can therefore Fourier transform the kernels in the original zero field Brillouin zone. In this representation, $\widetilde O(\fvec k) = e^{-i \fvec k \cdot \hat{\fvec r}} \widetilde O e^{i\fvec k \cdot \hat{\fvec r}}$ and $[\hat r_\mu, \widetilde O] \rightarrow i \pd_\mu \widetilde O(\fvec k)$. Here $\fvec k$ labels the Fourier representation of the gauge-invariant kernels, rather than ordinary Bloch eigenstates of the finite-field Hamiltonian.

The remaining field dependence is contained in the EM gauge invariant operators $\widetilde H$, $\widetilde P$, and $\widetilde Q$. For a tight binding Hamiltonian whose orbital coupling to the magnetic field enters only through the Peierls substitution, the EM gauge invariant Hamiltonian satisfies $\widetilde H(B) = H_0$, where $H_0$ is the zero field Hamiltonian. The field dependence of the projectors can instead be obtained from the gauge invariant Green function. To linear order in $B$~\cite{LeePRB2011, RaouxPRB2015, KangPRL2026, YeArXiv2026},
\begin{align}
    \left. \frac{\rmd\widetilde G(z)}{\rmd B} \right|_{B=0} = \frac{e \hbar}{2 i} \epsilon_{\mu\nu} G_0(z) \hat V^\mu G_0(z) \hat V^\nu G_0(z) \ , \label{Eq:dGdB}
\end{align}
where $G_0(z) = (z - H_0)^{-1}$ is the zero field Green function and $\hat V^\mu = (i\hbar)^{-1} [\hat r_\mu, \hat H_0]$ is the  corresponding velocity operator. Note that the momentum $\fvec k$ is dropped for notational convenience. For a translationally invariant system, its momentum space representation is $\hat V^\mu(\fvec k)=\hbar^{-1} \pd_\mu H_0(\fvec k)$.

The single particle density matrix can be expressed as $\hat P = \oint_{\cal C}\frac{\rmd z}{2\pi i} n_F(z) G(z)$, where $n_F(z)$ is the Fermi distribution function and the contour $\cal C$ encloses the whole real axis. At $T = 0$, $n_F(E) = \Theta(\mu - E)$, and $\hat P$ reduces to the projector onto the occupied states. However, at finite field its EM gauge invariant counterpart $\widetilde P$ is generally not a projector under ordinary operator multiplication because $\widetilde P^2 \neq \widetilde P$. For an insulator, contour integration of Eq.~\eqref{Eq:dGdB} gives~\cite{VanderbiltPRB2010, sm} two sectors: block diagonal terms associated with the Berry curvature correction to phase space weight, and off-diagonal terms describing mixing across the gap:
\begin{align}
    \left. \frac{\rmd\widetilde P}{\rmd B} \right|_{B=0} & = \frac{e}{2 i \hbar } \epsilon_{\mu\nu} \left\{ - \big( \pd_{\mu} \hat P \big) \hat Q \big( \pd_{\nu} \hat P \big) + \big( \pd_{\mu} \hat Q \big) \hat P \big( \pd_{\nu} \hat Q \big) \right. \nonumber \\
    & \quad  + \hat Q \calL^{-1}\left( V^{\mu} \big( \pd_{\nu} \hat Q \big) + \big( \pd_{\mu} \hat P \big) V^{\nu} \right) \hat P \nonumber \\
    & \quad \left. - \hat P \calL^{-1}\left( V^{\mu} \big( \pd_{\nu} \hat P \big) + \big( \pd_{\mu} \hat Q \big) V^{\nu} \right) \hat Q \right\}  \label{Eqn:dPdB}
\end{align}
Here, $\calL(\hat O)=\hbar^{-1}[\hat H,\hat O]$ denotes the Liouvillian operator. Its inverse is defined for matrix elements connecting states with different energies. For the eigenstate projectors $\hat P_n = |u_n \rangle \langle u_n |$, with $\hat H | u_n \rangle = E_n | u_n \rangle$,
\begin{align}
    \hat P_n\calL^{-1}(\hat O)\hat P_m = \hbar \frac{\hat P_n\hat O\hat P_m}{E_n-E_m} \ ,   \qquad   E_n\neq E_m.
\end{align}

The first two terms in Eq.~\eqref{Eqn:dPdB} act within the occupied and empty subspaces and, after contraction with $\epsilon_{\mu\nu}$, are governed by their non-Abelian Berry curvatures~\cite{ShindouNPB2005}. For a single occupied band, the occupied sector term gives the field correction to the phase space density of states~\cite{NiuRMP2010}. The last two terms mix the occupied and empty subspaces~\cite{GaoPRL2014, GaoPRB2015}; a self-consistent HF extension modifies this off-diagonal sector~\cite{KangPRL2026}. Eq.~\eqref{Eqn:dPdB} applies directly to insulators, while additional Fermi surface contributions are required in metals~\cite{sm}.

\section{MCD susceptibility in insulators}
For an insulator at $T=0$, the responses of the AB phase and the electronic states give rise to three contributions: a quantum metric contribution weighted by the inverse mass tensor, a Berry curvature contribution, and a contribution from interband mixing across the gap. Their sum gives the insulating MCD susceptibility~\cite{sm}:
\begin{align}
    \chi_{\rm MCD}^{\rm ins} & \equiv \left. \frac{\pd M_{\rm MCD}}{\pd B} \right|_{\mu, B = 0} = \chi_{\rm TR}^{(1)} + \chi_{\rm TR}^{(2)}  + \chi_{\rm TR}^{(3)} \label{Eqn:ChiMCDIns} \ .
\end{align}
Here $\rm TR$ labels the Fermi sea contribution associated with interband transitions. The three components $\chi_{\rm TR}^{(i)}$ are
\begin{align}
    \chi_{\rm TR}^{(1)} & = -\frac{e^2}{4} \epsilon_{\mu\nu} \epsilon_{\alpha\beta} \sum_{\fvec k} \oRe\mtr\left[ \calM^{-1}_{\mu\beta}   \big( \pd_{\nu} \hat P \big) \big( \pd_{\alpha} \hat Q \big) \right] \nonumber \\
    & =  \frac{e^2}{4} \epsilon_{\mu\nu} \epsilon_{\alpha\beta} \sum_{\fvec k} \oRe\mtr\left[ \calM^{-1}_{\mu\beta}  \left(   \big( \pd_{\nu} \hat P \big) \hat Q \big( \pd_{\alpha} \hat P \big)  + \right. \right. \nonumber \\
    & \qquad \left. \left. \big( \pd_{\nu} \hat Q \big) \hat P \big( \pd_{\alpha} \hat Q \big) \right)      \right]  \label{Eqn:Chi1MCDIns} \\
    \chi_{\rm TR}^{(2)} & = - \frac{e^2}{2 \hbar} \epsilon_{\mu\nu} \epsilon_{\alpha\beta} \sum_{\fvec k} \oRe\mtr\left[ V^{\mu} \big( \pd_{\alpha} \hat P \big) \hat Q \big( \pd_{\beta} \hat P \big) \big( \pd_{\nu} \hat Q \big) \right. \nonumber \\
        & \qquad \left. + V^{\mu} \big( \pd_{\alpha} \hat Q \big) \hat P \big( \pd_{\beta} \hat Q \big) \big( \pd_{\nu} \hat P \big)  \right]  \label{Eqn:Chi2MCDIns} \\
    \chi_{\rm TR}^{(3)} & = \frac{e^2}{2 \hbar} \sum_{\fvec k} \oRe\mtr\left[ \hat P \calL^{-1}\left( \hat O_1 \right) \hat Q \hat O_2 \right] \label{Eqn:Chi3MCDIns}
\end{align}

where $\calM^{-1}$ is the inverse mass tensor, defined as
\begin{align}
    \hat \calM^{-1}_{\mu\nu} & = \left( \frac1{i\hbar} \right)^2 [ \hat r_{\mu}, [\hat r_{\nu}, \hat H_{ 0}]  ]  =  \frac1{\hbar^2} \frac{\pd^2 H_0}{\pd k_\mu \pd k_\nu} \  .  
\end{align}
The contribution $\chi_{\rm TR}^{(1)}$ originates from the field derivative of the AB phase. The projector structure in Eq.~\eqref{Eqn:Chi1MCDIns} can be decomposed into the occupied and empty subspace components of the non-Abelian quantum geometric tensor~\cite{sm}. After contraction with $\epsilon_{\mu\nu}\epsilon_{\alpha\beta}$ and taking the real part, only the metric components contribute. Therefore, $\chi_{\rm TR}^{(1)}$ can be interpreted as a non-Abelian quantum metric contribution weighted by the inverse mass tensor.

The contributions $\chi_{\rm TR}^{(2)}$ and $\chi_{\rm TR}^{(3)}$ arise from the field dependence of $\widetilde P$ and $\widetilde Q$ in Eq.~\eqref{Eqn:MMCDGaugeInv}. In particular, $\chi_{\rm TR}^{(2)}$ originates from the block diagonal, Berry curvature related part of Eq.~\eqref{Eqn:dPdB}. Its structure resembles a product of the Berry curvature and orbital magnetic moment, consistent with the magnetic field induced change of the phase space density of states. The operators entering $\chi_{\rm TR}^{(3)}$ are
\begin{align}
    \hat O_1 & = \epsilon_{\alpha\beta} \left( V^{\alpha} \big( \pd_{\beta} \hat P \big) + \big( \pd_{\alpha} \hat Q \big) V^{\beta} \right) \\
    \hat O_2 & = \epsilon_{\alpha \beta} \left( V^{\alpha} \big( \pd_{\beta} \hat Q \big) - \big( \pd_{\alpha} \hat P \big) V^{\beta} \right) \ .
\end{align}
This term arises from magnetic field induced mixing between the occupied and empty subspaces~\cite{GaoPRB2015}. Because $\calL^{-1}$ introduces explicit cross gap energy denominators, $\chi_{\rm TR}^{(3)}$ represents an interband contribution associated with the magnetic deformation of the occupied subspace, analogous in structure to Van Vleck type interband magnetic response.

Thus, $\chi_{\rm MCD}^{\rm ins}$ extends the sum rule relation between integrated MCD and self-rotation orbital magnetization~\cite{SouzaVanderbiltPRB2008, RestaPRR2020} to weak magnetic fields. It combines the response of the AB phase with the field induced deformation of the occupied subspace to describe how the local orbital circulation encoded in $m_{\rm SR}$ evolves with the field.

\section{MCD susceptibility in metals}
In metals, a magnetic field modifies both the electronic states and their occupations near the Fermi surface. At $T=0$, even an arbitrarily weak field produces Landau quantization and quantum oscillations, making the response nonanalytic at $B=0$. The limits $B\rightarrow0$ and $T\rightarrow0$ therefore do not generally commute. Thus, directly differentiating the $T=0$ expression for $M_{\rm MCD}$, even with the additional Fermi surface term included in $\rmd\widetilde P/\rmd B$~\cite{sm}, does not generally yield the smooth weak field response.

To extract the smooth magnetic response, we take the zero-field limit at finite temperature before sending $T\rightarrow0$. We therefore define
\begin{align}
    \chi_{\rm MCD} & = \lim_{T\rightarrow0}
    \left. \frac{\pd M_{\rm MCD}(B,T)}{\pd B}
    \right|_{\mu, B=0} \ .
    \label{Eqn:ChiMCDMetalDefinition}
\end{align}
Finite temperature smooths the Landau level occupations and suppresses quantum oscillations in the weak field limit, allowing the derivative at $B=0$ to be evaluated. Reversing the order of limits generally preserves the nonanalytic quantum oscillation response.

To express the result obtained after taking the prescribed zero temperature limit, we separate the momentum dependence of the eigenstates from that of their occupation factors. We define the Fermi sea derivative by holding the occupation factors fixed:
\begin{align}
    \bpd_\mu \hat P(\fvec k) \equiv \sum_{n \in \rm occ} \left( | \pd_\mu u_n \rangle \langle u_n | + | u_n \rangle \langle \pd_\mu u_n | \right) \ .
    \label{Eqn:FermiSeaDerivative}
\end{align}
The sum runs over occupied states at momentum $\fvec k$. For a band $n$ crossing the Fermi surface, assumed nondegenerate in its vicinity, we also define the remaining occupied and empty projectors:
\begin{align}
    \hat P_{\sn}(\fvec k) & = \sum_{\substack{m \neq n \\ m \in \rm occ}} | u_m \rangle \langle u_m | \ ,   &
    \hat Q_{\sn}(\fvec k) & = \sum_{\substack{m \neq n \\ m \in \rm emp}} | u_m \rangle \langle u_m | \ .
\end{align}

To obtain the smooth magnetic response of a metal, we return to the optical definition of $M_{\rm MCD}$ in Eq.~\eqref{Eqn:MMCDDef} and use a finite temperature Matsubara Green function formulation. We evaluate the magnetic field derivative at $B=0$ while keeping $T$ finite, and take $T\rightarrow0$ only afterward, as required by Eq.~\eqref{Eqn:ChiMCDMetalDefinition}. This approach treats the Fermi sea and Fermi surface contributions on an equal footing. The derivation is given in the SM~\cite{sm}.

The resulting smooth MCD susceptibility separates into a Fermi sea contribution and a Fermi surface contribution:
\begin{align}
    \chi_{\rm MCD} & = \chi_{\rm MCD}^{\rm TR} + \chi_{\rm MCD}^{\rm FS} \ .  \label{Eqn:ChiMCD}
\end{align}
The contribution $\chi_{\rm MCD}^{\rm TR}$ has the same structure as the insulating result $\chi_{\rm MCD}^{\rm ins}$, with all derivatives of the occupied and empty projectors replaced by their Fermi sea counterparts. Thus, the quantum geometric quantities identified for insulators remain relevant in metals, but are evaluated separately within each Fermi sea. The explicit expression for $\chi_{\rm MCD}^{\rm TR}$ is provided in the SM~\cite{sm}.

The Fermi surface contribution describes how states near the Fermi surface modify the interband optical response under a magnetic field. It  contains an AB loop term $\chi_{\rm FS}^{(1)}$, a velocity weighted geometric and state deformation term $\chi_{\rm FS}^{(2)}$, and an orbital moment term $\chi_{\rm FS}^{(3)}$. The third has no explicit band velocity factor and can become important near stationary points with large orbital moments:
\begin{align}
    \chi_{\rm MCD}^{\rm FS}  & =  \chi_{\rm FS}^{(1)} + \chi_{\rm FS}^{(2)} + \chi_{\rm FS}^{(3)}  \\
    \chi_{\rm FS}^{(1)} & = \frac{e^2}4 \epsilon_{\mu\nu} \epsilon_{\alpha\beta} \sum_a n_F'(E_a) \big( \pd_{\alpha} E_a \big) \nonumber \\
    & \qquad \times  \oRe\mtr\left( \calM^{-1}_{\beta \mu}  \big( \hat Q_{\sa} -  \hat P_{\sa} \big)  \big( \pd_{\nu} \hat P_a \big) \right) \\
    \chi_{\rm FS}^{(2)} & =  \frac{e^2}{2 \hbar}  \epsilon_{\mu\nu} \epsilon_{\alpha\beta} \sum_a n_F'(E_a) \big( \pd_{\alpha} E_a \big) \nonumber \\ 
    & \qquad \times  \oRe\mtr\left( \big( \pd_{\beta} \hat P_a \big)  \hat Q_a \hat V^{\mu}  \big( \hat Q_{\sa} - \hat P_{\sa} \big) \big( \pd_{\nu} \hat P_a \big) \right)  \\
    \chi_{\rm FS}^{(3)} & =   - \frac{e}2 \epsilon_{\mu\nu} \sum_a n_F'(E_a) m_a \oIm\mtr\left( \big( \pd_{\nu}  \hat P_a \big) \big( \hat Q_{\sa} - \hat P_{\sa} \big) V^{\mu} \right) \  . 
\end{align}
Here $a$ labels a band, and the summation over the Brillouin zone is implicit in the Fermi surface expressions. We use $\hat P_a = | u_a \rangle \langle u_a |$ and $\hat Q_a = 1 - \hat P_a$; the latter includes all other bands, whereas $\hat Q_{\sa}$ contains only the remaining empty states. The orbital magnetic moment of $| u_a  \rangle$ is
\begin{align}
    m_a = \frac{e}{2 i \hbar } \epsilon_{\mu \nu} \langle \pd_{\mu} u_a | (\hat H - E_a) | \pd_{\nu} u_a \rangle 
\end{align}
Since every term in $\chi_{\rm MCD}^{\rm FS}$ is proportional to $n_F'(E_a)$, only states at the Fermi surface contribute as $T \rightarrow 0$. 

Interestingly, all three Fermi surface contributions contain the common structure $\hat Q_{\sa}-\hat P_{\sa}$, which is absent from the thermodynamical orbital susceptibility $\chi_{\rm orb}$~\cite{sm} and reflects the difference between interband processes involving empty and occupied states. When $E_a - \mu < 0$, the state $a$ is occupied and the relevant processes connect it to the empty states described by $\hat Q_{\sa}$. In contrast, when $E_a > \mu$, the state $a$ is empty and the relevant processes connect the occupied states described by $\hat P_{\sa}$ to the state $a$. The opposite signs of these two processes give rise to the combination $\hat Q_{\sa} -\hat P_{\sa}$ in $\chi_{\rm MCD}^{\rm FS}$.

Near band extrema and van Hove saddle points, the explicit velocity factor $\pd_\alpha E_a$  suppresses the local contributions to $\chi_{\rm FS}^{(1)}$ and $\chi_{\rm FS}^{(2)}$, whereas $\chi_{\rm FS}^{(3)}$ has no such factor. In a two-band system, the trace factor in $\chi_{\rm FS}^{(3)}$ is proportional to $m_a$, yielding an exact $m_a^2$ dependence~\cite{sm}. In a multiband system, replacing $\hat Q_{\sa}-\hat P_{\sa}$ by $\hat Q_{\sa}+\hat P_{\sa}$ and contracting with $\epsilon_{\mu\nu}$ likewise gives a quantity proportional to $m_a$, suggesting an approximate $m_a^2$ scaling. Since the orbital magnetic moment can be strongly enhanced near a small band gap, this contribution can become particularly important near such stationary points.

Before turning to the benchmark and application, we briefly discuss the chemical potential sum rule and related corrections. For a bounded band dispersion, $\chi_{\rm MCD}(\mu)$ vanishes when all states are empty or occupied, including the limits $\mu\rightarrow\pm\infty$. Unlike the thermodynamic orbital susceptibility $\chi_{\rm orb}$~\cite{StauberPRL2011, StauberPRB2016, RaouxPRL2014, RaouxPRB2015}, its chemical potential integral does not generally vanish, i.e.~$\int_{-\infty}^{\infty}\rmd\mu\,\chi_{\rm MCD}(\mu) \neq 0$. Additional symmetries,  such as particle hole symmetry, can enforce a vanishing integral. Numerical integration over the bounded spectrum of a particle hole asymmetric two-band QWZ model gives a finite result, explicitly demonstrating the absence of a generic sum rule~\cite{sm}.

Our main discussion focuses on the orbital response of the integrated MCD signal to a magnetic field, which enters through the vector potential. For completeness, we also derive in the SM the additional correction arising from the Zeeman coupling~\cite{sm}. This contribution vanishes for the models considered below because the two spin sectors are degenerate and give equal contributions with opposite signs. In the SM, we also present a projector based expression for the zero temperature thermodynamic orbital susceptibility and compare it with $\chi_{\rm MCD}$ in both insulating and metallic regimes~\cite{sm}.

\section{Benchmark and Application}
To test the general expressions derived above, we consider a two-dimensional massive Dirac model whose Landau level energies and wavefunctions are analytically solved for a perpendicular magnetic field~\cite{sm}. This allows an independent calculation of $M_{\rm MCD}(B)$, whose zero field intercept and weak field slope can be compared with $M_{\rm MCD}(0)$ and $\chi_{\rm MCD}$, respectively.

\begin{figure}[htb]
    \centering
    \includegraphics[width=0.9\columnwidth]{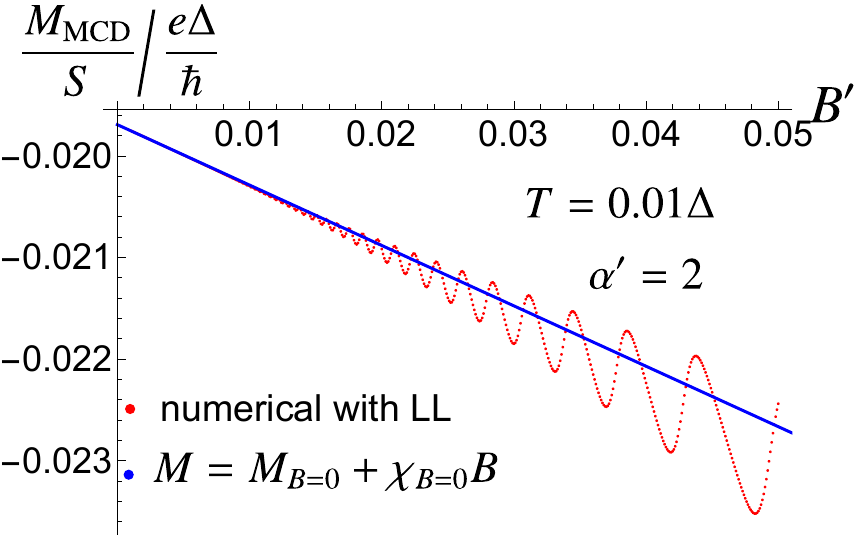}
    \caption{$M_{\rm MCD}$ per unit area as a function of magnetic field for the two-dimensional massive Dirac model at $\mu = 0$. The dimensionless model parameter is $\alpha' = \alpha \Delta / \beta^2 = 2$, and $B'= e B \beta^2/ (\hbar\Delta^2)$ is the dimensionless field. The red curve is the regular interband Landau level result obtained at $T = 0.01 \Delta$~\cite{sm}. The blue line is $M_{\rm MCD}(0) + \chi_{\rm MCD}B$, with the expressions for $M_{\rm MCD}(0)$ and $\chi_{\rm MCD}$ given by Eqs.~\eqref{Eqn:MMCDZeroT} and \eqref{Eqn:ChiMCD}, respectively. Quantum oscillations become more pronounced at larger $B'$.}
    \label{Fig:MCDDirac}
\end{figure}

The model is described by $H(\fvec k) = \begin{pmatrix} \alpha k^2 + \Delta & \beta k_- \\ \beta k_+ & \alpha k^2-\Delta\end{pmatrix}$, where $k_\pm = k_x \pm i k_y$. Its spectrum is bounded from below. Consistent with Eq.~\eqref{Eqn:MMCDDef}, we keep only optical transitions between the two Landau-level branches associated with the different zero field bands. Figure~\ref{Fig:MCDDirac} compares this regular integrated response, evaluated at $T = 0.01 \Delta$, with the linear expression $M_{\rm MCD}(0) + \chi_{\rm MCD} B$. The agreement at weak field benchmarks the zero field moment and linear response, while quantum oscillations become more pronounced at larger fields.

\begin{figure*}[t]
    \centering
    \includegraphics[width=\textwidth]{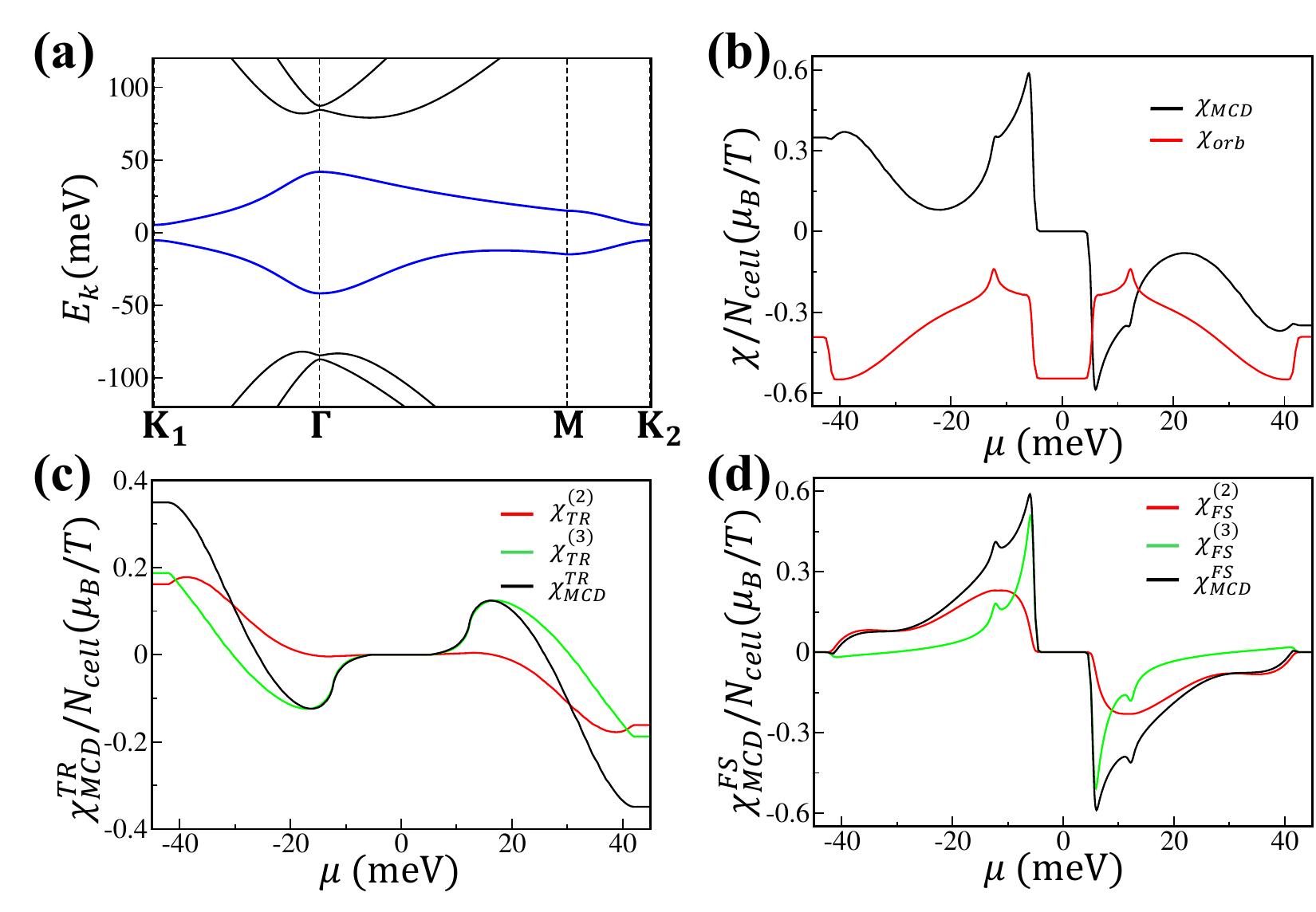}
    \caption{Electronic structure and magnetic responses of twisted bilayer graphene at $\theta = 1.4^\circ$. (a) Low energy band dispersion with the two central bands highlighted in blue and neighboring remote bands in black. (b) MCD susceptibility $\chi_{\rm MCD}$ compared with the thermodynamic orbital susceptibility $\chi_{\rm orb}$. (c) Fermi sea contributions $\chi_{\rm TR}^{(2)}$ and $\chi_{\rm TR}^{(3)}$, and their sum $\chi_{\rm MCD}^{\rm TR}$. (d) Fermi surface contributions $\chi_{\rm FS}^{(2)}$ and $\chi_{\rm FS}^{(3)}$, and their sum $\chi_{\rm MCD}^{\rm FS}$. The first contributions vanish, $\chi_{\rm TR}^{(1)}=\chi_{\rm FS}^{(1)}=0$, for the implemented continuum Hamiltonian with $\pd^2_{\alpha \beta} H = 0$. Responses are calculated on a $120\times120$ momentum mesh with Gaussian Fermi surface broadening $\eta = 0.5$~meV and $n_{\rm cut,TR}=n_{\rm cut,FS}=100$~\cite{sm}, and are given per moir\'e cell for one spin and one valley.}
    \label{Fig:TBG_responses}
\end{figure*}

We now examine the distinct orbital processes reflected in $\chi_{\rm MCD}$ using the noninteracting Bistritzer-MacDonald (BM) continuum model for twisted bilayer graphene (TBG)~\cite{mac_2011}, supplemented by an hBN-induced sublattice potential. We take the twist angle $\theta=1.4^\circ$, intra- and inter-sublattice tunneling amplitudes $u_0=80$~meV and $u_1=110$~meV, and a staggered sublattice potential $\Delta\sigma_z$ with $\Delta=10$~meV on both layers. The small layer-dependent rotations of the Dirac matrices are neglected. The Hamiltonian, conventions, and numerical details are given in the SM~\cite{sm}. Both $\chi_{\rm MCD}$ and $\chi_{\rm orb}$ are evaluated at fixed chemical potential and reported per moir\'e cell for one spin and one valley.

The clearest distinction between $\chi_{\rm MCD}$ and $\chi_{\rm orb}$ appears in the gap at charge neutrality point (CNP). Fig.~\ref{Fig:TBG_responses}(a) shows the two central moir\'e bands, separated by a gap of approximately $10$~meV, and neighboring remote bands. 
In the BM model, particle hole symmetry remains preserved even in the presence of the staggered sublattice potential, but the susceptibilities have different parity: $\chi_{\rm MCD}(-\mu)=-\chi_{\rm MCD}(\mu)$ and $\chi_{\rm orb}(-\mu)=\chi_{\rm orb}(\mu)$. At zero temperature, the occupied subspace is unchanged as $\mu$ varies within the gap, so $\chi_{\rm MCD}$ is constant there; its odd parity forces this constant to vanish. A sizable diamagnetic $\chi_{\rm orb}$ remains, as expected for gapped Dirac systems. Outside the CNP gap, Fig.~\ref{Fig:TBG_responses}(b) shows that $\chi_{\rm MCD}$ and $\chi_{\rm orb}$ can have either the same
or opposite signs, depending on the chemical potential. Fig.~\ref{Fig:TBG_responses}(b) illustrates this contrast using the finite Fermi surface broadening $\eta=0.5$~meV~\cite{sm}, which smooths the behavior near the gap edges.

Away from the CNP gap, the decomposition of $\chi_{\rm MCD}$ shows that the prominent structures near the band edges and van Hove energies have different microscopic origins. 
Since the Hamiltonian is linear in momentum, $\pd_{\alpha\beta}^2 H = 0$, and both inverse mass contributions $\chi_{\rm TR}^{(1)}$ and $\chi_{\rm FS}^{(1)}$ vanish. The remaining Fermi sea terms, $\chi_{\rm TR}^{(2)}$ and $\chi_{\rm TR}^{(3)}$, are shown separately in Fig.~\ref{Fig:TBG_responses}(c). They describe the Berry curvature related change of the occupied and empty subspaces and field-induced interband mixing, respectively. 

The strongest peaks just outside the CNP gap are dominated by the orbital magnetic moment Fermi surface terms. As illustrated in Fig.~\ref{Fig:TBG_responses}(d), near $\mu = \pm6$~meV, $\chi_{\rm FS}^{(3)}$ accounts for approximately $85\%$ of the total Fermi surface response, while $\chi_{\rm MCD}^{\rm TR}$ is comparatively small.  
The small band velocity near the gap edge extrema suppresses the explicit velocity factor in $\chi_{\rm FS}^{(2)}$. By contrast, $\chi_{\rm FS}^{(3)}$ contains the orbital moment, enhanced by small interband gaps, leading to a large MCD response even when the band dispersion is weak.

The smaller features near $\mu\simeq\pm12$~meV occur near saddle points of the central moir\'e bands, where the enhanced density of states amplifies the Fermi surface response. The two energy scales therefore emphasize different dominant enhancement factors within the response: a large orbital moment weight near the gap edges and a large density of states near the van Hove singularities (vHs). These factors need not be mutually exclusive.

The numerical parameters and the band-cutoff and momentum-mesh dependence of $\chi_{\rm MCD}^{\rm TR}$ are discussed in the SM~\cite{sm}. Taken together, the TBG results demonstrate that the integrated MCD response resolves a combination of interband and Fermi surface processes distinct from the thermodynamic orbital response, even for the same noninteracting bands.

\section{Discussion and outlook}
We have established how the orbital moment encoded in frequency integrated magnetic circular dichroism responds to a weak magnetic field. This response is distinct from the thermodynamic orbital susceptibility. It arises from both the explicit field dependence of the AB loop phase and field induced changes of the electronic states. Our EM gauge invariant formulation expresses the smooth weak field response entirely using zero-field electronic structure. For insulators, $\chi_{\rm MCD}$ separates into contributions associated with the quantum metric, Berry curvature, and field-induced interband mixing. Metals acquire additional Fermi surface terms that encode the field dependence of the occupations. These terms contain information distinct from that entering the thermodynamic orbital susceptibility. The massive Dirac benchmark compares the zero-field moment and its linear response with an independent Landau level calculation.

The application to TBG further illustrates the physical content of $\chi_{\rm MCD}$ in a multiband moir\'e system. Although $\chi_{\rm MCD}$ and $\chi_{\rm orb}$ are evaluated from the same electronic structure, they exhibit qualitatively different dependence on chemical potential, most clearly inside the charge neutrality gap. The decomposition of $\chi_{\rm MCD}$ further distinguishes interband geometric contributions from Fermi surface enhancements associated with large orbital moments near band edges and the density of states near saddle points. Integrated MCD thus provides a complementary probe of orbital magnetic response, sensitive to quantum geometry, orbital moments, and Fermi surface structure.

\section*{ACKNOWLEDGMENTS}
We thank W.-Y.~He, R.-W.~Jiang, Y.-Y.~Zhao, and Y.-Z.~Zhao for helpful discussions. J.~K.~acknowledges the support from the NSFC Grant No.~12574159, and the start-up grant of ShanghaiTech University. YZ is supported in part by the Shanghai Science and Technology Innovation Action Plan (Grant No. 24LZ1400800) and National Natural Science Foundation of China (NSFC) Grants No.~12274279 and No.~12674186. ZW is supported by the U.S. Department of Energy, Basic Energy Sciences, Grant No. DE FG02-99ER45747.

\bibliography{MCDChi}

\begin{thebibliography}{73}%
\makeatletter
\providecommand \@ifxundefined [1]{%
 \@ifx{#1\undefined}
}%
\providecommand \@ifnum [1]{%
 \ifnum #1\expandafter \@firstoftwo
 \else \expandafter \@secondoftwo
 \fi
}%
\providecommand \@ifx [1]{%
 \ifx #1\expandafter \@firstoftwo
 \else \expandafter \@secondoftwo
 \fi
}%
\providecommand \natexlab [1]{#1}%
\providecommand \enquote  [1]{``#1''}%
\providecommand \bibnamefont  [1]{#1}%
\providecommand \bibfnamefont [1]{#1}%
\providecommand \citenamefont [1]{#1}%
\providecommand \href@noop [0]{\@secondoftwo}%
\providecommand \href [0]{\begingroup \@sanitize@url \@href}%
\providecommand \@href[1]{\@@startlink{#1}\@@href}%
\providecommand \@@href[1]{\endgroup#1\@@endlink}%
\providecommand \@sanitize@url [0]{\catcode `\\12\catcode `\$12\catcode
  `\&12\catcode `\#12\catcode `\^12\catcode `\_12\catcode `\%12\relax}%
\providecommand \@@startlink[1]{}%
\providecommand \@@endlink[0]{}%
\providecommand \url  [0]{\begingroup\@sanitize@url \@url }%
\providecommand \@url [1]{\endgroup\@href {#1}{\urlprefix }}%
\providecommand \urlprefix  [0]{URL }%
\providecommand \Eprint [0]{\href }%
\providecommand \doibase [0]{https://doi.org/}%
\providecommand \selectlanguage [0]{\@gobble}%
\providecommand \bibinfo  [0]{\@secondoftwo}%
\providecommand \bibfield  [0]{\@secondoftwo}%
\providecommand \translation [1]{[#1]}%
\providecommand \BibitemOpen [0]{}%
\providecommand \bibitemStop [0]{}%
\providecommand \bibitemNoStop [0]{.\EOS\space}%
\providecommand \EOS [0]{\spacefactor3000\relax}%
\providecommand \BibitemShut  [1]{\csname bibitem#1\endcsname}%
\let\auto@bib@innerbib\@empty
\bibitem [{\citenamefont {Xiao}\ \emph {et~al.}(2010)\citenamefont {Xiao},
  \citenamefont {Chang},\ and\ \citenamefont {Niu}}]{NiuRMP2010}%
  \BibitemOpen
  \bibfield  {author} {\bibinfo {author} {\bibfnamefont {D.}~\bibnamefont
  {Xiao}}, \bibinfo {author} {\bibfnamefont {M.-C.}\ \bibnamefont {Chang}},\
  and\ \bibinfo {author} {\bibfnamefont {Q.}~\bibnamefont {Niu}},\ }\bibfield
  {title} {\bibinfo {title} {Berry phase effects on electronic properties},\
  }\href {https://doi.org/10.1103/RevModPhys.82.1959} {\bibfield  {journal}
  {\bibinfo  {journal} {Rev. Mod. Phys.}\ }\textbf {\bibinfo {volume} {82}},\
  \bibinfo {pages} {1959} (\bibinfo {year} {2010})}\BibitemShut {NoStop}%
\bibitem [{\citenamefont {Liu}\ and\ \citenamefont {Dai}(2021)}]{DaiNRP2021}%
  \BibitemOpen
  \bibfield  {author} {\bibinfo {author} {\bibfnamefont {J.}~\bibnamefont
  {Liu}}\ and\ \bibinfo {author} {\bibfnamefont {X.}~\bibnamefont {Dai}},\
  }\bibfield  {title} {\bibinfo {title} {Orbital magnetic states in moir{\'e}
  graphene systems},\ }\href {https://doi.org/10.1038/s42254-021-00297-3}
  {\bibfield  {journal} {\bibinfo  {journal} {Nature Reviews Physics}\ }\textbf
  {\bibinfo {volume} {3}},\ \bibinfo {pages} {367} (\bibinfo {year}
  {2021})}\BibitemShut {NoStop}%
\bibitem [{\citenamefont {Haldane}(1988)}]{HaldanePRL1988}%
  \BibitemOpen
  \bibfield  {author} {\bibinfo {author} {\bibfnamefont {F.~D.~M.}\
  \bibnamefont {Haldane}},\ }\bibfield  {title} {\bibinfo {title} {Model for a
  quantum hall effect without landau levels: Condensed-matter realization of
  the "parity anomaly"},\ }\href {https://doi.org/10.1103/PhysRevLett.61.2015}
  {\bibfield  {journal} {\bibinfo  {journal} {Phys. Rev. Lett.}\ }\textbf
  {\bibinfo {volume} {61}},\ \bibinfo {pages} {2015} (\bibinfo {year}
  {1988})}\BibitemShut {NoStop}%
\bibitem [{\citenamefont {Cao}\ \emph {et~al.}(2018{\natexlab{a}})\citenamefont
  {Cao}, \citenamefont {Fatemi}, \citenamefont {Demir}, \citenamefont {Fang},
  \citenamefont {Tomarken}, \citenamefont {Luo}, \citenamefont
  {Sanchez-Yamagishi}, \citenamefont {Watanabe}, \citenamefont {Taniguchi},
  \citenamefont {Kaxiras}, \citenamefont {Ashoori},\ and\ \citenamefont
  {Jarillo-Herrero}}]{PabloNature2018_1}%
  \BibitemOpen
  \bibfield  {author} {\bibinfo {author} {\bibfnamefont {Y.}~\bibnamefont
  {Cao}}, \bibinfo {author} {\bibfnamefont {V.}~\bibnamefont {Fatemi}},
  \bibinfo {author} {\bibfnamefont {A.}~\bibnamefont {Demir}}, \bibinfo
  {author} {\bibfnamefont {S.}~\bibnamefont {Fang}}, \bibinfo {author}
  {\bibfnamefont {S.~L.}\ \bibnamefont {Tomarken}}, \bibinfo {author}
  {\bibfnamefont {J.~Y.}\ \bibnamefont {Luo}}, \bibinfo {author} {\bibfnamefont
  {J.~D.}\ \bibnamefont {Sanchez-Yamagishi}}, \bibinfo {author} {\bibfnamefont
  {K.}~\bibnamefont {Watanabe}}, \bibinfo {author} {\bibfnamefont
  {T.}~\bibnamefont {Taniguchi}}, \bibinfo {author} {\bibfnamefont
  {E.}~\bibnamefont {Kaxiras}}, \bibinfo {author} {\bibfnamefont {R.~C.}\
  \bibnamefont {Ashoori}},\ and\ \bibinfo {author} {\bibfnamefont
  {P.}~\bibnamefont {Jarillo-Herrero}},\ }\bibfield  {title} {\bibinfo {title}
  {Correlated insulator behaviour at half-filling in magic-angle graphene
  superlattices},\ }\href {https://doi.org/10.1038/nature26154} {\bibfield
  {journal} {\bibinfo  {journal} {Nature}\ }\textbf {\bibinfo {volume} {556}},\
  \bibinfo {pages} {80} (\bibinfo {year} {2018}{\natexlab{a}})}\BibitemShut
  {NoStop}%
\bibitem [{\citenamefont {Cao}\ \emph {et~al.}(2018{\natexlab{b}})\citenamefont
  {Cao}, \citenamefont {Fatemi}, \citenamefont {Fang}, \citenamefont
  {Watanabe}, \citenamefont {Taniguchi}, \citenamefont {Kaxiras},\ and\
  \citenamefont {Jarillo-Herrero}}]{PabloNature2018_2}%
  \BibitemOpen
  \bibfield  {author} {\bibinfo {author} {\bibfnamefont {Y.}~\bibnamefont
  {Cao}}, \bibinfo {author} {\bibfnamefont {V.}~\bibnamefont {Fatemi}},
  \bibinfo {author} {\bibfnamefont {S.}~\bibnamefont {Fang}}, \bibinfo {author}
  {\bibfnamefont {K.}~\bibnamefont {Watanabe}}, \bibinfo {author}
  {\bibfnamefont {T.}~\bibnamefont {Taniguchi}}, \bibinfo {author}
  {\bibfnamefont {E.}~\bibnamefont {Kaxiras}},\ and\ \bibinfo {author}
  {\bibfnamefont {P.}~\bibnamefont {Jarillo-Herrero}},\ }\bibfield  {title}
  {\bibinfo {title} {Unconventional superconductivity in magic-angle graphene
  superlattices},\ }\href {https://doi.org/10.1038/nature26160} {\bibfield
  {journal} {\bibinfo  {journal} {Nature}\ }\textbf {\bibinfo {volume} {556}},\
  \bibinfo {pages} {43} (\bibinfo {year} {2018}{\natexlab{b}})}\BibitemShut
  {NoStop}%
\bibitem [{\citenamefont {Sharpe}\ \emph {et~al.}(2019)\citenamefont {Sharpe},
  \citenamefont {Fox}, \citenamefont {Barnard}, \citenamefont {Finney},
  \citenamefont {Watanabe}, \citenamefont {Taniguchi}, \citenamefont
  {Kastner},\ and\ \citenamefont {Goldhaber-Gordon}}]{DavidGGScience2019}%
  \BibitemOpen
  \bibfield  {author} {\bibinfo {author} {\bibfnamefont {A.~L.}\ \bibnamefont
  {Sharpe}}, \bibinfo {author} {\bibfnamefont {E.~J.}\ \bibnamefont {Fox}},
  \bibinfo {author} {\bibfnamefont {A.~W.}\ \bibnamefont {Barnard}}, \bibinfo
  {author} {\bibfnamefont {J.}~\bibnamefont {Finney}}, \bibinfo {author}
  {\bibfnamefont {K.}~\bibnamefont {Watanabe}}, \bibinfo {author}
  {\bibfnamefont {T.}~\bibnamefont {Taniguchi}}, \bibinfo {author}
  {\bibfnamefont {M.~A.}\ \bibnamefont {Kastner}},\ and\ \bibinfo {author}
  {\bibfnamefont {D.}~\bibnamefont {Goldhaber-Gordon}},\ }\bibfield  {title}
  {\bibinfo {title} {Emergent ferromagnetism near three-quarters filling in
  twisted bilayer graphene},\ }\href {https://doi.org/10.1126/science.aaw3780}
  {\bibfield  {journal} {\bibinfo  {journal} {Science}\ }\textbf {\bibinfo
  {volume} {365}},\ \bibinfo {pages} {605} (\bibinfo {year}
  {2019})}\BibitemShut {NoStop}%
\bibitem [{\citenamefont {Chen}\ \emph {et~al.}(2020)\citenamefont {Chen},
  \citenamefont {Sharpe}, \citenamefont {Fox}, \citenamefont {Zhang},
  \citenamefont {Wang}, \citenamefont {Jiang}, \citenamefont {Lyu},
  \citenamefont {Li}, \citenamefont {Watanabe}, \citenamefont {Taniguchi},
  \citenamefont {Shi}, \citenamefont {Senthil}, \citenamefont
  {Goldhaber-Gordon}, \citenamefont {Zhang},\ and\ \citenamefont
  {Wang}}]{WangNature2020}%
  \BibitemOpen
  \bibfield  {author} {\bibinfo {author} {\bibfnamefont {G.}~\bibnamefont
  {Chen}}, \bibinfo {author} {\bibfnamefont {A.~L.}\ \bibnamefont {Sharpe}},
  \bibinfo {author} {\bibfnamefont {E.~J.}\ \bibnamefont {Fox}}, \bibinfo
  {author} {\bibfnamefont {Y.-H.}\ \bibnamefont {Zhang}}, \bibinfo {author}
  {\bibfnamefont {S.}~\bibnamefont {Wang}}, \bibinfo {author} {\bibfnamefont
  {L.}~\bibnamefont {Jiang}}, \bibinfo {author} {\bibfnamefont
  {B.}~\bibnamefont {Lyu}}, \bibinfo {author} {\bibfnamefont {H.}~\bibnamefont
  {Li}}, \bibinfo {author} {\bibfnamefont {K.}~\bibnamefont {Watanabe}},
  \bibinfo {author} {\bibfnamefont {T.}~\bibnamefont {Taniguchi}}, \bibinfo
  {author} {\bibfnamefont {Z.}~\bibnamefont {Shi}}, \bibinfo {author}
  {\bibfnamefont {T.}~\bibnamefont {Senthil}}, \bibinfo {author} {\bibfnamefont
  {D.}~\bibnamefont {Goldhaber-Gordon}}, \bibinfo {author} {\bibfnamefont
  {Y.}~\bibnamefont {Zhang}},\ and\ \bibinfo {author} {\bibfnamefont
  {F.}~\bibnamefont {Wang}},\ }\bibfield  {title} {\bibinfo {title} {Tunable
  correlated chern insulator and ferromagnetism in a moir{\'e} superlattice},\
  }\href {https://doi.org/10.1038/s41586-020-2049-7} {\bibfield  {journal}
  {\bibinfo  {journal} {Nature}\ }\textbf {\bibinfo {volume} {579}},\ \bibinfo
  {pages} {56} (\bibinfo {year} {2020})}\BibitemShut {NoStop}%
\bibitem [{\citenamefont {Nuckolls}\ \emph {et~al.}(2020)\citenamefont
  {Nuckolls}, \citenamefont {Oh}, \citenamefont {Wong}, \citenamefont {Lian},
  \citenamefont {Watanabe}, \citenamefont {Taniguchi}, \citenamefont
  {Bernevig},\ and\ \citenamefont {Yazdani}}]{YazdaniNature2020}%
  \BibitemOpen
  \bibfield  {author} {\bibinfo {author} {\bibfnamefont {K.~P.}\ \bibnamefont
  {Nuckolls}}, \bibinfo {author} {\bibfnamefont {M.}~\bibnamefont {Oh}},
  \bibinfo {author} {\bibfnamefont {D.}~\bibnamefont {Wong}}, \bibinfo {author}
  {\bibfnamefont {B.}~\bibnamefont {Lian}}, \bibinfo {author} {\bibfnamefont
  {K.}~\bibnamefont {Watanabe}}, \bibinfo {author} {\bibfnamefont
  {T.}~\bibnamefont {Taniguchi}}, \bibinfo {author} {\bibfnamefont {B.~A.}\
  \bibnamefont {Bernevig}},\ and\ \bibinfo {author} {\bibfnamefont
  {A.}~\bibnamefont {Yazdani}},\ }\bibfield  {title} {\bibinfo {title}
  {Strongly correlated chern insulators in magic-angle twisted bilayer
  graphene},\ }\href {https://doi.org/10.1038/s41586-020-3028-8} {\bibfield
  {journal} {\bibinfo  {journal} {Nature}\ }\textbf {\bibinfo {volume} {588}},\
  \bibinfo {pages} {610} (\bibinfo {year} {2020})}\BibitemShut {NoStop}%
\bibitem [{\citenamefont {He}\ \emph {et~al.}(2021)\citenamefont {He},
  \citenamefont {Zhang}, \citenamefont {Li}, \citenamefont {Fei}, \citenamefont
  {Watanabe}, \citenamefont {Taniguchi}, \citenamefont {Xu},\ and\
  \citenamefont {Yankowitz}}]{YankowitzNC2021}%
  \BibitemOpen
  \bibfield  {author} {\bibinfo {author} {\bibfnamefont {M.}~\bibnamefont
  {He}}, \bibinfo {author} {\bibfnamefont {Y.-H.}\ \bibnamefont {Zhang}},
  \bibinfo {author} {\bibfnamefont {Y.}~\bibnamefont {Li}}, \bibinfo {author}
  {\bibfnamefont {Z.}~\bibnamefont {Fei}}, \bibinfo {author} {\bibfnamefont
  {K.}~\bibnamefont {Watanabe}}, \bibinfo {author} {\bibfnamefont
  {T.}~\bibnamefont {Taniguchi}}, \bibinfo {author} {\bibfnamefont
  {X.}~\bibnamefont {Xu}},\ and\ \bibinfo {author} {\bibfnamefont
  {M.}~\bibnamefont {Yankowitz}},\ }\bibfield  {title} {\bibinfo {title}
  {Competing correlated states and abundant orbital magnetism in twisted
  monolayer-bilayer graphene},\ }\href
  {https://doi.org/10.1038/s41467-021-25044-1} {\bibfield  {journal} {\bibinfo
  {journal} {Nature Communications}\ }\textbf {\bibinfo {volume} {12}},\
  \bibinfo {pages} {4727} (\bibinfo {year} {2021})}\BibitemShut {NoStop}%
\bibitem [{\citenamefont {Polshyn}\ \emph {et~al.}(2020)\citenamefont
  {Polshyn}, \citenamefont {Zhu}, \citenamefont {Kumar}, \citenamefont {Zhang},
  \citenamefont {Yang}, \citenamefont {Tschirhart}, \citenamefont {Serlin},
  \citenamefont {Watanabe}, \citenamefont {Taniguchi}, \citenamefont
  {MacDonald},\ and\ \citenamefont {Young}}]{YoungNature2020}%
  \BibitemOpen
  \bibfield  {author} {\bibinfo {author} {\bibfnamefont {H.}~\bibnamefont
  {Polshyn}}, \bibinfo {author} {\bibfnamefont {J.}~\bibnamefont {Zhu}},
  \bibinfo {author} {\bibfnamefont {M.~A.}\ \bibnamefont {Kumar}}, \bibinfo
  {author} {\bibfnamefont {Y.}~\bibnamefont {Zhang}}, \bibinfo {author}
  {\bibfnamefont {F.}~\bibnamefont {Yang}}, \bibinfo {author} {\bibfnamefont
  {C.~L.}\ \bibnamefont {Tschirhart}}, \bibinfo {author} {\bibfnamefont
  {M.}~\bibnamefont {Serlin}}, \bibinfo {author} {\bibfnamefont
  {K.}~\bibnamefont {Watanabe}}, \bibinfo {author} {\bibfnamefont
  {T.}~\bibnamefont {Taniguchi}}, \bibinfo {author} {\bibfnamefont {A.~H.}\
  \bibnamefont {MacDonald}},\ and\ \bibinfo {author} {\bibfnamefont {A.~F.}\
  \bibnamefont {Young}},\ }\bibfield  {title} {\bibinfo {title} {Electrical
  switching of magnetic order in an orbital chern insulator},\ }\href
  {https://doi.org/10.1038/s41586-020-2963-8} {\bibfield  {journal} {\bibinfo
  {journal} {Nature}\ }\textbf {\bibinfo {volume} {588}},\ \bibinfo {pages}
  {66} (\bibinfo {year} {2020})}\BibitemShut {NoStop}%
\bibitem [{\citenamefont {Zeng}\ \emph {et~al.}(2023)\citenamefont {Zeng},
  \citenamefont {Xia}, \citenamefont {Kang}, \citenamefont {Zhu}, \citenamefont
  {Kn{\"u}ppel}, \citenamefont {Vaswani}, \citenamefont {Watanabe},
  \citenamefont {Taniguchi}, \citenamefont {Mak},\ and\ \citenamefont
  {Shan}}]{KinFaiMakNature2023}%
  \BibitemOpen
  \bibfield  {author} {\bibinfo {author} {\bibfnamefont {Y.}~\bibnamefont
  {Zeng}}, \bibinfo {author} {\bibfnamefont {Z.}~\bibnamefont {Xia}}, \bibinfo
  {author} {\bibfnamefont {K.}~\bibnamefont {Kang}}, \bibinfo {author}
  {\bibfnamefont {J.}~\bibnamefont {Zhu}}, \bibinfo {author} {\bibfnamefont
  {P.}~\bibnamefont {Kn{\"u}ppel}}, \bibinfo {author} {\bibfnamefont
  {C.}~\bibnamefont {Vaswani}}, \bibinfo {author} {\bibfnamefont
  {K.}~\bibnamefont {Watanabe}}, \bibinfo {author} {\bibfnamefont
  {T.}~\bibnamefont {Taniguchi}}, \bibinfo {author} {\bibfnamefont {K.~F.}\
  \bibnamefont {Mak}},\ and\ \bibinfo {author} {\bibfnamefont {J.}~\bibnamefont
  {Shan}},\ }\bibfield  {title} {\bibinfo {title} {Thermodynamic evidence of
  fractional chern insulator in moir{\'e} mote2},\ }\href
  {https://doi.org/10.1038/s41586-023-06452-3} {\bibfield  {journal} {\bibinfo
  {journal} {Nature}\ }\textbf {\bibinfo {volume} {622}},\ \bibinfo {pages}
  {69} (\bibinfo {year} {2023})}\BibitemShut {NoStop}%
\bibitem [{\citenamefont {Redekop}\ \emph {et~al.}(2024)\citenamefont
  {Redekop}, \citenamefont {Zhang}, \citenamefont {Park}, \citenamefont {Cai},
  \citenamefont {Anderson}, \citenamefont {Sheekey}, \citenamefont {Arp},
  \citenamefont {Babikyan}, \citenamefont {Salters}, \citenamefont {Watanabe},
  \citenamefont {Taniguchi}, \citenamefont {Huber}, \citenamefont {Xu},\ and\
  \citenamefont {Young}}]{YoungNature2024}%
  \BibitemOpen
  \bibfield  {author} {\bibinfo {author} {\bibfnamefont {E.}~\bibnamefont
  {Redekop}}, \bibinfo {author} {\bibfnamefont {C.}~\bibnamefont {Zhang}},
  \bibinfo {author} {\bibfnamefont {H.}~\bibnamefont {Park}}, \bibinfo {author}
  {\bibfnamefont {J.}~\bibnamefont {Cai}}, \bibinfo {author} {\bibfnamefont
  {E.}~\bibnamefont {Anderson}}, \bibinfo {author} {\bibfnamefont
  {O.}~\bibnamefont {Sheekey}}, \bibinfo {author} {\bibfnamefont
  {T.}~\bibnamefont {Arp}}, \bibinfo {author} {\bibfnamefont {G.}~\bibnamefont
  {Babikyan}}, \bibinfo {author} {\bibfnamefont {S.}~\bibnamefont {Salters}},
  \bibinfo {author} {\bibfnamefont {K.}~\bibnamefont {Watanabe}}, \bibinfo
  {author} {\bibfnamefont {T.}~\bibnamefont {Taniguchi}}, \bibinfo {author}
  {\bibfnamefont {M.~E.}\ \bibnamefont {Huber}}, \bibinfo {author}
  {\bibfnamefont {X.}~\bibnamefont {Xu}},\ and\ \bibinfo {author}
  {\bibfnamefont {A.~F.}\ \bibnamefont {Young}},\ }\bibfield  {title} {\bibinfo
  {title} {Direct magnetic imaging of fractional chern insulators in twisted
  mote2},\ }\href {https://doi.org/10.1038/s41586-024-08153-x} {\bibfield
  {journal} {\bibinfo  {journal} {Nature}\ }\textbf {\bibinfo {volume} {635}},\
  \bibinfo {pages} {584} (\bibinfo {year} {2024})}\BibitemShut {NoStop}%
\bibitem [{\citenamefont {Lu}\ \emph {et~al.}(2019)\citenamefont {Lu},
  \citenamefont {Stepanov}, \citenamefont {Yang}, \citenamefont {Xie},
  \citenamefont {Aamir}, \citenamefont {Das}, \citenamefont {Urgell},
  \citenamefont {Watanabe}, \citenamefont {Taniguchi}, \citenamefont {Zhang},
  \citenamefont {Bachtold}, \citenamefont {MacDonald},\ and\ \citenamefont
  {Efetov}}]{Efretov2019}%
  \BibitemOpen
  \bibfield  {author} {\bibinfo {author} {\bibfnamefont {X.}~\bibnamefont
  {Lu}}, \bibinfo {author} {\bibfnamefont {P.}~\bibnamefont {Stepanov}},
  \bibinfo {author} {\bibfnamefont {W.}~\bibnamefont {Yang}}, \bibinfo {author}
  {\bibfnamefont {M.}~\bibnamefont {Xie}}, \bibinfo {author} {\bibfnamefont
  {M.~A.}\ \bibnamefont {Aamir}}, \bibinfo {author} {\bibfnamefont
  {I.}~\bibnamefont {Das}}, \bibinfo {author} {\bibfnamefont {C.}~\bibnamefont
  {Urgell}}, \bibinfo {author} {\bibfnamefont {K.}~\bibnamefont {Watanabe}},
  \bibinfo {author} {\bibfnamefont {T.}~\bibnamefont {Taniguchi}}, \bibinfo
  {author} {\bibfnamefont {G.}~\bibnamefont {Zhang}}, \bibinfo {author}
  {\bibfnamefont {A.}~\bibnamefont {Bachtold}}, \bibinfo {author}
  {\bibfnamefont {A.~H.}\ \bibnamefont {MacDonald}},\ and\ \bibinfo {author}
  {\bibfnamefont {D.~K.}\ \bibnamefont {Efetov}},\ }\bibfield  {title}
  {\bibinfo {title} {Superconductors, orbital magnets and correlated states in
  magic-angle bilayer graphene},\ }\href
  {https://doi.org/10.1038/s41586-019-1695-0} {\bibfield  {journal} {\bibinfo
  {journal} {Nature}\ }\textbf {\bibinfo {volume} {574}},\ \bibinfo {pages}
  {653} (\bibinfo {year} {2019})}\BibitemShut {NoStop}%
\bibitem [{\citenamefont {Song}\ \emph {et~al.}(2021)\citenamefont {Song},
  \citenamefont {Sun}, \citenamefont {Anderson}, \citenamefont {Wang},
  \citenamefont {Qian}, \citenamefont {Taniguchi}, \citenamefont {Watanabe},
  \citenamefont {McGuire}, \citenamefont {St\"{o}hr}, \citenamefont {Xiao},
  \citenamefont {Cao}, \citenamefont {Wrachtrup},\ and\ \citenamefont
  {Xu}}]{Xiaodong21}%
  \BibitemOpen
  \bibfield  {author} {\bibinfo {author} {\bibfnamefont {T.}~\bibnamefont
  {Song}}, \bibinfo {author} {\bibfnamefont {Q.-C.}\ \bibnamefont {Sun}},
  \bibinfo {author} {\bibfnamefont {E.}~\bibnamefont {Anderson}}, \bibinfo
  {author} {\bibfnamefont {C.}~\bibnamefont {Wang}}, \bibinfo {author}
  {\bibfnamefont {J.}~\bibnamefont {Qian}}, \bibinfo {author} {\bibfnamefont
  {T.}~\bibnamefont {Taniguchi}}, \bibinfo {author} {\bibfnamefont
  {K.}~\bibnamefont {Watanabe}}, \bibinfo {author} {\bibfnamefont {M.~A.}\
  \bibnamefont {McGuire}}, \bibinfo {author} {\bibfnamefont {R.}~\bibnamefont
  {St\"{o}hr}}, \bibinfo {author} {\bibfnamefont {D.}~\bibnamefont {Xiao}},
  \bibinfo {author} {\bibfnamefont {T.}~\bibnamefont {Cao}}, \bibinfo {author}
  {\bibfnamefont {J.}~\bibnamefont {Wrachtrup}},\ and\ \bibinfo {author}
  {\bibfnamefont {X.}~\bibnamefont {Xu}},\ }\bibfield  {title} {\bibinfo
  {title} {Direct visualization of magnetic domains and moir\'{e} magnetism in
  twisted 2d magnets},\ }\href {https://doi.org/10.1126/science.abj7478}
  {\bibfield  {journal} {\bibinfo  {journal} {Science}\ }\textbf {\bibinfo
  {volume} {374}},\ \bibinfo {pages} {1140} (\bibinfo {year}
  {2021})}\BibitemShut {NoStop}%
\bibitem [{\citenamefont {Han}\ \emph {et~al.}(2025)\citenamefont {Han},
  \citenamefont {Lu}, \citenamefont {Hadjri}, \citenamefont {Shi},
  \citenamefont {Wu}, \citenamefont {Xu}, \citenamefont {Yao}, \citenamefont
  {Cotten}, \citenamefont {Sharifi~Sedeh}, \citenamefont {Weldeyesus},
  \citenamefont {Yang}, \citenamefont {Seo}, \citenamefont {Ye}, \citenamefont
  {Zhou}, \citenamefont {Liu}, \citenamefont {Shi}, \citenamefont {Hua},
  \citenamefont {Watanabe}, \citenamefont {Taniguchi}, \citenamefont {Xiong},
  \citenamefont {Zumb{\"u}hl}, \citenamefont {Fu},\ and\ \citenamefont
  {Ju}}]{JuLongChiralSC}%
  \BibitemOpen
  \bibfield  {author} {\bibinfo {author} {\bibfnamefont {T.}~\bibnamefont
  {Han}}, \bibinfo {author} {\bibfnamefont {Z.}~\bibnamefont {Lu}}, \bibinfo
  {author} {\bibfnamefont {Z.}~\bibnamefont {Hadjri}}, \bibinfo {author}
  {\bibfnamefont {L.}~\bibnamefont {Shi}}, \bibinfo {author} {\bibfnamefont
  {Z.}~\bibnamefont {Wu}}, \bibinfo {author} {\bibfnamefont {W.}~\bibnamefont
  {Xu}}, \bibinfo {author} {\bibfnamefont {Y.}~\bibnamefont {Yao}}, \bibinfo
  {author} {\bibfnamefont {A.~A.}\ \bibnamefont {Cotten}}, \bibinfo {author}
  {\bibfnamefont {O.}~\bibnamefont {Sharifi~Sedeh}}, \bibinfo {author}
  {\bibfnamefont {H.}~\bibnamefont {Weldeyesus}}, \bibinfo {author}
  {\bibfnamefont {J.}~\bibnamefont {Yang}}, \bibinfo {author} {\bibfnamefont
  {J.}~\bibnamefont {Seo}}, \bibinfo {author} {\bibfnamefont {S.}~\bibnamefont
  {Ye}}, \bibinfo {author} {\bibfnamefont {M.}~\bibnamefont {Zhou}}, \bibinfo
  {author} {\bibfnamefont {H.}~\bibnamefont {Liu}}, \bibinfo {author}
  {\bibfnamefont {G.}~\bibnamefont {Shi}}, \bibinfo {author} {\bibfnamefont
  {Z.}~\bibnamefont {Hua}}, \bibinfo {author} {\bibfnamefont {K.}~\bibnamefont
  {Watanabe}}, \bibinfo {author} {\bibfnamefont {T.}~\bibnamefont {Taniguchi}},
  \bibinfo {author} {\bibfnamefont {P.}~\bibnamefont {Xiong}}, \bibinfo
  {author} {\bibfnamefont {D.~M.}\ \bibnamefont {Zumb{\"u}hl}}, \bibinfo
  {author} {\bibfnamefont {L.}~\bibnamefont {Fu}},\ and\ \bibinfo {author}
  {\bibfnamefont {L.}~\bibnamefont {Ju}},\ }\bibfield  {title} {\bibinfo
  {title} {Signatures of chiral superconductivity in rhombohedral graphene},\
  }\href {https://doi.org/10.1038/s41586-025-09169-7} {\bibfield  {journal}
  {\bibinfo  {journal} {Nature}\ }\textbf {\bibinfo {volume} {643}},\ \bibinfo
  {pages} {654} (\bibinfo {year} {2025})}\BibitemShut {NoStop}%
\bibitem [{\citenamefont {Yu}\ \emph {et~al.}(2022)\citenamefont {Yu},
  \citenamefont {Foutty}, \citenamefont {Han}, \citenamefont {Barber},
  \citenamefont {Schattner}, \citenamefont {Watanabe}, \citenamefont
  {Taniguchi}, \citenamefont {Phillips}, \citenamefont {Shen}, \citenamefont
  {Kivelson},\ and\ \citenamefont {Feldman}}]{FeldmanHofstadter}%
  \BibitemOpen
  \bibfield  {author} {\bibinfo {author} {\bibfnamefont {J.}~\bibnamefont
  {Yu}}, \bibinfo {author} {\bibfnamefont {B.~A.}\ \bibnamefont {Foutty}},
  \bibinfo {author} {\bibfnamefont {Z.}~\bibnamefont {Han}}, \bibinfo {author}
  {\bibfnamefont {M.~E.}\ \bibnamefont {Barber}}, \bibinfo {author}
  {\bibfnamefont {Y.}~\bibnamefont {Schattner}}, \bibinfo {author}
  {\bibfnamefont {K.}~\bibnamefont {Watanabe}}, \bibinfo {author}
  {\bibfnamefont {T.}~\bibnamefont {Taniguchi}}, \bibinfo {author}
  {\bibfnamefont {P.}~\bibnamefont {Phillips}}, \bibinfo {author}
  {\bibfnamefont {Z.-X.}\ \bibnamefont {Shen}}, \bibinfo {author}
  {\bibfnamefont {S.~A.}\ \bibnamefont {Kivelson}},\ and\ \bibinfo {author}
  {\bibfnamefont {B.~E.}\ \bibnamefont {Feldman}},\ }\bibfield  {title}
  {\bibinfo {title} {Correlated hofstadter spectrum and flavour phase diagram
  in magic-angle twisted bilayer graphene},\ }\href
  {https://doi.org/10.1038/s41567-022-01589-w} {\bibfield  {journal} {\bibinfo
  {journal} {Nature Physics}\ }\textbf {\bibinfo {volume} {18}},\ \bibinfo
  {pages} {825} (\bibinfo {year} {2022})}\BibitemShut {NoStop}%
\bibitem [{\citenamefont {Wu}\ \emph {et~al.}(2021)\citenamefont {Wu},
  \citenamefont {Zhang}, \citenamefont {Watanabe}, \citenamefont {Taniguchi},\
  and\ \citenamefont {Andrei}}]{wus_2021}%
  \BibitemOpen
  \bibfield  {author} {\bibinfo {author} {\bibfnamefont {S.}~\bibnamefont
  {Wu}}, \bibinfo {author} {\bibfnamefont {Z.}~\bibnamefont {Zhang}}, \bibinfo
  {author} {\bibfnamefont {K.}~\bibnamefont {Watanabe}}, \bibinfo {author}
  {\bibfnamefont {T.}~\bibnamefont {Taniguchi}},\ and\ \bibinfo {author}
  {\bibfnamefont {E.~Y.}\ \bibnamefont {Andrei}},\ }\bibfield  {title}
  {\bibinfo {title} {Chern insulators, van hove singularities and topological
  flat bands in magic-angle twisted bilayer graphene},\ }\href
  {https://doi.org/10.1038/s41563-020-00911-2} {\bibfield  {journal} {\bibinfo
  {journal} {Nature Materials}\ }\textbf {\bibinfo {volume} {20}},\ \bibinfo
  {pages} {488} (\bibinfo {year} {2021})}\BibitemShut {NoStop}%
\bibitem [{\citenamefont {Das}\ \emph {et~al.}(2021)\citenamefont {Das},
  \citenamefont {Lu}, \citenamefont {Herzog-Arbeitman}, \citenamefont {Song},
  \citenamefont {Watanabe}, \citenamefont {Taniguchi}, \citenamefont
  {Bernevig},\ and\ \citenamefont {Efetov}}]{das_2021}%
  \BibitemOpen
  \bibfield  {author} {\bibinfo {author} {\bibfnamefont {I.}~\bibnamefont
  {Das}}, \bibinfo {author} {\bibfnamefont {X.}~\bibnamefont {Lu}}, \bibinfo
  {author} {\bibfnamefont {J.}~\bibnamefont {Herzog-Arbeitman}}, \bibinfo
  {author} {\bibfnamefont {Z.-D.}\ \bibnamefont {Song}}, \bibinfo {author}
  {\bibfnamefont {K.}~\bibnamefont {Watanabe}}, \bibinfo {author}
  {\bibfnamefont {T.}~\bibnamefont {Taniguchi}}, \bibinfo {author}
  {\bibfnamefont {B.~A.}\ \bibnamefont {Bernevig}},\ and\ \bibinfo {author}
  {\bibfnamefont {D.~K.}\ \bibnamefont {Efetov}},\ }\bibfield  {title}
  {\bibinfo {title} {Symmetry-broken chern insulators and rashba-like
  landau-level crossings in magic-angle bilayer graphene},\ }\href
  {https://doi.org/10.1038/s41567-021-01186-3} {\bibfield  {journal} {\bibinfo
  {journal} {Nature Physics}\ }\textbf {\bibinfo {volume} {17}},\ \bibinfo
  {pages} {710} (\bibinfo {year} {2021})}\BibitemShut {NoStop}%
\bibitem [{\citenamefont {Stepanov}\ \emph {et~al.}(2021)\citenamefont
  {Stepanov}, \citenamefont {Xie}, \citenamefont {Taniguchi}, \citenamefont
  {Watanabe}, \citenamefont {Lu}, \citenamefont {MacDonald}, \citenamefont
  {Bernevig},\ and\ \citenamefont {Efetov}}]{ste_2021}%
  \BibitemOpen
  \bibfield  {author} {\bibinfo {author} {\bibfnamefont {P.}~\bibnamefont
  {Stepanov}}, \bibinfo {author} {\bibfnamefont {M.}~\bibnamefont {Xie}},
  \bibinfo {author} {\bibfnamefont {T.}~\bibnamefont {Taniguchi}}, \bibinfo
  {author} {\bibfnamefont {K.}~\bibnamefont {Watanabe}}, \bibinfo {author}
  {\bibfnamefont {X.}~\bibnamefont {Lu}}, \bibinfo {author} {\bibfnamefont
  {A.~H.}\ \bibnamefont {MacDonald}}, \bibinfo {author} {\bibfnamefont {B.~A.}\
  \bibnamefont {Bernevig}},\ and\ \bibinfo {author} {\bibfnamefont {D.~K.}\
  \bibnamefont {Efetov}},\ }\bibfield  {title} {\bibinfo {title} {Competing
  zero-field chern insulators in superconducting twisted bilayer graphene},\
  }\href {https://doi.org/10.1103/PhysRevLett.127.197701} {\bibfield  {journal}
  {\bibinfo  {journal} {Phys. Rev. Lett.}\ }\textbf {\bibinfo {volume} {127}},\
  \bibinfo {pages} {197701} (\bibinfo {year} {2021})}\BibitemShut {NoStop}%
\bibitem [{\citenamefont {Xie}\ \emph {et~al.}(2021)\citenamefont {Xie},
  \citenamefont {Pierce}, \citenamefont {Park}, \citenamefont {Parker},
  \citenamefont {Khalaf}, \citenamefont {Ledwith}, \citenamefont {Cao},
  \citenamefont {Lee}, \citenamefont {Chen}, \citenamefont {Forrester},
  \citenamefont {Watanabe}, \citenamefont {Taniguchi}, \citenamefont
  {Vishwanath}, \citenamefont {Jarillo-Herrero},\ and\ \citenamefont
  {Yacoby}}]{xie2021}%
  \BibitemOpen
  \bibfield  {author} {\bibinfo {author} {\bibfnamefont {Y.}~\bibnamefont
  {Xie}}, \bibinfo {author} {\bibfnamefont {A.~T.}\ \bibnamefont {Pierce}},
  \bibinfo {author} {\bibfnamefont {J.~M.}\ \bibnamefont {Park}}, \bibinfo
  {author} {\bibfnamefont {D.~E.}\ \bibnamefont {Parker}}, \bibinfo {author}
  {\bibfnamefont {E.}~\bibnamefont {Khalaf}}, \bibinfo {author} {\bibfnamefont
  {P.}~\bibnamefont {Ledwith}}, \bibinfo {author} {\bibfnamefont
  {Y.}~\bibnamefont {Cao}}, \bibinfo {author} {\bibfnamefont {S.~H.}\
  \bibnamefont {Lee}}, \bibinfo {author} {\bibfnamefont {S.}~\bibnamefont
  {Chen}}, \bibinfo {author} {\bibfnamefont {P.~R.}\ \bibnamefont {Forrester}},
  \bibinfo {author} {\bibfnamefont {K.}~\bibnamefont {Watanabe}}, \bibinfo
  {author} {\bibfnamefont {T.}~\bibnamefont {Taniguchi}}, \bibinfo {author}
  {\bibfnamefont {A.}~\bibnamefont {Vishwanath}}, \bibinfo {author}
  {\bibfnamefont {P.}~\bibnamefont {Jarillo-Herrero}},\ and\ \bibinfo {author}
  {\bibfnamefont {A.}~\bibnamefont {Yacoby}},\ }\bibfield  {title} {\bibinfo
  {title} {Fractional chern insulators in magic-angle twisted bilayer
  graphene},\ }\href {https://doi.org/10.1038/s41586-021-04002-3} {\bibfield
  {journal} {\bibinfo  {journal} {Nature}\ }\textbf {\bibinfo {volume} {600}},\
  \bibinfo {pages} {439} (\bibinfo {year} {2021})}\BibitemShut {NoStop}%
\bibitem [{\citenamefont {Kuiri}\ \emph {et~al.}(2022)\citenamefont {Kuiri},
  \citenamefont {Coleman}, \citenamefont {Gao}, \citenamefont {Vishnuradhan},
  \citenamefont {Watanabe}, \citenamefont {Taniguchi}, \citenamefont {Zhu},
  \citenamefont {MacDonald},\ and\ \citenamefont {Folk}}]{FolkNC2022}%
  \BibitemOpen
  \bibfield  {author} {\bibinfo {author} {\bibfnamefont {M.}~\bibnamefont
  {Kuiri}}, \bibinfo {author} {\bibfnamefont {C.}~\bibnamefont {Coleman}},
  \bibinfo {author} {\bibfnamefont {Z.}~\bibnamefont {Gao}}, \bibinfo {author}
  {\bibfnamefont {A.}~\bibnamefont {Vishnuradhan}}, \bibinfo {author}
  {\bibfnamefont {K.}~\bibnamefont {Watanabe}}, \bibinfo {author}
  {\bibfnamefont {T.}~\bibnamefont {Taniguchi}}, \bibinfo {author}
  {\bibfnamefont {J.}~\bibnamefont {Zhu}}, \bibinfo {author} {\bibfnamefont
  {A.~H.}\ \bibnamefont {MacDonald}},\ and\ \bibinfo {author} {\bibfnamefont
  {J.}~\bibnamefont {Folk}},\ }\bibfield  {title} {\bibinfo {title}
  {Spontaneous time-reversal symmetry breaking in twisted double bilayer
  graphene},\ }\href {https://doi.org/10.1038/s41467-022-34192-x} {\bibfield
  {journal} {\bibinfo  {journal} {Nature Communications}\ }\textbf {\bibinfo
  {volume} {13}},\ \bibinfo {pages} {6468} (\bibinfo {year}
  {2022})}\BibitemShut {NoStop}%
\bibitem [{\citenamefont {Kang}\ and\ \citenamefont
  {Vafek}(2019)}]{KangPRL2019}%
  \BibitemOpen
  \bibfield  {author} {\bibinfo {author} {\bibfnamefont {J.}~\bibnamefont
  {Kang}}\ and\ \bibinfo {author} {\bibfnamefont {O.}~\bibnamefont {Vafek}},\
  }\bibfield  {title} {\bibinfo {title} {Strong coupling phases of partially
  filled twisted bilayer graphene narrow bands},\ }\href
  {https://doi.org/10.1103/PhysRevLett.122.246401} {\bibfield  {journal}
  {\bibinfo  {journal} {Phys. Rev. Lett.}\ }\textbf {\bibinfo {volume} {122}},\
  \bibinfo {pages} {246401} (\bibinfo {year} {2019})}\BibitemShut {NoStop}%
\bibitem [{\citenamefont {Liu}\ \emph {et~al.}(2019)\citenamefont {Liu},
  \citenamefont {Ma}, \citenamefont {Gao},\ and\ \citenamefont
  {Dai}}]{DaiPRX2019}%
  \BibitemOpen
  \bibfield  {author} {\bibinfo {author} {\bibfnamefont {J.}~\bibnamefont
  {Liu}}, \bibinfo {author} {\bibfnamefont {Z.}~\bibnamefont {Ma}}, \bibinfo
  {author} {\bibfnamefont {J.}~\bibnamefont {Gao}},\ and\ \bibinfo {author}
  {\bibfnamefont {X.}~\bibnamefont {Dai}},\ }\bibfield  {title} {\bibinfo
  {title} {Quantum valley hall effect, orbital magnetism, and anomalous hall
  effect in twisted multilayer graphene systems},\ }\href
  {https://doi.org/10.1103/PhysRevX.9.031021} {\bibfield  {journal} {\bibinfo
  {journal} {Phys. Rev. X}\ }\textbf {\bibinfo {volume} {9}},\ \bibinfo {pages}
  {031021} (\bibinfo {year} {2019})}\BibitemShut {NoStop}%
\bibitem [{\citenamefont {Wu}\ and\ \citenamefont
  {Das~Sarma}(2020)}]{WuDasSarmaPRL2020}%
  \BibitemOpen
  \bibfield  {author} {\bibinfo {author} {\bibfnamefont {F.}~\bibnamefont
  {Wu}}\ and\ \bibinfo {author} {\bibfnamefont {S.}~\bibnamefont {Das~Sarma}},\
  }\bibfield  {title} {\bibinfo {title} {Collective excitations of quantum
  anomalous hall ferromagnets in twisted bilayer graphene},\ }\href
  {https://doi.org/10.1103/PhysRevLett.124.046403} {\bibfield  {journal}
  {\bibinfo  {journal} {Phys. Rev. Lett.}\ }\textbf {\bibinfo {volume} {124}},\
  \bibinfo {pages} {046403} (\bibinfo {year} {2020})}\BibitemShut {NoStop}%
\bibitem [{\citenamefont {Chen}\ \emph {et~al.}(2021)\citenamefont {Chen},
  \citenamefont {Liao}, \citenamefont {Chen}, \citenamefont {Vafek},
  \citenamefont {Kang}, \citenamefont {Li},\ and\ \citenamefont
  {Meng}}]{ChenBBNC2020}%
  \BibitemOpen
  \bibfield  {author} {\bibinfo {author} {\bibfnamefont {B.-B.}\ \bibnamefont
  {Chen}}, \bibinfo {author} {\bibfnamefont {Y.~D.}\ \bibnamefont {Liao}},
  \bibinfo {author} {\bibfnamefont {Z.}~\bibnamefont {Chen}}, \bibinfo {author}
  {\bibfnamefont {O.}~\bibnamefont {Vafek}}, \bibinfo {author} {\bibfnamefont
  {J.}~\bibnamefont {Kang}}, \bibinfo {author} {\bibfnamefont {W.}~\bibnamefont
  {Li}},\ and\ \bibinfo {author} {\bibfnamefont {Z.~Y.}\ \bibnamefont {Meng}},\
  }\bibfield  {title} {\bibinfo {title} {Realization of topological mott
  insulator in a twisted bilayer graphene lattice model},\ }\href
  {https://doi.org/10.1038/s41467-021-25438-1} {\bibfield  {journal} {\bibinfo
  {journal} {Nature Communications}\ }\textbf {\bibinfo {volume} {12}},\
  \bibinfo {pages} {5480} (\bibinfo {year} {2021})}\BibitemShut {NoStop}%
\bibitem [{\citenamefont {Repellin}\ \emph {et~al.}(2020)\citenamefont
  {Repellin}, \citenamefont {Dong}, \citenamefont {Zhang},\ and\ \citenamefont
  {Senthil}}]{SenthilPRL2020}%
  \BibitemOpen
  \bibfield  {author} {\bibinfo {author} {\bibfnamefont {C.}~\bibnamefont
  {Repellin}}, \bibinfo {author} {\bibfnamefont {Z.}~\bibnamefont {Dong}},
  \bibinfo {author} {\bibfnamefont {Y.-H.}\ \bibnamefont {Zhang}},\ and\
  \bibinfo {author} {\bibfnamefont {T.}~\bibnamefont {Senthil}},\ }\bibfield
  {title} {\bibinfo {title} {Ferromagnetism in narrow bands of moir\'e
  superlattices},\ }\href {https://doi.org/10.1103/PhysRevLett.124.187601}
  {\bibfield  {journal} {\bibinfo  {journal} {Phys. Rev. Lett.}\ }\textbf
  {\bibinfo {volume} {124}},\ \bibinfo {pages} {187601} (\bibinfo {year}
  {2020})}\BibitemShut {NoStop}%
\bibitem [{\citenamefont {Xie}\ \emph {et~al.}(2022)\citenamefont {Xie},
  \citenamefont {Zhang}, \citenamefont {Hu}, \citenamefont {Mak},\ and\
  \citenamefont {Law}}]{KTLawPRL2022}%
  \BibitemOpen
  \bibfield  {author} {\bibinfo {author} {\bibfnamefont {Y.-M.}\ \bibnamefont
  {Xie}}, \bibinfo {author} {\bibfnamefont {C.-P.}\ \bibnamefont {Zhang}},
  \bibinfo {author} {\bibfnamefont {J.-X.}\ \bibnamefont {Hu}}, \bibinfo
  {author} {\bibfnamefont {K.~F.}\ \bibnamefont {Mak}},\ and\ \bibinfo {author}
  {\bibfnamefont {K.~T.}\ \bibnamefont {Law}},\ }\bibfield  {title} {\bibinfo
  {title} {Valley-polarized quantum anomalous hall state in moir\'e
  ${\mathrm{mote}}_{2}/{\mathrm{wse}}_{2}$ heterobilayers},\ }\href
  {https://doi.org/10.1103/PhysRevLett.128.026402} {\bibfield  {journal}
  {\bibinfo  {journal} {Phys. Rev. Lett.}\ }\textbf {\bibinfo {volume} {128}},\
  \bibinfo {pages} {026402} (\bibinfo {year} {2022})}\BibitemShut {NoStop}%
\bibitem [{\citenamefont {Devakul}\ and\ \citenamefont
  {Fu}(2022)}]{LiangFuPRX2022}%
  \BibitemOpen
  \bibfield  {author} {\bibinfo {author} {\bibfnamefont {T.}~\bibnamefont
  {Devakul}}\ and\ \bibinfo {author} {\bibfnamefont {L.}~\bibnamefont {Fu}},\
  }\bibfield  {title} {\bibinfo {title} {Quantum anomalous hall effect from
  inverted charge transfer gap},\ }\href
  {https://doi.org/10.1103/PhysRevX.12.021031} {\bibfield  {journal} {\bibinfo
  {journal} {Phys. Rev. X}\ }\textbf {\bibinfo {volume} {12}},\ \bibinfo
  {pages} {021031} (\bibinfo {year} {2022})}\BibitemShut {NoStop}%
\bibitem [{\citenamefont {Wang}\ and\ \citenamefont
  {Vafek}(2024)}]{WangPRX2024}%
  \BibitemOpen
  \bibfield  {author} {\bibinfo {author} {\bibfnamefont {X.}~\bibnamefont
  {Wang}}\ and\ \bibinfo {author} {\bibfnamefont {O.}~\bibnamefont {Vafek}},\
  }\bibfield  {title} {\bibinfo {title} {Theory of correlated chern insulators
  in twisted bilayer graphene},\ }\href
  {https://doi.org/10.1103/PhysRevX.14.021042} {\bibfield  {journal} {\bibinfo
  {journal} {Phys. Rev. X}\ }\textbf {\bibinfo {volume} {14}},\ \bibinfo
  {pages} {021042} (\bibinfo {year} {2024})}\BibitemShut {NoStop}%
\bibitem [{\citenamefont {Dong}\ \emph {et~al.}(2024)\citenamefont {Dong},
  \citenamefont {Patri},\ and\ \citenamefont {Senthil}}]{SenthilPRL2024}%
  \BibitemOpen
  \bibfield  {author} {\bibinfo {author} {\bibfnamefont {Z.}~\bibnamefont
  {Dong}}, \bibinfo {author} {\bibfnamefont {A.~S.}\ \bibnamefont {Patri}},\
  and\ \bibinfo {author} {\bibfnamefont {T.}~\bibnamefont {Senthil}},\
  }\bibfield  {title} {\bibinfo {title} {Theory of quantum anomalous hall
  phases in pentalayer rhombohedral graphene moir\'e structures},\ }\href
  {https://doi.org/10.1103/PhysRevLett.133.206502} {\bibfield  {journal}
  {\bibinfo  {journal} {Phys. Rev. Lett.}\ }\textbf {\bibinfo {volume} {133}},\
  \bibinfo {pages} {206502} (\bibinfo {year} {2024})}\BibitemShut {NoStop}%
\bibitem [{\citenamefont {Wang}\ \emph {et~al.}(2024)\citenamefont {Wang},
  \citenamefont {Wang},\ and\ \citenamefont {Vafek}}]{MinxuanPRB2024}%
  \BibitemOpen
  \bibfield  {author} {\bibinfo {author} {\bibfnamefont {M.}~\bibnamefont
  {Wang}}, \bibinfo {author} {\bibfnamefont {X.}~\bibnamefont {Wang}},\ and\
  \bibinfo {author} {\bibfnamefont {O.}~\bibnamefont {Vafek}},\ }\bibfield
  {title} {\bibinfo {title} {Phase diagram of twisted bilayer
  ${\mathrm{mote}}_{2}$ in a magnetic field with an account for the
  electron-electron interaction},\ }\href
  {https://doi.org/10.1103/PhysRevB.110.L201107} {\bibfield  {journal}
  {\bibinfo  {journal} {Phys. Rev. B}\ }\textbf {\bibinfo {volume} {110}},\
  \bibinfo {pages} {L201107} (\bibinfo {year} {2024})}\BibitemShut {NoStop}%
\bibitem [{\citenamefont {Zhang}\ \emph {et~al.}(2020)\citenamefont {Zhang},
  \citenamefont {Jiang}, \citenamefont {Wang},\ and\ \citenamefont
  {Zhang}}]{zhang_2020}%
  \BibitemOpen
  \bibfield  {author} {\bibinfo {author} {\bibfnamefont {Y.}~\bibnamefont
  {Zhang}}, \bibinfo {author} {\bibfnamefont {K.}~\bibnamefont {Jiang}},
  \bibinfo {author} {\bibfnamefont {Z.}~\bibnamefont {Wang}},\ and\ \bibinfo
  {author} {\bibfnamefont {F.}~\bibnamefont {Zhang}},\ }\bibfield  {title}
  {\bibinfo {title} {Correlated insulating phases of twisted bilayer graphene
  at commensurate filling fractions: A hartree-fock study},\ }\href
  {https://doi.org/10.1103/PhysRevB.102.035136} {\bibfield  {journal} {\bibinfo
   {journal} {Phys. Rev. B}\ }\textbf {\bibinfo {volume} {102}},\ \bibinfo
  {pages} {035136} (\bibinfo {year} {2020})}\BibitemShut {NoStop}%
\bibitem [{\citenamefont {Serlin}\ \emph {et~al.}(2020)\citenamefont {Serlin},
  \citenamefont {Tschirhart}, \citenamefont {Polshyn}, \citenamefont {Zhang},
  \citenamefont {Zhu}, \citenamefont {Watanabe}, \citenamefont {Taniguchi},
  \citenamefont {Balents},\ and\ \citenamefont {Young}}]{YoungScience2020}%
  \BibitemOpen
  \bibfield  {author} {\bibinfo {author} {\bibfnamefont {M.}~\bibnamefont
  {Serlin}}, \bibinfo {author} {\bibfnamefont {C.~L.}\ \bibnamefont
  {Tschirhart}}, \bibinfo {author} {\bibfnamefont {H.}~\bibnamefont {Polshyn}},
  \bibinfo {author} {\bibfnamefont {Y.}~\bibnamefont {Zhang}}, \bibinfo
  {author} {\bibfnamefont {J.}~\bibnamefont {Zhu}}, \bibinfo {author}
  {\bibfnamefont {K.}~\bibnamefont {Watanabe}}, \bibinfo {author}
  {\bibfnamefont {T.}~\bibnamefont {Taniguchi}}, \bibinfo {author}
  {\bibfnamefont {L.}~\bibnamefont {Balents}},\ and\ \bibinfo {author}
  {\bibfnamefont {A.~F.}\ \bibnamefont {Young}},\ }\bibfield  {title} {\bibinfo
  {title} {Intrinsic quantized anomalous hall effect in a moiré
  heterostructure},\ }\href {https://doi.org/10.1126/science.aay5533}
  {\bibfield  {journal} {\bibinfo  {journal} {Science}\ }\textbf {\bibinfo
  {volume} {367}},\ \bibinfo {pages} {900} (\bibinfo {year}
  {2020})}\BibitemShut {NoStop}%
\bibitem [{\citenamefont {Li}\ \emph {et~al.}(2021)\citenamefont {Li},
  \citenamefont {Jiang}, \citenamefont {Shen}, \citenamefont {Zhang},
  \citenamefont {Li}, \citenamefont {Tao}, \citenamefont {Devakul},
  \citenamefont {Watanabe}, \citenamefont {Taniguchi}, \citenamefont {Fu},
  \citenamefont {Shan},\ and\ \citenamefont {Mak}}]{KinFaiMakNature2021}%
  \BibitemOpen
  \bibfield  {author} {\bibinfo {author} {\bibfnamefont {T.}~\bibnamefont
  {Li}}, \bibinfo {author} {\bibfnamefont {S.}~\bibnamefont {Jiang}}, \bibinfo
  {author} {\bibfnamefont {B.}~\bibnamefont {Shen}}, \bibinfo {author}
  {\bibfnamefont {Y.}~\bibnamefont {Zhang}}, \bibinfo {author} {\bibfnamefont
  {L.}~\bibnamefont {Li}}, \bibinfo {author} {\bibfnamefont {Z.}~\bibnamefont
  {Tao}}, \bibinfo {author} {\bibfnamefont {T.}~\bibnamefont {Devakul}},
  \bibinfo {author} {\bibfnamefont {K.}~\bibnamefont {Watanabe}}, \bibinfo
  {author} {\bibfnamefont {T.}~\bibnamefont {Taniguchi}}, \bibinfo {author}
  {\bibfnamefont {L.}~\bibnamefont {Fu}}, \bibinfo {author} {\bibfnamefont
  {J.}~\bibnamefont {Shan}},\ and\ \bibinfo {author} {\bibfnamefont {K.~F.}\
  \bibnamefont {Mak}},\ }\bibfield  {title} {\bibinfo {title} {Quantum
  anomalous hall effect from intertwined moir{\'e} bands},\ }\href
  {https://doi.org/10.1038/s41586-021-04171-1} {\bibfield  {journal} {\bibinfo
  {journal} {Nature}\ }\textbf {\bibinfo {volume} {600}},\ \bibinfo {pages}
  {641} (\bibinfo {year} {2021})}\BibitemShut {NoStop}%
\bibitem [{\citenamefont {Lu}\ \emph {et~al.}(2024)\citenamefont {Lu},
  \citenamefont {Han}, \citenamefont {Yao}, \citenamefont {Reddy},
  \citenamefont {Yang}, \citenamefont {Seo}, \citenamefont {Watanabe},
  \citenamefont {Taniguchi}, \citenamefont {Fu},\ and\ \citenamefont
  {Ju}}]{JuLongNature2024}%
  \BibitemOpen
  \bibfield  {author} {\bibinfo {author} {\bibfnamefont {Z.}~\bibnamefont
  {Lu}}, \bibinfo {author} {\bibfnamefont {T.}~\bibnamefont {Han}}, \bibinfo
  {author} {\bibfnamefont {Y.}~\bibnamefont {Yao}}, \bibinfo {author}
  {\bibfnamefont {A.~P.}\ \bibnamefont {Reddy}}, \bibinfo {author}
  {\bibfnamefont {J.}~\bibnamefont {Yang}}, \bibinfo {author} {\bibfnamefont
  {J.}~\bibnamefont {Seo}}, \bibinfo {author} {\bibfnamefont {K.}~\bibnamefont
  {Watanabe}}, \bibinfo {author} {\bibfnamefont {T.}~\bibnamefont {Taniguchi}},
  \bibinfo {author} {\bibfnamefont {L.}~\bibnamefont {Fu}},\ and\ \bibinfo
  {author} {\bibfnamefont {L.}~\bibnamefont {Ju}},\ }\bibfield  {title}
  {\bibinfo {title} {Fractional quantum anomalous hall effect in multilayer
  graphene},\ }\href {https://doi.org/10.1038/s41586-023-07010-7} {\bibfield
  {journal} {\bibinfo  {journal} {Nature}\ }\textbf {\bibinfo {volume} {626}},\
  \bibinfo {pages} {759} (\bibinfo {year} {2024})}\BibitemShut {NoStop}%
\bibitem [{\citenamefont {Tseng}\ \emph {et~al.}(2022)\citenamefont {Tseng},
  \citenamefont {Ma}, \citenamefont {Liu}, \citenamefont {Watanabe},
  \citenamefont {Taniguchi}, \citenamefont {Chu},\ and\ \citenamefont
  {Yankowitz}}]{YankowitzNP2022}%
  \BibitemOpen
  \bibfield  {author} {\bibinfo {author} {\bibfnamefont {C.-C.}\ \bibnamefont
  {Tseng}}, \bibinfo {author} {\bibfnamefont {X.}~\bibnamefont {Ma}}, \bibinfo
  {author} {\bibfnamefont {Z.}~\bibnamefont {Liu}}, \bibinfo {author}
  {\bibfnamefont {K.}~\bibnamefont {Watanabe}}, \bibinfo {author}
  {\bibfnamefont {T.}~\bibnamefont {Taniguchi}}, \bibinfo {author}
  {\bibfnamefont {J.-H.}\ \bibnamefont {Chu}},\ and\ \bibinfo {author}
  {\bibfnamefont {M.}~\bibnamefont {Yankowitz}},\ }\bibfield  {title} {\bibinfo
  {title} {Anomalous hall effect at half filling in twisted bilayer graphene},\
  }\href {https://doi.org/10.1038/s41567-022-01697-7} {\bibfield  {journal}
  {\bibinfo  {journal} {Nature Physics}\ }\textbf {\bibinfo {volume} {18}},\
  \bibinfo {pages} {1038} (\bibinfo {year} {2022})}\BibitemShut {NoStop}%
\bibitem [{\citenamefont {Sharpe}\ \emph {et~al.}(2021)\citenamefont {Sharpe},
  \citenamefont {Fox}, \citenamefont {Barnard}, \citenamefont {Finney},
  \citenamefont {Watanabe}, \citenamefont {Taniguchi}, \citenamefont
  {Kastner},\ and\ \citenamefont {Goldhaber-Gordon}}]{DavidGGNL2021}%
  \BibitemOpen
  \bibfield  {author} {\bibinfo {author} {\bibfnamefont {A.~L.}\ \bibnamefont
  {Sharpe}}, \bibinfo {author} {\bibfnamefont {E.~J.}\ \bibnamefont {Fox}},
  \bibinfo {author} {\bibfnamefont {A.~W.}\ \bibnamefont {Barnard}}, \bibinfo
  {author} {\bibfnamefont {J.}~\bibnamefont {Finney}}, \bibinfo {author}
  {\bibfnamefont {K.}~\bibnamefont {Watanabe}}, \bibinfo {author}
  {\bibfnamefont {T.}~\bibnamefont {Taniguchi}}, \bibinfo {author}
  {\bibfnamefont {M.~A.}\ \bibnamefont {Kastner}},\ and\ \bibinfo {author}
  {\bibfnamefont {D.}~\bibnamefont {Goldhaber-Gordon}},\ }\bibfield  {title}
  {\bibinfo {title} {Evidence of orbital ferromagnetism in twisted bilayer
  graphene aligned to hexagonal boron nitride},\ }\href
  {https://doi.org/10.1021/acs.nanolett.1c00696} {\bibfield  {journal}
  {\bibinfo  {journal} {Nano Letters}\ }\textbf {\bibinfo {volume} {21}},\
  \bibinfo {pages} {4299} (\bibinfo {year} {2021})}\BibitemShut {NoStop}%
\bibitem [{\citenamefont {Tschirhart}\ \emph {et~al.}(2021)\citenamefont
  {Tschirhart}, \citenamefont {Serlin}, \citenamefont {Polshyn}, \citenamefont
  {Shragai}, \citenamefont {Xia}, \citenamefont {Zhu}, \citenamefont {Zhang},
  \citenamefont {Watanabe}, \citenamefont {Taniguchi}, \citenamefont {Huber},\
  and\ \citenamefont {Young}}]{YoungScience2021}%
  \BibitemOpen
  \bibfield  {author} {\bibinfo {author} {\bibfnamefont {C.~L.}\ \bibnamefont
  {Tschirhart}}, \bibinfo {author} {\bibfnamefont {M.}~\bibnamefont {Serlin}},
  \bibinfo {author} {\bibfnamefont {H.}~\bibnamefont {Polshyn}}, \bibinfo
  {author} {\bibfnamefont {A.}~\bibnamefont {Shragai}}, \bibinfo {author}
  {\bibfnamefont {Z.}~\bibnamefont {Xia}}, \bibinfo {author} {\bibfnamefont
  {J.}~\bibnamefont {Zhu}}, \bibinfo {author} {\bibfnamefont {Y.}~\bibnamefont
  {Zhang}}, \bibinfo {author} {\bibfnamefont {K.}~\bibnamefont {Watanabe}},
  \bibinfo {author} {\bibfnamefont {T.}~\bibnamefont {Taniguchi}}, \bibinfo
  {author} {\bibfnamefont {M.~E.}\ \bibnamefont {Huber}},\ and\ \bibinfo
  {author} {\bibfnamefont {A.~F.}\ \bibnamefont {Young}},\ }\bibfield  {title}
  {\bibinfo {title} {Imaging orbital ferromagnetism in a moiré chern
  insulator},\ }\href {https://doi.org/10.1126/science.abd3190} {\bibfield
  {journal} {\bibinfo  {journal} {Science}\ }\textbf {\bibinfo {volume}
  {372}},\ \bibinfo {pages} {1323} (\bibinfo {year} {2021})}\BibitemShut
  {NoStop}%
\bibitem [{\citenamefont {Cai}\ \emph {et~al.}(2023)\citenamefont {Cai},
  \citenamefont {Anderson}, \citenamefont {Wang}, \citenamefont {Zhang},
  \citenamefont {Liu}, \citenamefont {Holtzmann}, \citenamefont {Zhang},
  \citenamefont {Fan}, \citenamefont {Taniguchi}, \citenamefont {Watanabe},
  \citenamefont {Ran}, \citenamefont {Cao}, \citenamefont {Fu}, \citenamefont
  {Xiao}, \citenamefont {Yao},\ and\ \citenamefont
  {Xu}}]{XiaodongCaiNature2023}%
  \BibitemOpen
  \bibfield  {author} {\bibinfo {author} {\bibfnamefont {J.}~\bibnamefont
  {Cai}}, \bibinfo {author} {\bibfnamefont {E.}~\bibnamefont {Anderson}},
  \bibinfo {author} {\bibfnamefont {C.}~\bibnamefont {Wang}}, \bibinfo {author}
  {\bibfnamefont {X.}~\bibnamefont {Zhang}}, \bibinfo {author} {\bibfnamefont
  {X.}~\bibnamefont {Liu}}, \bibinfo {author} {\bibfnamefont {W.}~\bibnamefont
  {Holtzmann}}, \bibinfo {author} {\bibfnamefont {Y.}~\bibnamefont {Zhang}},
  \bibinfo {author} {\bibfnamefont {F.}~\bibnamefont {Fan}}, \bibinfo {author}
  {\bibfnamefont {T.}~\bibnamefont {Taniguchi}}, \bibinfo {author}
  {\bibfnamefont {K.}~\bibnamefont {Watanabe}}, \bibinfo {author}
  {\bibfnamefont {Y.}~\bibnamefont {Ran}}, \bibinfo {author} {\bibfnamefont
  {T.}~\bibnamefont {Cao}}, \bibinfo {author} {\bibfnamefont {L.}~\bibnamefont
  {Fu}}, \bibinfo {author} {\bibfnamefont {D.}~\bibnamefont {Xiao}}, \bibinfo
  {author} {\bibfnamefont {W.}~\bibnamefont {Yao}},\ and\ \bibinfo {author}
  {\bibfnamefont {X.}~\bibnamefont {Xu}},\ }\bibfield  {title} {\bibinfo
  {title} {Signatures of fractional quantum anomalous hall states in twisted
  mote2},\ }\href {https://doi.org/10.1038/s41586-023-06289-w} {\bibfield
  {journal} {\bibinfo  {journal} {Nature}\ }\textbf {\bibinfo {volume} {622}},\
  \bibinfo {pages} {63} (\bibinfo {year} {2023})}\BibitemShut {NoStop}%
\bibitem [{\citenamefont {Park}\ \emph {et~al.}(2023)\citenamefont {Park},
  \citenamefont {Cai}, \citenamefont {Anderson}, \citenamefont {Zhang},
  \citenamefont {Zhu}, \citenamefont {Liu}, \citenamefont {Wang}, \citenamefont
  {Holtzmann}, \citenamefont {Hu}, \citenamefont {Liu}, \citenamefont
  {Taniguchi}, \citenamefont {Watanabe}, \citenamefont {Chu}, \citenamefont
  {Cao}, \citenamefont {Fu}, \citenamefont {Yao}, \citenamefont {Chang},
  \citenamefont {Cobden}, \citenamefont {Xiao},\ and\ \citenamefont
  {Xu}}]{XiaodongNature2023}%
  \BibitemOpen
  \bibfield  {author} {\bibinfo {author} {\bibfnamefont {H.}~\bibnamefont
  {Park}}, \bibinfo {author} {\bibfnamefont {J.}~\bibnamefont {Cai}}, \bibinfo
  {author} {\bibfnamefont {E.}~\bibnamefont {Anderson}}, \bibinfo {author}
  {\bibfnamefont {Y.}~\bibnamefont {Zhang}}, \bibinfo {author} {\bibfnamefont
  {J.}~\bibnamefont {Zhu}}, \bibinfo {author} {\bibfnamefont {X.}~\bibnamefont
  {Liu}}, \bibinfo {author} {\bibfnamefont {C.}~\bibnamefont {Wang}}, \bibinfo
  {author} {\bibfnamefont {W.}~\bibnamefont {Holtzmann}}, \bibinfo {author}
  {\bibfnamefont {C.}~\bibnamefont {Hu}}, \bibinfo {author} {\bibfnamefont
  {Z.}~\bibnamefont {Liu}}, \bibinfo {author} {\bibfnamefont {T.}~\bibnamefont
  {Taniguchi}}, \bibinfo {author} {\bibfnamefont {K.}~\bibnamefont {Watanabe}},
  \bibinfo {author} {\bibfnamefont {J.-H.}\ \bibnamefont {Chu}}, \bibinfo
  {author} {\bibfnamefont {T.}~\bibnamefont {Cao}}, \bibinfo {author}
  {\bibfnamefont {L.}~\bibnamefont {Fu}}, \bibinfo {author} {\bibfnamefont
  {W.}~\bibnamefont {Yao}}, \bibinfo {author} {\bibfnamefont {C.-Z.}\
  \bibnamefont {Chang}}, \bibinfo {author} {\bibfnamefont {D.}~\bibnamefont
  {Cobden}}, \bibinfo {author} {\bibfnamefont {D.}~\bibnamefont {Xiao}},\ and\
  \bibinfo {author} {\bibfnamefont {X.}~\bibnamefont {Xu}},\ }\bibfield
  {title} {\bibinfo {title} {Observation of fractionally quantized anomalous
  hall effect},\ }\href {https://doi.org/10.1038/s41586-023-06536-0} {\bibfield
   {journal} {\bibinfo  {journal} {Nature}\ }\textbf {\bibinfo {volume}
  {622}},\ \bibinfo {pages} {74} (\bibinfo {year} {2023})}\BibitemShut
  {NoStop}%
\bibitem [{\citenamefont {Xu}\ \emph {et~al.}(2022)\citenamefont {Xu},
  \citenamefont {Ni}, \citenamefont {Liu}, \citenamefont {Ortiz}, \citenamefont
  {Deng}, \citenamefont {Wilson}, \citenamefont {Yan}, \citenamefont
  {Balents},\ and\ \citenamefont {Wu}}]{xu_2022}%
  \BibitemOpen
  \bibfield  {author} {\bibinfo {author} {\bibfnamefont {Y.}~\bibnamefont
  {Xu}}, \bibinfo {author} {\bibfnamefont {Z.}~\bibnamefont {Ni}}, \bibinfo
  {author} {\bibfnamefont {Y.}~\bibnamefont {Liu}}, \bibinfo {author}
  {\bibfnamefont {B.~R.}\ \bibnamefont {Ortiz}}, \bibinfo {author}
  {\bibfnamefont {Q.}~\bibnamefont {Deng}}, \bibinfo {author} {\bibfnamefont
  {S.~D.}\ \bibnamefont {Wilson}}, \bibinfo {author} {\bibfnamefont
  {B.}~\bibnamefont {Yan}}, \bibinfo {author} {\bibfnamefont {L.}~\bibnamefont
  {Balents}},\ and\ \bibinfo {author} {\bibfnamefont {L.}~\bibnamefont {Wu}},\
  }\bibfield  {title} {\bibinfo {title} {Three-state nematicity and
  magneto-optical kerr effect in the charge density waves in kagome
  superconductors},\ }\href {https://doi.org/10.1038/s41567-022-01805-7}
  {\bibfield  {journal} {\bibinfo  {journal} {Nature Physics}\ }\textbf
  {\bibinfo {volume} {18}},\ \bibinfo {pages} {1470} (\bibinfo {year}
  {2022})}\BibitemShut {NoStop}%
\bibitem [{\citenamefont {Thonhauser}\ \emph {et~al.}(2005)\citenamefont
  {Thonhauser}, \citenamefont {Ceresoli}, \citenamefont {Vanderbilt},\ and\
  \citenamefont {Resta}}]{RestaPRL2005}%
  \BibitemOpen
  \bibfield  {author} {\bibinfo {author} {\bibfnamefont {T.}~\bibnamefont
  {Thonhauser}}, \bibinfo {author} {\bibfnamefont {D.}~\bibnamefont
  {Ceresoli}}, \bibinfo {author} {\bibfnamefont {D.}~\bibnamefont
  {Vanderbilt}},\ and\ \bibinfo {author} {\bibfnamefont {R.}~\bibnamefont
  {Resta}},\ }\bibfield  {title} {\bibinfo {title} {Orbital magnetization in
  periodic insulators},\ }\href {https://doi.org/10.1103/PhysRevLett.95.137205}
  {\bibfield  {journal} {\bibinfo  {journal} {Phys. Rev. Lett.}\ }\textbf
  {\bibinfo {volume} {95}},\ \bibinfo {pages} {137205} (\bibinfo {year}
  {2005})}\BibitemShut {NoStop}%
\bibitem [{\citenamefont {Ceresoli}\ \emph {et~al.}(2006)\citenamefont
  {Ceresoli}, \citenamefont {Thonhauser}, \citenamefont {Vanderbilt},\ and\
  \citenamefont {Resta}}]{RestaPRB2006}%
  \BibitemOpen
  \bibfield  {author} {\bibinfo {author} {\bibfnamefont {D.}~\bibnamefont
  {Ceresoli}}, \bibinfo {author} {\bibfnamefont {T.}~\bibnamefont
  {Thonhauser}}, \bibinfo {author} {\bibfnamefont {D.}~\bibnamefont
  {Vanderbilt}},\ and\ \bibinfo {author} {\bibfnamefont {R.}~\bibnamefont
  {Resta}},\ }\bibfield  {title} {\bibinfo {title} {Orbital magnetization in
  crystalline solids: Multi-band insulators, chern insulators, and metals},\
  }\href {https://doi.org/10.1103/PhysRevB.74.024408} {\bibfield  {journal}
  {\bibinfo  {journal} {Phys. Rev. B}\ }\textbf {\bibinfo {volume} {74}},\
  \bibinfo {pages} {024408} (\bibinfo {year} {2006})}\BibitemShut {NoStop}%
\bibitem [{\citenamefont {Shi}\ \emph {et~al.}(2007)\citenamefont {Shi},
  \citenamefont {Vignale}, \citenamefont {Xiao},\ and\ \citenamefont
  {Niu}}]{JunrenShiPRL2007}%
  \BibitemOpen
  \bibfield  {author} {\bibinfo {author} {\bibfnamefont {J.}~\bibnamefont
  {Shi}}, \bibinfo {author} {\bibfnamefont {G.}~\bibnamefont {Vignale}},
  \bibinfo {author} {\bibfnamefont {D.}~\bibnamefont {Xiao}},\ and\ \bibinfo
  {author} {\bibfnamefont {Q.}~\bibnamefont {Niu}},\ }\bibfield  {title}
  {\bibinfo {title} {Quantum theory of orbital magnetization and its
  generalization to interacting systems},\ }\href
  {https://doi.org/10.1103/PhysRevLett.99.197202} {\bibfield  {journal}
  {\bibinfo  {journal} {Phys. Rev. Lett.}\ }\textbf {\bibinfo {volume} {99}},\
  \bibinfo {pages} {197202} (\bibinfo {year} {2007})}\BibitemShut {NoStop}%
\bibitem [{\citenamefont {Raoux}\ \emph {et~al.}(2015)\citenamefont {Raoux},
  \citenamefont {Pi\'echon}, \citenamefont {Fuchs},\ and\ \citenamefont
  {Montambaux}}]{RaouxPRB2015}%
  \BibitemOpen
  \bibfield  {author} {\bibinfo {author} {\bibfnamefont {A.}~\bibnamefont
  {Raoux}}, \bibinfo {author} {\bibfnamefont {F.}~\bibnamefont {Pi\'echon}},
  \bibinfo {author} {\bibfnamefont {J.-N.}\ \bibnamefont {Fuchs}},\ and\
  \bibinfo {author} {\bibfnamefont {G.}~\bibnamefont {Montambaux}},\ }\bibfield
   {title} {\bibinfo {title} {Orbital magnetism in coupled-bands models},\
  }\href {https://doi.org/10.1103/PhysRevB.91.085120} {\bibfield  {journal}
  {\bibinfo  {journal} {Phys. Rev. B}\ }\textbf {\bibinfo {volume} {91}},\
  \bibinfo {pages} {085120} (\bibinfo {year} {2015})}\BibitemShut {NoStop}%
\bibitem [{\citenamefont {Pi\'echon}\ \emph {et~al.}(2016)\citenamefont
  {Pi\'echon}, \citenamefont {Raoux}, \citenamefont {Fuchs},\ and\
  \citenamefont {Montambaux}}]{MontambauxPRL2016}%
  \BibitemOpen
  \bibfield  {author} {\bibinfo {author} {\bibfnamefont {F.}~\bibnamefont
  {Pi\'echon}}, \bibinfo {author} {\bibfnamefont {A.}~\bibnamefont {Raoux}},
  \bibinfo {author} {\bibfnamefont {J.-N.}\ \bibnamefont {Fuchs}},\ and\
  \bibinfo {author} {\bibfnamefont {G.}~\bibnamefont {Montambaux}},\ }\bibfield
   {title} {\bibinfo {title} {Geometric orbital susceptibility: Quantum metric
  without berry curvature},\ }\href
  {https://doi.org/10.1103/PhysRevB.94.134423} {\bibfield  {journal} {\bibinfo
  {journal} {Phys. Rev. B}\ }\textbf {\bibinfo {volume} {94}},\ \bibinfo
  {pages} {134423} (\bibinfo {year} {2016})}\BibitemShut {NoStop}%
\bibitem [{\citenamefont {Tang}\ \emph {et~al.}(2020)\citenamefont {Tang},
  \citenamefont {Li}, \citenamefont {Li}, \citenamefont {Xu}, \citenamefont
  {Liu}, \citenamefont {Barmak}, \citenamefont {Watanabe}, \citenamefont
  {Taniguchi}, \citenamefont {MacDonald}, \citenamefont {Shan},\ and\
  \citenamefont {Mak}}]{KinFaiMakNature2020}%
  \BibitemOpen
  \bibfield  {author} {\bibinfo {author} {\bibfnamefont {Y.}~\bibnamefont
  {Tang}}, \bibinfo {author} {\bibfnamefont {L.}~\bibnamefont {Li}}, \bibinfo
  {author} {\bibfnamefont {T.}~\bibnamefont {Li}}, \bibinfo {author}
  {\bibfnamefont {Y.}~\bibnamefont {Xu}}, \bibinfo {author} {\bibfnamefont
  {S.}~\bibnamefont {Liu}}, \bibinfo {author} {\bibfnamefont {K.}~\bibnamefont
  {Barmak}}, \bibinfo {author} {\bibfnamefont {K.}~\bibnamefont {Watanabe}},
  \bibinfo {author} {\bibfnamefont {T.}~\bibnamefont {Taniguchi}}, \bibinfo
  {author} {\bibfnamefont {A.~H.}\ \bibnamefont {MacDonald}}, \bibinfo {author}
  {\bibfnamefont {J.}~\bibnamefont {Shan}},\ and\ \bibinfo {author}
  {\bibfnamefont {K.~F.}\ \bibnamefont {Mak}},\ }\bibfield  {title} {\bibinfo
  {title} {Simulation of hubbard model physics in wse2/ws2 moir{\'e}
  superlattices},\ }\href {https://doi.org/10.1038/s41586-020-2085-3}
  {\bibfield  {journal} {\bibinfo  {journal} {Nature}\ }\textbf {\bibinfo
  {volume} {579}},\ \bibinfo {pages} {353} (\bibinfo {year}
  {2020})}\BibitemShut {NoStop}%
\bibitem [{\citenamefont {Wang}\ \emph {et~al.}(2022)\citenamefont {Wang},
  \citenamefont {Xiao}, \citenamefont {Park}, \citenamefont {Zhu},
  \citenamefont {Wang}, \citenamefont {Taniguchi}, \citenamefont {Watanabe},
  \citenamefont {Yan}, \citenamefont {Xiao}, \citenamefont {Gamelin},
  \citenamefont {Yao},\ and\ \citenamefont {Xu}}]{XiaodongNature2022}%
  \BibitemOpen
  \bibfield  {author} {\bibinfo {author} {\bibfnamefont {X.}~\bibnamefont
  {Wang}}, \bibinfo {author} {\bibfnamefont {C.}~\bibnamefont {Xiao}}, \bibinfo
  {author} {\bibfnamefont {H.}~\bibnamefont {Park}}, \bibinfo {author}
  {\bibfnamefont {J.}~\bibnamefont {Zhu}}, \bibinfo {author} {\bibfnamefont
  {C.}~\bibnamefont {Wang}}, \bibinfo {author} {\bibfnamefont {T.}~\bibnamefont
  {Taniguchi}}, \bibinfo {author} {\bibfnamefont {K.}~\bibnamefont {Watanabe}},
  \bibinfo {author} {\bibfnamefont {J.}~\bibnamefont {Yan}}, \bibinfo {author}
  {\bibfnamefont {D.}~\bibnamefont {Xiao}}, \bibinfo {author} {\bibfnamefont
  {D.~R.}\ \bibnamefont {Gamelin}}, \bibinfo {author} {\bibfnamefont
  {W.}~\bibnamefont {Yao}},\ and\ \bibinfo {author} {\bibfnamefont
  {X.}~\bibnamefont {Xu}},\ }\bibfield  {title} {\bibinfo {title}
  {Light-induced ferromagnetism in moir{\'e} superlattices},\ }\href
  {https://doi.org/10.1038/s41586-022-04472-z} {\bibfield  {journal} {\bibinfo
  {journal} {Nature}\ }\textbf {\bibinfo {volume} {604}},\ \bibinfo {pages}
  {468} (\bibinfo {year} {2022})}\BibitemShut {NoStop}%
\bibitem [{\citenamefont {Kn{\"u}ppel}\ \emph {et~al.}(2025)\citenamefont
  {Kn{\"u}ppel}, \citenamefont {Zhu}, \citenamefont {Xia}, \citenamefont {Xia},
  \citenamefont {Han}, \citenamefont {Zeng}, \citenamefont {Watanabe},
  \citenamefont {Taniguchi}, \citenamefont {Shan},\ and\ \citenamefont
  {Mak}}]{KinFaiMakNC2025}%
  \BibitemOpen
  \bibfield  {author} {\bibinfo {author} {\bibfnamefont {P.}~\bibnamefont
  {Kn{\"u}ppel}}, \bibinfo {author} {\bibfnamefont {J.}~\bibnamefont {Zhu}},
  \bibinfo {author} {\bibfnamefont {Y.}~\bibnamefont {Xia}}, \bibinfo {author}
  {\bibfnamefont {Z.}~\bibnamefont {Xia}}, \bibinfo {author} {\bibfnamefont
  {Z.}~\bibnamefont {Han}}, \bibinfo {author} {\bibfnamefont {Y.}~\bibnamefont
  {Zeng}}, \bibinfo {author} {\bibfnamefont {K.}~\bibnamefont {Watanabe}},
  \bibinfo {author} {\bibfnamefont {T.}~\bibnamefont {Taniguchi}}, \bibinfo
  {author} {\bibfnamefont {J.}~\bibnamefont {Shan}},\ and\ \bibinfo {author}
  {\bibfnamefont {K.~F.}\ \bibnamefont {Mak}},\ }\bibfield  {title} {\bibinfo
  {title} {Correlated states controlled by a tunable van hove singularity in
  moir{\'e} wse2 bilayers},\ }\href
  {https://doi.org/10.1038/s41467-025-57235-5} {\bibfield  {journal} {\bibinfo
  {journal} {Nature Communications}\ }\textbf {\bibinfo {volume} {16}},\
  \bibinfo {pages} {1959} (\bibinfo {year} {2025})}\BibitemShut {NoStop}%
\bibitem [{\citenamefont {Xia}\ \emph {et~al.}(2026)\citenamefont {Xia},
  \citenamefont {Han}, \citenamefont {Zhu}, \citenamefont {Zhang},
  \citenamefont {Kn{\"u}ppel}, \citenamefont {Watanabe}, \citenamefont
  {Taniguchi}, \citenamefont {Mak},\ and\ \citenamefont
  {Shan}}]{KinFaiMakNature2026}%
  \BibitemOpen
  \bibfield  {author} {\bibinfo {author} {\bibfnamefont {Y.}~\bibnamefont
  {Xia}}, \bibinfo {author} {\bibfnamefont {Z.}~\bibnamefont {Han}}, \bibinfo
  {author} {\bibfnamefont {J.}~\bibnamefont {Zhu}}, \bibinfo {author}
  {\bibfnamefont {Y.}~\bibnamefont {Zhang}}, \bibinfo {author} {\bibfnamefont
  {P.}~\bibnamefont {Kn{\"u}ppel}}, \bibinfo {author} {\bibfnamefont
  {K.}~\bibnamefont {Watanabe}}, \bibinfo {author} {\bibfnamefont
  {T.}~\bibnamefont {Taniguchi}}, \bibinfo {author} {\bibfnamefont {K.~F.}\
  \bibnamefont {Mak}},\ and\ \bibinfo {author} {\bibfnamefont {J.}~\bibnamefont
  {Shan}},\ }\bibfield  {title} {\bibinfo {title} {Bandwidth-tuned mott
  transition and superconductivity in moir{\'e} wse2},\ }\href
  {https://doi.org/10.1038/s41586-025-10049-3} {\bibfield  {journal} {\bibinfo
  {journal} {Nature}\ }\textbf {\bibinfo {volume} {650}},\ \bibinfo {pages}
  {585} (\bibinfo {year} {2026})}\BibitemShut {NoStop}%
\bibitem [{\citenamefont {Jiao}\ \emph {et~al.}(2026)\citenamefont {Jiao},
  \citenamefont {Qian}, \citenamefont {Mao}, \citenamefont {Chang},
  \citenamefont {Xiao}, \citenamefont {Liu}, \citenamefont {Wang},
  \citenamefont {Wu}, \citenamefont {Peng}, \citenamefont {Xu}, \citenamefont
  {Dong}, \citenamefont {Zheng}, \citenamefont {Wu}, \citenamefont {Zheng},
  \citenamefont {Watanabe}, \citenamefont {Taniguchi}, \citenamefont {Jia},
  \citenamefont {Liu}, \citenamefont {Shi}, \citenamefont {Wang}, \citenamefont
  {Chen}, \citenamefont {Li}, \citenamefont {Zhong}, \citenamefont {Zhang},
  \citenamefont {Qian}, \citenamefont {Chen},\ and\ \citenamefont
  {Jiang}}]{JiangPRX2026}%
  \BibitemOpen
  \bibfield  {author} {\bibinfo {author} {\bibfnamefont {P.}~\bibnamefont
  {Jiao}}, \bibinfo {author} {\bibfnamefont {C.}~\bibnamefont {Qian}}, \bibinfo
  {author} {\bibfnamefont {N.}~\bibnamefont {Mao}}, \bibinfo {author}
  {\bibfnamefont {X.}~\bibnamefont {Chang}}, \bibinfo {author} {\bibfnamefont
  {J.}~\bibnamefont {Xiao}}, \bibinfo {author} {\bibfnamefont {F.}~\bibnamefont
  {Liu}}, \bibinfo {author} {\bibfnamefont {S.}~\bibnamefont {Wang}}, \bibinfo
  {author} {\bibfnamefont {X.}~\bibnamefont {Wu}}, \bibinfo {author}
  {\bibfnamefont {D.}~\bibnamefont {Peng}}, \bibinfo {author} {\bibfnamefont
  {C.}~\bibnamefont {Xu}}, \bibinfo {author} {\bibfnamefont {H.}~\bibnamefont
  {Dong}}, \bibinfo {author} {\bibfnamefont {Y.}~\bibnamefont {Zheng}},
  \bibinfo {author} {\bibfnamefont {J.}~\bibnamefont {Wu}}, \bibinfo {author}
  {\bibfnamefont {T.}~\bibnamefont {Zheng}}, \bibinfo {author} {\bibfnamefont
  {K.}~\bibnamefont {Watanabe}}, \bibinfo {author} {\bibfnamefont
  {T.}~\bibnamefont {Taniguchi}}, \bibinfo {author} {\bibfnamefont
  {J.}~\bibnamefont {Jia}}, \bibinfo {author} {\bibfnamefont {X.}~\bibnamefont
  {Liu}}, \bibinfo {author} {\bibfnamefont {Z.}~\bibnamefont {Shi}}, \bibinfo
  {author} {\bibfnamefont {S.}~\bibnamefont {Wang}}, \bibinfo {author}
  {\bibfnamefont {G.}~\bibnamefont {Chen}}, \bibinfo {author} {\bibfnamefont
  {T.}~\bibnamefont {Li}}, \bibinfo {author} {\bibfnamefont {R.}~\bibnamefont
  {Zhong}}, \bibinfo {author} {\bibfnamefont {Y.}~\bibnamefont {Zhang}},
  \bibinfo {author} {\bibfnamefont {D.}~\bibnamefont {Qian}}, \bibinfo {author}
  {\bibfnamefont {Z.}~\bibnamefont {Chen}},\ and\ \bibinfo {author}
  {\bibfnamefont {S.}~\bibnamefont {Jiang}},\ }\bibfield  {title} {\bibinfo
  {title} {Hydrostatic pressure-enhanced correlated magnetism and chern
  insulator in moir\'e ${\mathrm{wse}}_{2}$},\ }\href
  {https://doi.org/10.1103/th65-gff6} {\bibfield  {journal} {\bibinfo
  {journal} {Phys. Rev. X}\ }\textbf {\bibinfo {volume} {16}},\ \bibinfo
  {pages} {011068} (\bibinfo {year} {2026})}\BibitemShut {NoStop}%
\bibitem [{\citenamefont {Souza}\ and\ \citenamefont
  {Vanderbilt}(2008)}]{SouzaVanderbiltPRB2008}%
  \BibitemOpen
  \bibfield  {author} {\bibinfo {author} {\bibfnamefont {I.}~\bibnamefont
  {Souza}}\ and\ \bibinfo {author} {\bibfnamefont {D.}~\bibnamefont
  {Vanderbilt}},\ }\bibfield  {title} {\bibinfo {title} {Dichroic $f$-sum rule
  and the orbital magnetization of crystals},\ }\href
  {https://doi.org/10.1103/PhysRevB.77.054438} {\bibfield  {journal} {\bibinfo
  {journal} {Phys. Rev. B}\ }\textbf {\bibinfo {volume} {77}},\ \bibinfo
  {pages} {054438} (\bibinfo {year} {2008})}\BibitemShut {NoStop}%
\bibitem [{\citenamefont {Resta}(2020)}]{RestaPRR2020}%
  \BibitemOpen
  \bibfield  {author} {\bibinfo {author} {\bibfnamefont {R.}~\bibnamefont
  {Resta}},\ }\bibfield  {title} {\bibinfo {title} {Magnetic circular dichroism
  versus orbital magnetization},\ }\href
  {https://doi.org/10.1103/PhysRevResearch.2.023139} {\bibfield  {journal}
  {\bibinfo  {journal} {Phys. Rev. Res.}\ }\textbf {\bibinfo {volume} {2}},\
  \bibinfo {pages} {023139} (\bibinfo {year} {2020})}\BibitemShut {NoStop}%
\bibitem [{\citenamefont {Ji}\ \emph {et~al.}(2025)\citenamefont {Ji},
  \citenamefont {Palomino}, \citenamefont {Goldman}, \citenamefont {Ozawa},
  \citenamefont {Riseborough}, \citenamefont {Wang},\ and\ \citenamefont
  {Mera}}]{BrunoArXiv2507}%
  \BibitemOpen
  \bibfield  {author} {\bibinfo {author} {\bibfnamefont {G.}~\bibnamefont
  {Ji}}, \bibinfo {author} {\bibfnamefont {D.~E.}\ \bibnamefont {Palomino}},
  \bibinfo {author} {\bibfnamefont {N.}~\bibnamefont {Goldman}}, \bibinfo
  {author} {\bibfnamefont {T.}~\bibnamefont {Ozawa}}, \bibinfo {author}
  {\bibfnamefont {P.}~\bibnamefont {Riseborough}}, \bibinfo {author}
  {\bibfnamefont {J.}~\bibnamefont {Wang}},\ and\ \bibinfo {author}
  {\bibfnamefont {B.}~\bibnamefont {Mera}},\ }\href
  {https://arxiv.org/abs/2507.14028} {\bibinfo {title} {Density matrix geometry
  and sum rules}} (\bibinfo {year} {2025}),\ \Eprint
  {https://arxiv.org/abs/2507.14028} {arXiv:2507.14028 [cond-mat.mes-hall]}
  \BibitemShut {NoStop}%
\bibitem [{sm()}]{sm}%
  \BibitemOpen
  \href@noop {} {}\bibinfo {note} {See the Supplemental Material for details
  on: (1) the derivation of the integrated MCD; (2) the derivation of
  $\rmd\widetilde{P}/\rmd B$; (3) the projector and Green function based
  derivations of $\chi_{\rm MCD}$; (4) the analysis of a generic two-band model
  and the massive Dirac model; (5) the numerical demonstration of the absence
  of a generic chemical potential sum rule for $\chi_{\rm MCD}$; (6) the
  contribution to $M_{\rm MCD}$ from Zeeman coupling; (7) the projector
  formulation of the thermodynamic orbital susceptibility $\chi_{\rm orb}$; (8)
  the gauge choice for generic tight-binding models; and (9) the continuum
  model and numerical evaluation of $\chi_{\rm MCD}$ and $\chi_{\rm orb}$ for
  twisted bilayer graphene.}\BibitemShut {Stop}%
\bibitem [{\citenamefont {Thole}\ \emph {et~al.}(1992)\citenamefont {Thole},
  \citenamefont {Carra}, \citenamefont {Sette},\ and\ \citenamefont {van~der
  Laan}}]{LaanPRL1992}%
  \BibitemOpen
  \bibfield  {author} {\bibinfo {author} {\bibfnamefont {B.~T.}\ \bibnamefont
  {Thole}}, \bibinfo {author} {\bibfnamefont {P.}~\bibnamefont {Carra}},
  \bibinfo {author} {\bibfnamefont {F.}~\bibnamefont {Sette}},\ and\ \bibinfo
  {author} {\bibfnamefont {G.}~\bibnamefont {van~der Laan}},\ }\bibfield
  {title} {\bibinfo {title} {X-ray circular dichroism as a probe of orbital
  magnetization},\ }\href {https://doi.org/10.1103/PhysRevLett.68.1943}
  {\bibfield  {journal} {\bibinfo  {journal} {Phys. Rev. Lett.}\ }\textbf
  {\bibinfo {volume} {68}},\ \bibinfo {pages} {1943} (\bibinfo {year}
  {1992})}\BibitemShut {NoStop}%
\bibitem [{\citenamefont {Atencia}\ \emph {et~al.}(2024)\citenamefont
  {Atencia}, \citenamefont {Agarwal},\ and\ \citenamefont
  {Culcer}}]{BurgosAP2024}%
  \BibitemOpen
  \bibfield  {author} {\bibinfo {author} {\bibfnamefont {R.~B.}\ \bibnamefont
  {Atencia}}, \bibinfo {author} {\bibfnamefont {A.}~\bibnamefont {Agarwal}},\
  and\ \bibinfo {author} {\bibfnamefont {D.}~\bibnamefont {Culcer}},\
  }\bibfield  {title} {\bibinfo {title} {Orbital angular momentum of bloch
  electrons: equilibrium formulation, magneto-electric phenomena, and the
  orbital hall effect},\ }\href {https://doi.org/10.1080/23746149.2024.2371972}
  {\bibfield  {journal} {\bibinfo  {journal} {Advances in Physics: X}\ }\textbf
  {\bibinfo {volume} {9}},\ \bibinfo {pages} {2371972} (\bibinfo {year}
  {2024})}\BibitemShut {NoStop}%
\bibitem [{\citenamefont {G\'omez-Santos}\ and\ \citenamefont
  {Stauber}(2011)}]{StauberPRL2011}%
  \BibitemOpen
  \bibfield  {author} {\bibinfo {author} {\bibfnamefont {G.}~\bibnamefont
  {G\'omez-Santos}}\ and\ \bibinfo {author} {\bibfnamefont {T.}~\bibnamefont
  {Stauber}},\ }\bibfield  {title} {\bibinfo {title} {Measurable lattice
  effects on the charge and magnetic response in graphene},\ }\href
  {https://doi.org/10.1103/PhysRevLett.106.045504} {\bibfield  {journal}
  {\bibinfo  {journal} {Phys. Rev. Lett.}\ }\textbf {\bibinfo {volume} {106}},\
  \bibinfo {pages} {045504} (\bibinfo {year} {2011})}\BibitemShut {NoStop}%
\bibitem [{\citenamefont {Gao}\ \emph {et~al.}(2015)\citenamefont {Gao},
  \citenamefont {Yang},\ and\ \citenamefont {Niu}}]{GaoPRB2015}%
  \BibitemOpen
  \bibfield  {author} {\bibinfo {author} {\bibfnamefont {Y.}~\bibnamefont
  {Gao}}, \bibinfo {author} {\bibfnamefont {S.~A.}\ \bibnamefont {Yang}},\ and\
  \bibinfo {author} {\bibfnamefont {Q.}~\bibnamefont {Niu}},\ }\bibfield
  {title} {\bibinfo {title} {Geometrical effects in orbital magnetic
  susceptibility},\ }\href {https://doi.org/10.1103/PhysRevB.91.214405}
  {\bibfield  {journal} {\bibinfo  {journal} {Phys. Rev. B}\ }\textbf {\bibinfo
  {volume} {91}},\ \bibinfo {pages} {214405} (\bibinfo {year}
  {2015})}\BibitemShut {NoStop}%
\bibitem [{\citenamefont {Kang}\ \emph {et~al.}(2026)\citenamefont {Kang},
  \citenamefont {Wang},\ and\ \citenamefont {Vafek}}]{KangPRL2026}%
  \BibitemOpen
  \bibfield  {author} {\bibinfo {author} {\bibfnamefont {J.}~\bibnamefont
  {Kang}}, \bibinfo {author} {\bibfnamefont {M.}~\bibnamefont {Wang}},\ and\
  \bibinfo {author} {\bibfnamefont {O.}~\bibnamefont {Vafek}},\ }\bibfield
  {title} {\bibinfo {title} {Orbital magnetization and magnetic susceptibility
  of interacting electrons},\ }\href {https://doi.org/10.1103/9l9c-mp3t}
  {\bibfield  {journal} {\bibinfo  {journal} {Phys. Rev. Lett.}\ }\textbf
  {\bibinfo {volume} {137}},\ \bibinfo {pages} {106703} (\bibinfo {year}
  {2026})}\BibitemShut {NoStop}%
\bibitem [{\citenamefont {Huang}\ and\ \citenamefont
  {Li}(2026)}]{XiaoArXiv2026}%
  \BibitemOpen
  \bibfield  {author} {\bibinfo {author} {\bibfnamefont {K.}~\bibnamefont
  {Huang}}\ and\ \bibinfo {author} {\bibfnamefont {X.}~\bibnamefont {Li}},\
  }\href {https://arxiv.org/abs/2607.27112} {\bibinfo {title}
  {Interaction-induced sign reversal of the orbital magnetic susceptibility in
  chern insulators}} (\bibinfo {year} {2026}),\ \Eprint
  {https://arxiv.org/abs/2607.27112} {arXiv:2607.27112 [cond-mat.str-el]}
  \BibitemShut {NoStop}%
\bibitem [{\citenamefont {Li}\ \emph {et~al.}(2026)\citenamefont {Li},
  \citenamefont {Jiang}, \citenamefont {Wang}, \citenamefont {Kang},\ and\
  \citenamefont {Zhang}}]{ZhangArXiv2026}%
  \BibitemOpen
  \bibfield  {author} {\bibinfo {author} {\bibfnamefont {P.}~\bibnamefont
  {Li}}, \bibinfo {author} {\bibfnamefont {K.}~\bibnamefont {Jiang}}, \bibinfo
  {author} {\bibfnamefont {Z.}~\bibnamefont {Wang}}, \bibinfo {author}
  {\bibfnamefont {J.}~\bibnamefont {Kang}},\ and\ \bibinfo {author}
  {\bibfnamefont {Y.}~\bibnamefont {Zhang}},\ }\href
  {https://arxiv.org/abs/2603.03702} {\bibinfo {title} {Contribution of remote
  bands to orbital magnetization in twisted bilayer graphene}} (\bibinfo {year}
  {2026}),\ \Eprint {https://arxiv.org/abs/2603.03702} {arXiv:2603.03702
  [cond-mat.str-el]} \BibitemShut {NoStop}%
\bibitem [{\citenamefont {Bychkov}\ and\ \citenamefont
  {Gor'kov}(1962)}]{Gorkov62}%
  \BibitemOpen
  \bibfield  {author} {\bibinfo {author} {\bibfnamefont {Y.~A.}\ \bibnamefont
  {Bychkov}}\ and\ \bibinfo {author} {\bibfnamefont {L.~P.}\ \bibnamefont
  {Gor'kov}},\ }\bibfield  {title} {\bibinfo {title} {Quantum oscillations of
  the thermodynamic quantities of a metal in a magnetic field according to the
  fermi-liquid model},\ }\href@noop {} {\bibfield  {journal} {\bibinfo
  {journal} {Sov. Phys. JETP.}\ }\textbf {\bibinfo {volume} {14}},\ \bibinfo
  {pages} {1132} (\bibinfo {year} {1962})}\BibitemShut {NoStop}%
\bibitem [{\citenamefont {Chen}\ and\ \citenamefont {Lee}(2011)}]{LeePRB2011}%
  \BibitemOpen
  \bibfield  {author} {\bibinfo {author} {\bibfnamefont {K.-T.}\ \bibnamefont
  {Chen}}\ and\ \bibinfo {author} {\bibfnamefont {P.~A.}\ \bibnamefont {Lee}},\
  }\bibfield  {title} {\bibinfo {title} {Unified formalism for calculating
  polarization, magnetization, and more in a periodic insulator},\ }\href
  {https://doi.org/10.1103/PhysRevB.84.205137} {\bibfield  {journal} {\bibinfo
  {journal} {Phys. Rev. B}\ }\textbf {\bibinfo {volume} {84}},\ \bibinfo
  {pages} {205137} (\bibinfo {year} {2011})}\BibitemShut {NoStop}%
\bibitem [{\citenamefont {Ye}(2026)}]{YeArXiv2026}%
  \BibitemOpen
  \bibfield  {author} {\bibinfo {author} {\bibfnamefont {M.}~\bibnamefont
  {Ye}},\ }\href {https://arxiv.org/abs/2601.14372} {\bibinfo {title} {A
  quantum many-body approach for orbital magnetism in correlated multiband
  electron systems}} (\bibinfo {year} {2026}),\ \Eprint
  {https://arxiv.org/abs/2601.14372} {arXiv:2601.14372 [cond-mat.str-el]}
  \BibitemShut {NoStop}%
\bibitem [{\citenamefont {Essin}\ \emph {et~al.}(2010)\citenamefont {Essin},
  \citenamefont {Turner}, \citenamefont {Moore},\ and\ \citenamefont
  {Vanderbilt}}]{VanderbiltPRB2010}%
  \BibitemOpen
  \bibfield  {author} {\bibinfo {author} {\bibfnamefont {A.~M.}\ \bibnamefont
  {Essin}}, \bibinfo {author} {\bibfnamefont {A.~M.}\ \bibnamefont {Turner}},
  \bibinfo {author} {\bibfnamefont {J.~E.}\ \bibnamefont {Moore}},\ and\
  \bibinfo {author} {\bibfnamefont {D.}~\bibnamefont {Vanderbilt}},\ }\bibfield
   {title} {\bibinfo {title} {Orbital magnetoelectric coupling in band
  insulators},\ }\href {https://doi.org/10.1103/PhysRevB.81.205104} {\bibfield
  {journal} {\bibinfo  {journal} {Phys. Rev. B}\ }\textbf {\bibinfo {volume}
  {81}},\ \bibinfo {pages} {205104} (\bibinfo {year} {2010})}\BibitemShut
  {NoStop}%
\bibitem [{\citenamefont {Shindou}\ and\ \citenamefont
  {Imura}(2005)}]{ShindouNPB2005}%
  \BibitemOpen
  \bibfield  {author} {\bibinfo {author} {\bibfnamefont {R.}~\bibnamefont
  {Shindou}}\ and\ \bibinfo {author} {\bibfnamefont {K.-I.}\ \bibnamefont
  {Imura}},\ }\bibfield  {title} {\bibinfo {title} {Noncommutative geometry and
  non-abelian berry phase in the wave-packet dynamics of bloch electrons},\
  }\href {https://doi.org/https://doi.org/10.1016/j.nuclphysb.2005.05.019}
  {\bibfield  {journal} {\bibinfo  {journal} {Nuclear Physics B}\ }\textbf
  {\bibinfo {volume} {720}},\ \bibinfo {pages} {399} (\bibinfo {year}
  {2005})}\BibitemShut {NoStop}%
\bibitem [{\citenamefont {Gao}\ \emph {et~al.}(2014)\citenamefont {Gao},
  \citenamefont {Yang},\ and\ \citenamefont {Niu}}]{GaoPRL2014}%
  \BibitemOpen
  \bibfield  {author} {\bibinfo {author} {\bibfnamefont {Y.}~\bibnamefont
  {Gao}}, \bibinfo {author} {\bibfnamefont {S.~A.}\ \bibnamefont {Yang}},\ and\
  \bibinfo {author} {\bibfnamefont {Q.}~\bibnamefont {Niu}},\ }\bibfield
  {title} {\bibinfo {title} {Field induced positional shift of bloch electrons
  and its dynamical implications},\ }\href
  {https://doi.org/10.1103/PhysRevLett.112.166601} {\bibfield  {journal}
  {\bibinfo  {journal} {Phys. Rev. Lett.}\ }\textbf {\bibinfo {volume} {112}},\
  \bibinfo {pages} {166601} (\bibinfo {year} {2014})}\BibitemShut {NoStop}%
\bibitem [{\citenamefont {Guti\'errez-Rubio}\ \emph {et~al.}(2016)\citenamefont
  {Guti\'errez-Rubio}, \citenamefont {Stauber}, \citenamefont {G\'omez-Santos},
  \citenamefont {Asgari},\ and\ \citenamefont {Guinea}}]{StauberPRB2016}%
  \BibitemOpen
  \bibfield  {author} {\bibinfo {author} {\bibfnamefont {A.}~\bibnamefont
  {Guti\'errez-Rubio}}, \bibinfo {author} {\bibfnamefont {T.}~\bibnamefont
  {Stauber}}, \bibinfo {author} {\bibfnamefont {G.}~\bibnamefont
  {G\'omez-Santos}}, \bibinfo {author} {\bibfnamefont {R.}~\bibnamefont
  {Asgari}},\ and\ \bibinfo {author} {\bibfnamefont {F.}~\bibnamefont
  {Guinea}},\ }\bibfield  {title} {\bibinfo {title} {Orbital magnetic
  susceptibility of graphene and ${\mathrm{mos}}_{2}$},\ }\href
  {https://doi.org/10.1103/PhysRevB.93.085133} {\bibfield  {journal} {\bibinfo
  {journal} {Phys. Rev. B}\ }\textbf {\bibinfo {volume} {93}},\ \bibinfo
  {pages} {085133} (\bibinfo {year} {2016})}\BibitemShut {NoStop}%
\bibitem [{\citenamefont {Raoux}\ \emph {et~al.}(2014)\citenamefont {Raoux},
  \citenamefont {Morigi}, \citenamefont {Fuchs}, \citenamefont {Pi\'echon},\
  and\ \citenamefont {Montambaux}}]{RaouxPRL2014}%
  \BibitemOpen
  \bibfield  {author} {\bibinfo {author} {\bibfnamefont {A.}~\bibnamefont
  {Raoux}}, \bibinfo {author} {\bibfnamefont {M.}~\bibnamefont {Morigi}},
  \bibinfo {author} {\bibfnamefont {J.-N.}\ \bibnamefont {Fuchs}}, \bibinfo
  {author} {\bibfnamefont {F.}~\bibnamefont {Pi\'echon}},\ and\ \bibinfo
  {author} {\bibfnamefont {G.}~\bibnamefont {Montambaux}},\ }\bibfield  {title}
  {\bibinfo {title} {From dia- to paramagnetic orbital susceptibility of
  massless fermions},\ }\href {https://doi.org/10.1103/PhysRevLett.112.026402}
  {\bibfield  {journal} {\bibinfo  {journal} {Phys. Rev. Lett.}\ }\textbf
  {\bibinfo {volume} {112}},\ \bibinfo {pages} {026402} (\bibinfo {year}
  {2014})}\BibitemShut {NoStop}%
\bibitem [{\citenamefont {Bistritzer}\ and\ \citenamefont
  {MacDonald}(2011)}]{mac_2011}%
  \BibitemOpen
  \bibfield  {author} {\bibinfo {author} {\bibfnamefont {R.}~\bibnamefont
  {Bistritzer}}\ and\ \bibinfo {author} {\bibfnamefont {A.~H.}\ \bibnamefont
  {MacDonald}},\ }\bibfield  {title} {\bibinfo {title} {Moiré bands in twisted
  double-layer graphene},\ }\href {https://doi.org/10.1073/pnas.1108174108}
  {\bibfield  {journal} {\bibinfo  {journal} {Proceedings of the National
  Academy of Sciences}\ }\textbf {\bibinfo {volume} {108}},\ \bibinfo {pages}
  {12233} (\bibinfo {year} {2011})}\BibitemShut {NoStop}%
\bibitem [{\citenamefont {Venderbos}(2025)}]{VenderbosArXiv2512}%
  \BibitemOpen
  \bibfield  {author} {\bibinfo {author} {\bibfnamefont {J.~W.~F.}\
  \bibnamefont {Venderbos}},\ }\href {https://arxiv.org/abs/2512.19380}
  {\bibinfo {title} {Berry phase polarization and orbital magnetization
  responses of insulators: Formulas for generalized polarizabilities and their
  application}} (\bibinfo {year} {2025}),\ \Eprint
  {https://arxiv.org/abs/2512.19380} {arXiv:2512.19380 [cond-mat.other]}
  \BibitemShut {NoStop}%
\bibitem [{\citenamefont {Mizoguchi}\ and\ \citenamefont
  {Okuma}(2024)}]{OkumaJPSJ2024}%
  \BibitemOpen
  \bibfield  {author} {\bibinfo {author} {\bibfnamefont {T.}~\bibnamefont
  {Mizoguchi}}\ and\ \bibinfo {author} {\bibfnamefont {N.}~\bibnamefont
  {Okuma}},\ }\bibfield  {title} {\bibinfo {title} {On equivalence of two
  formulas of orbital magnetic susceptibility for tight-binding models},\
  }\href {https://doi.org/10.7566/JPSJ.93.095002} {\bibfield  {journal}
  {\bibinfo  {journal} {Journal of the Physical Society of Japan}\ }\textbf
  {\bibinfo {volume} {93}},\ \bibinfo {pages} {095002} (\bibinfo {year}
  {2024})}\BibitemShut {NoStop}%
\end{thebibliography}%


\newpage

\clearpage
\onecolumngrid
\begin{center}
\textbf{\large Supplemental Material: Field Response of the Orbital Moment Probed by Integrated Magnetic Circular Dichroism}
\end{center}

\setcounter{equation}{0}
\setcounter{figure}{0}
\setcounter{table}{0}
\setcounter{page}{1}
\setcounter{section}{0}
\makeatletter
\renewcommand{\theequation}{S\arabic{equation}}
\renewcommand{\thefigure}{S\arabic{figure}}
\renewcommand{\bibnumfmt}[1]{[S#1]}
\renewcommand{\citenumfont}[1]{S#1}

\section{Notational convention}
We introduce a ``Fermi sea'' derivative, denoted by $\bpd_\lambda$, that acts only on the eigenstates while keeping their occupation factors fixed. For
\begin{align}
    \hat P = \sum_n n_F(E_n) \hat P_n  \ ,  \qquad \hat P_n = | u_n \rangle \langle u_n| \ ,
\end{align}
where $n_F(E_n) = \Theta(\mu - E_n)$ is the zero temperature Fermi distribution function. The Fermi-sea derivative is defined as
\begin{align}
    \bpd_{\lambda} \hat P \equiv \sum_{n \in \rm occ} \left( | \pd_{\lambda} u_n \rangle \langle u_n |  +  | u_n \rangle \langle \pd_{\lambda} u_n |  \right)
\end{align}
At zero temperature, the sum is restricted to the occupied states. The full derivative of $\hat P$ is then
\begin{align}
    \pd_{\lambda} \hat P = \bpd_{\lambda} \hat P + \sum_n n_F'(E_n) \big( \pd_{\lambda} E_n \big) \hat P_n
\end{align}
Only bands that cross the Fermi surface contribute to the second term. Similarly, for $\hat Q = \hat I - \hat P$, we have
\begin{align}
    \pd_{\lambda} \hat Q = \bpd_{\lambda} \hat Q - \sum_n n_F'(E_n) \big( \pd_{\lambda} E_n \big) \hat P_n
\end{align}

We next consider a band $n$ that crosses the Fermi surface. Assuming that this band is nondegenerate in the vicinity of the Fermi surface, we define the projectors onto the remaining occupied and empty subspaces as
\begin{align}
    \hat P_{\sn} & = \sum_{\substack{m \neq n \\ m \in \rm occ}} | u_m \rangle \langle u_m | \ ,   &
    \hat Q_{\sn} & = \sum_{\substack{m \neq n \\ m \in \rm emp}} | u_m \rangle \langle u_m | \ .
\end{align}
These projectors provide the decomposition $\hat P_n + \hat P_{\sn} + \hat Q_{\sn} = \hat I$. They are mutually orthogonal:
\begin{align}
    \hat P_n\hat P_{\sn} & = \hat P_n \hat Q_{\sn}  = \hat Q_{\sn} \hat P_{\sn} = 0  \ .
\end{align}

\section{MCD orbital magnetization}
In this section, we discuss how self-rotation part of the orbital magnetization can be extracted from magnetic circular dichroism (MCD)~\cite{RestaPRR2020}. We begin by considering the optical conductivity for incident light with polarization vector $\hat{\varepsilon}$. Throughout this discussion, we work within the electric dipole approximation and choose the gauge $\phi=0$ and $\fvec A = \fvec A(t)$, such that the vector potential is spatially uniform and time dependent. The corresponding electric field is
\begin{align}
    \fvec E(t)  = - \frac{\pd \fvec A}{\pd t} \ ,   \qquad  \fvec E(\omega)  = i\omega A(\omega)\hat{\varepsilon} \  .
\end{align}
The velocity operator is defined as
\begin{align}
    \hat V_{\mu}  = \frac1{\hbar} \frac{\pd H_0}{\pd k_{\mu}} = \frac1{i\hbar}[\hat r_{\mu}\ , \hat H_0]  \ ,
\end{align}
where the commutator representation remains valid for general systems, including aperiodic systems. Here $H_0$ denotes the noninteracting Hamiltonian. 

For a non-interacting system, the optical conductivity for light with
polarization $\hat{\varepsilon}$ can be expressed in terms of the
single-particle Green functions as
\begin{align}
    \sigma(\omega)  = -\frac{e^2}{i\omega}  \sum_{\alpha \beta} \left. \oint_{\calC} \frac{\rmd z}{2\pi i} n_F(z) G_{\alpha}(z) G_{\beta}(z + i \omega_n ) \right|_{i \omega_n \rightarrow \hbar\omega + i 0^+} ( \hat{\varepsilon} \cdot \fvec V)_{\alpha\beta}  ( \hat{\varepsilon}^* \cdot \fvec V)_{\beta\alpha}   ,
\end{align}
Here, $\calC$ encloses the real axis. The indices $\alpha$ and $\beta$ label single particle eigenstates. Evaluating the contour integral and performing the analytic continuation yield
\begin{align}
    \sigma(\omega) = \frac{i e^2}{\omega}  \sum_{\alpha,\beta} \frac{ ( \hat{\varepsilon} \cdot \fvec V)_{\alpha\beta}
    (\hat{\varepsilon}^{*}\cdot\fvec V)_{\beta\alpha} }{ \hbar\omega + E_{\alpha} - E_{\beta} + i 0^+ } \big[ n_F(E_{\alpha}) - n_F(E_{\beta}) \big]  \  .
\end{align}
MCD measures the difference in absorption between right and left circularly polarized light. We therefore focus on the dissipative part of the optical conductivity. At positive frequencies, it is given by
\begin{align}
    \operatorname{Re} \sigma(\omega) = \frac{\pi e^2}{\omega } \sum_{\alpha, \beta} (\hat{\varepsilon} \cdot \fvec V)_{\alpha\beta} ( \hat{\varepsilon}^* \cdot \fvec V)_{\beta \alpha} \delta(\hbar\omega + E_{\alpha} - E_{\beta})
    \left[ n_F(E_{\alpha})-n_F(E_{\beta})  \right]   \  .
\end{align}
Note that the zero frequency Drude contribution is not included here. At $T = 0$, only transitions from occupied to empty states
contribute. Integrating over positive frequencies yields
\begin{align}
    \int_{0^+}^{\infty} \rmd \omega\, \oRe \sigma(\omega)  = \pi e^2 \sum_{\alpha\in{\rm occ}} \sum_{\beta\in{\rm emp}}  (\hat{\varepsilon}\cdot\fvec V)_{\alpha\beta}  (\hat{\varepsilon}^{*}\cdot\fvec V)_{\beta\alpha} \frac1{E_{\beta}-E_{\alpha}} \ .
\end{align}

We choose the two circular polarization vectors as $\hat{\varepsilon}_{\pm} = \left( \hat x \pm i\hat y \right)/\sqrt{2}$, and the corresponding integrated optical conductivities become
\begin{align}
    \int_{0^+}^{\infty}\rmd\omega\oRe\sigma_{\pm}(\omega) = \frac{\pi e^2}{2} \sum_{\alpha\in\mathrm{occ}} \sum_{\beta\in\mathrm{emp}} \frac{1}{E_{\beta} - E_{\alpha}} \Big[ V^x_{\alpha\beta} V^x_{\beta\alpha} + V^y_{\alpha\beta} V^y_{\beta\alpha} \pm i \left( V^y_{\alpha\beta} V^x_{\beta\alpha} - V^x_{\alpha\beta} V^y_{\beta\alpha} \right) \Big]   \   .
\end{align}
Here, $V^\mu_{\alpha\beta} = \langle \alpha | \hat V_\mu | \beta \rangle$. The integrated MCD signal is defined as the difference between the two circularly polarized responses:
\begin{align}
 I_{\rm MCD} & = \int_{0^+}^{\infty} \rmd \omega\, \oRe\left[  \sigma_+(\omega) - \sigma_-(\omega) \right] = 2\pi e^2 \sum_{\alpha\in \rm occ} \sum_{\beta \in \rm emp} \frac{ \oIm \left(
V^x_{\alpha\beta} V^y_{\beta\alpha} \right) }{ E_{\beta}-E_{\alpha} }  \   . \label{Eqn:IMCD}
\end{align}

For a periodic system, the single particle eigenstates can be labeled by the crystal momentum $\fvec k$ and the band index $n$. For $m\neq n$, the velocity matrix elements obey
\begin{align}
    \langle u_m(\fvec k) | \hat V^{\mu} | u_n(\fvec k) \rangle = \hbar^{-1} (E_m-E_n) \langle \pd_{\mu} u_m | u_n \rangle \ .
\end{align}
where $\pd_{\mu} \equiv \pd/\pd k_{\mu}$. Substituting the velocity matrix elements into Eq.~\eqref{Eqn:IMCD} gives
\begin{align}
    I_{\rm MCD} & = \frac{2\pi e^2}{\hbar^2} \sum_{\fvec k} \sum_{n\in{\rm occ}} \sum_{m\in{\rm emp}} \oIm \left[ \langle \pd_x u_n | u_m \rangle \langle u_m | \pd_y u_n \rangle \right]
    (E_m - E_n)   \ .
\end{align}

We further introduce the projectors onto the occupied and empty subspaces,
\begin{align}
    P_{\fvec k} & = \sum_{n \in \rm occ} | u_{n\fvec k} \rangle \langle u_{n\fvec k} |  \ ,  &  Q_{\fvec k} & = I - P_{\fvec k}.
\end{align}
Using completeness within the empty subspace, we obtain
\begin{align}
    I_{\rm MCD} = \frac{2\pi e^2}{\hbar^2 } \sum_{\fvec k} \sum_{n \in \rm occ} \oIm \langle \pd_x u_{n \fvec k} | Q_{\fvec k} \left( \hat H_{\fvec k} - E_{n\fvec k} \right) Q_{\fvec k} | \pd_y u_{n\fvec k} \rangle  \  .
\end{align}
This result motivates the definition
\begin{align}
    M_{\rm MCD} & = \frac{\hbar}{2\pi e} I_{\rm MCD} = \frac{e}{\hbar } \sum_{\fvec k} \sum_{n \in \rm occ} \oIm \langle \pd_x u_{n\fvec k} | Q_{\fvec k} \left(\hat H_{\fvec k} - E_{n\fvec k} \right) Q_{\fvec k} | \pd_y u_{n\fvec k} \rangle \equiv M_{\rm SR} \ .
\end{align}
Thus, within the independent particle approximation, the frequency integrated MCD signal directly measures the self-rotation contribution to the orbital magnetization.

The MCD moment can be further written as
\begin{align}
    M_{\rm MCD} & = e \hbar \oIm \sum_{\alpha \in \rm occ} \sum_{\beta \in \rm emp} \frac{ \mtr\left( \hat P_{\alpha} \hat V_x \hat P_{\beta} \hat V_y \right) }{ E_{\beta} - E_{\alpha} } = \frac{e\hbar}{2 i} \epsilon_{\mu\nu} \sum_{\alpha \in \rm occ} \sum_{\beta \in \rm emp} \frac{ \mtr\left( \hat P_{\alpha} \hat V_{\mu} \hat P_{\beta} \hat V_{\nu} \right) }{ E_{\beta} -E_{\alpha} } \nonumber\\
    & = \frac{ie}{2 \hbar} \epsilon_{\mu\nu} \mtr\left( \hat P \hat r_{\mu} \hat Q[\hat r_{\nu},\hat H] \right) = \frac{ie}{2 \hbar} \epsilon_{\mu\nu} \mtr\left( \hat P\hat r_{\mu}\hat Q\hat r_{\nu}\hat H - \hat P\hat r_{\mu}\hat Q\hat H\hat r_{\nu} \right) \  .
\end{align}
Here, $\hat P_{\alpha} = | \alpha \rangle \langle \alpha|$, $\hat P = \sum_{\alpha\in\mathrm{occ}}\hat P_{\alpha}$, and $\hat Q=1-\hat P$.

The expression above is directly applicable to finite systems. For infinite or periodic systems, the position operator is not well defined. We therefore rewrite it in terms of commutators. Using
\begin{align}
    \epsilon_{\mu\nu} \hat r_{\mu} \hat r_{\nu} & = 0 \ , & [\hat H,\hat P] & = [\hat H, \hat Q] = 0  \ , & \hat P\hat Q & = 0  \  ,
\end{align}
together with the cyclic property of the trace, we find
\begin{align}
    M_{\rm MCD} = \frac{ie}{2 \hbar} \epsilon_{\mu\nu} \mtr\left[ \hat H [\hat r_{\mu},\hat P] [\hat r_{\nu},\hat Q] \right] \ .
\end{align}
Although $\hat{\fvec r}$ itself is not compatible with periodic
boundary conditions, the commutators $[\hat r_{\mu},\hat P]$ and $[\hat r_{\mu},\hat Q]$ remain well defined for an insulating system. The last expression, therefore, extends naturally to periodic systems.

In momentum space, $[\hat r_{\mu},\hat P]$ is represented by
$i\partial_{\mu}\hat P$. We thus obtain
\begin{align}
    M_{\rm MCD} & = \frac{e}{2 i \hbar} \epsilon_{\mu\nu} \mtr\left[ \hat H (\pd_{\mu}\hat P) (\pd_{\nu}\hat Q) \right] = \frac{e}{2\hbar} \epsilon_{\mu\nu} \oIm \mtr\left[ \hat H (\pd_{\mu} \hat P) (\pd_{\nu} \hat Q) \right] \  .
\end{align}
Here, the trace includes the Brillouin zone sum and the trace over internal degrees of freedom. This form will be useful below when evaluating $\chi_{\rm MCD}$ in insulating states.

\section{Translational invariance of EM Gauge Invariant Operators} 
In this section, we develop an EM $U(1)$ gauge invariant formulation. In the presence of an external magnetic field, the Hamiltonian, Green function, band projectors, and density matrices are generally not invariant under ordinary lattice translations. Instead, the appropriate symmetry is magnetic translation. For a lattice system with rational magnetic flux, this symmetry is commonly represented using an enlarged magnetic unit cell, whose particular realization depends on the gauge choice. EM gauge invariant quantities, however, cannot depend on this choice and must preserve the periodicity of the underlying zero field lattice.

For concreteness, consider a non-interacting Hamiltonian that, in the presence of a uniform magnetic field, can be written as
\begin{align}
    t_x = \pi_x + \frac{y}{l^2} \ , \quad t_y = \pi_y - \frac{x}{l^2} 
\end{align}
It is easy to check that $[t_i, \pi_j] = 0$. As a consequence, $\hat{t}_i$ can be identified as the generator of the magnetic translation. For simplicity, we choose the Landau gauge $\fvec A = (0, B x)$, then
\[   U_{\fvec L} = e^{-i \fvec t \cdot \fvec L}  \quad \Longrightarrow \quad U_{\fvec L} | \fvec r \rangle =   e^{-i \frac1{l^2}(r_y L_x + \half L_x L_y)} |\fvec r + \fvec L \rangle \quad \mbox{and} \quad U_{\fvec L} c^{\dagger}_{\fvec r} U_{\fvec L}^{\dagger} =   e^{-i \frac1{l^2}(r_y L_x + \half L_x L_y)} c_{\fvec r + \fvec L}^{\dagger} \]
where $\fvec L$ is the lattice vector. Assuming that the state in invariant under magnetic translation, we conclude that
\begin{align}
    G(\fvec r, \fvec r') = \langle c^{\dagger}_{\fvec r'} c_{\fvec r} \rangle = \langle U_{\fvec L} c^{\dagger}_{\fvec r'} c_{\fvec r} U_{\fvec L}^{\dagger} \rangle = G(\fvec r + \fvec L, \fvec r' + \fvec L)  \exp\left( -\frac{i}{l^2} (r'_y - r_y) L_x \right)
\end{align}
To recover the translational invariance of the Green function, we introduce the function:
\begin{align}
    \Phi(\fvec r, \fvec r') = \int_{\fvec r'}^{\fvec r} \fvec A(\fvec r'') \cdot \rmd \fvec r'' = \fvec A\left( \frac{\fvec r + \fvec r'}2 \right) \cdot (\fvec r - \fvec r') = \frac{B}2 (r_x + r'_x)(r_y - r'_y)
\end{align}
Now, consider the gauge-independent Green function $\widetilde{G}(\fvec r, \fvec r') = G(\fvec r, \fvec r') e^{i e \Phi(\fvec r, \fvec r')/\hbar}$. Under the magnetic translation, we have
\begin{align}
    & \quad \widetilde{G}(\fvec r, \fvec r') = G(\fvec r, \fvec r') \exp\left( \frac{i}{2 l^2} (r_x + r'_x)(r_y - r'_y) \right) \nonumber \\
    & = G(\fvec r + \fvec L, \fvec r' + \fvec L) \exp\left( \frac{i}{2l^2} \left( (r_x + r'_x)(r_y - r'_y) + 2 (r_y - r'_y)  L_x \right) \right) \nonumber \\
    & = \widetilde{G}(\fvec r + \fvec L, \fvec r' + \fvec L) \exp\left( \frac{i}{2l^2} \left( (r_x + r'_x)(r_y - r'_y) + 2 (r_y - r'_y)  L_x - (r_x + r'_x + 2 L_x)(r_y - r'_y) \right) \right) \nonumber \\
    & = \widetilde{G}(\fvec r + \fvec L, \fvec r' + \fvec L) 
\end{align}
We will see that this property is crucial for studying the orbital magnetism since $\widetilde{G}$, instead of $\fvec G$ can be expanded in terms of the eigenstates of the zero field Hamiltonian.

\section{$\rmd \widetilde{P} / \rmd B$ for insulating and Metallic systems}
Ref.~\cite{LeePRB2011,RaouxPRB2015,KangPRL2026} derived an expression for $\rmd\widetilde G/\rmd B$. Here, we use this result to obtain the corresponding expression for the occupied state projector $\widetilde P$. We assume that the chemical potential lies inside a gap. We then consider the first order change of $\widetilde P$ under a magnetic field:
\begin{align}
    \left. \frac{\rmd \widetilde{P}}{\rmd B} \right|_{B = 0} & = \frac{e \hbar}{2 i} \epsilon_{\mu\nu} \oint_{\calC'} \frac{\rmd z}{2\pi i } n_F(z) \hat G_0 \hat V^{\mu} G_0 \hat V^{\nu} G_0  = \frac{e \hbar}{2 i} \epsilon_{\mu\nu} \oint_{\calC'} \frac{\rmd z}{2\pi i} n_F(z)  \frac{\hat P_a \hat V^{\mu} \hat P_b \hat V^{\nu} \hat P_c}{(z - E_a) (z - E_b) (z - E_c)} 
\end{align}
where the contour $\calC'$ encloses the whole real axis.  The contour integral can be evaluated from its residues. We considr only the zero-temperature case and separate the result into the following cases.
\begin{enumerate}
    \item Only state $a$ is occupied, or only state $a$ is empty:
    \begin{align}
        & \quad \frac{e \hbar}{2 i } \epsilon_{\mu\nu} \frac{n_F(E_a) n_F(- E_b) n_F(- E_c) - n_F(-E_a) n_F(E_b) n_F(E_c)}{(E_a - E_b) (E_a - E_c)} \hat P_a \hat V^{\mu} \hat P_b \hat V^{\nu} \hat P_c \nonumber \\
        & =  \frac{e }{2 i \hbar } \epsilon_{\mu\nu} \big[ n_F(E_a) n_F(- E_b) n_F(- E_c) - n_F(- E_a) n_F(E_b) n_F(E_c) \big] (-) \hat P_a \calL^{-1}\left( \big( \pd_{\mu} \hat P_b \big) v^{\nu} \right) \hat P_c \nonumber \\
        & = \frac{e }{2 i \hbar } \epsilon_{\mu\nu}  \left[ - \hat P \calL^{-1}\left( \big( \pd_{\mu} \hat Q \big) \hat V^{\nu} \right) \hat Q + \hat Q \calL^{-1}\left(  \big( \pd_{\mu} \hat P \big) \hat V^{\nu} \right) \hat P    \right]
    \end{align}
    \item Only state $b$ is occupied, or only state $b$ is empty:
        \begin{align}
        & \quad \frac{e \hbar}{2 i } \epsilon_{\mu\nu} \frac{n_F(E_b) n_F(- E_a) n_F(- E_c) - n_F(-E_b) n_F(E_a) n_F(E_c)}{(E_b - E_a) (E_b - E_c)} \hat P_a \hat V^{\mu} \hat P_b \hat V^{\nu} \hat P_c \nonumber \\
        & =  \frac{e }{2 i\hbar } \epsilon_{\mu\nu} \Big[ n_F(E_b) n_F(- E_a) n_F(- E_c) - n_F(-E_b) n_F(E_a) n_F(E_c) \Big]  \big( \pd_{\mu} \hat P_a \big) \hat P_b \big( \pd_{\nu}  \hat P_c \big) \nonumber \\
        & = \frac{e }{2 i\hbar } \epsilon_{\mu\nu}  \left[ \big( \pd_{\mu} \hat Q \big) \hat P  \big( \pd_{\nu} \hat Q \big) - \big( \pd_{\mu} \hat P \big) \hat Q  \big( \pd_{\nu} \hat P \big) \right]
    \end{align}
    \item Only state $c$ is occupied, or only state $c$ is empty:
    \begin{align}
        & \quad \frac{e \hbar}{2 i } \epsilon_{\mu\nu} \frac{n_F(E_c) n_F(- E_b) n_F(- E_a) - n_F(-E_c) n_F(E_b) n_F(E_a)}{(E_c - E_b) (E_c - E_a)} \hat P_a \hat V^{\mu} \hat P_b \hat V^{\nu} \hat P_c \nonumber \\
        & =  \frac{e}{2 i \hbar } \epsilon_{\mu\nu} \left( n_F(E_c) n_F(- E_b) n_F(- E_a) - n_F(-E_c) n_F(E_b) n_F(E_a) \right)  \hat P_a \calL^{-1}\left( V^{\mu} \big( \pd_{\nu} \hat P_b \big) \right) \hat P_c \nonumber \\
        & = \frac{e}{2 i \hbar } \epsilon_{\mu\nu}  \left[ - \hat P \calL^{-1}\left( \hat V^{\mu} \big( \pd_{\nu} \hat P \big)  \right) \hat Q + \hat Q \calL^{-1}\left(  \hat V^{\mu} \big( \pd_{\nu} \hat Q \big) \right) \hat P    \right]
    \end{align}
\end{enumerate}
Therefore, for insulating systems, we obtained
\begin{align}
    \left. \frac{\rmd \widetilde{P}}{\rmd B} \right|_{B = 0} & = \frac{e}{2 i \hbar } \epsilon_{\mu\nu} \left\{ - \big( \pd_{\mu} \hat P \big) \hat Q \big( \pd_{\nu} \hat P \big) + \big( \pd_{\mu} \hat Q \big) \hat P \big( \pd_{\nu} \hat Q \big) + \hat Q \calL^{-1}\left( V^{\mu} \big( \pd_{\nu} \hat Q \big) + \big( \pd_{\mu} \hat P \big) V^{\nu} \right) \hat P - \hat P \calL^{-1}\left( V^{\mu} \big( \pd_{\nu} \hat P \big) + \big( \pd_{\mu} \hat Q \big) V^{\nu} \right) \hat Q \right\}  \label{EqS:dPdBIns}
\end{align}
The antisymmetric parts of the first two terms are related to the non-Abelian Berry curvatures of the occupied and empty subspaces. We define them as
\begin{align}
    \calF_{\mu\nu}^P & = i \hat P \left[ \pd_\mu \hat P , \pd_\nu \hat P \right] \hat P  \qquad \calF_{\mu\nu}^Q = i \hat Q \left[ \pd_\mu \hat Q , \pd_\nu \hat Q \right] \hat Q   \  .
\end{align}
Using these definitions, the result can be written as
\begin{align}
    \left. \frac{\rmd \widetilde{P}}{\rmd B} \right|_{B = 0} & = \frac{e}{4 \hbar} \epsilon_{\mu\nu} \left( \calF^P_{\mu\nu} - \calF^Q_{\mu\nu} \right) + \frac{e}{2 i \hbar} \epsilon_{\mu\nu} \left\{ \hat Q \calL^{-1}\left( V^{\mu} \big( \pd_{\nu} \hat Q \big) + \big( \pd_{\mu} \hat P \big) V^{\nu} \right) \hat P - \hat P \calL^{-1}\left( V^{\mu} \big( \pd_{\nu} \hat P \big) + \big( \pd_{\mu} \hat Q \big) V^{\nu} \right) \hat Q \right\}  \label{Eqn:dPdBCompact}
\end{align}
We will use this result to derive the response function $\chi_{\rm MCD}$. The curvature term in Eq.~\eqref{Eqn:dPdBCompact} captures the projector space geometric structure associated with the horizontal mixing discussed in Ref.~\cite{GaoPRB2015}. The two Liouvillian terms are off diagonal
with respect to the occupied empty decomposition and describe the corresponding vertical interband mixing. Eq.~\eqref{Eqn:dPdBCompact} therefore provides a generalization of the single-band wave-packet construction to multiband systems with non-Abelian band geometry.

For metallic systems, $\rmd \widetilde P/\rmd B$ also contains the following contributions from Fermi surfaces:
\begin{enumerate}    
    \item The poles satisfy $a=b\neq c$:
    \begin{align}
        & \quad \frac{e}{2 i} \epsilon_{\mu\nu} \big( \pd_{\mu} E_a \big) n_F'(E_a) \frac{V^{\nu}_{ac}}{E_a - E_c} = \frac{e}{2i \hbar } \epsilon_{\mu\nu}  \sum_a n_F'(E_a) \big( \pd_{\mu} E_a \big) \big( \pd_{\nu} \hat P_a \big) \hat Q_a 
    \end{align}
    \item The poles satisfy $b=c\neq a$:
    \begin{align}
        & \quad \frac{e}{2 i} \epsilon_{\mu\nu} \big( \pd_{\nu} E_b \big) n_F'(E_b) \frac{V^{\mu}_{ab}}{E_b - E_a}  = - \frac{e }{2 i \hbar } \epsilon_{\mu\nu}  \sum_a n_F'(E_a) \big( \pd_{\mu} E_a \big) \hat Q_a \big( \pd_{\nu} \hat P_a \big) 
    \end{align}
    \item The poles satisfy $a = c \neq b$:
    \begin{align}
        & \quad \frac{e \hbar}{2 i } \epsilon_{\mu\nu} n_F'(E_a) \frac{V^{\mu}_{ab} V^{\nu}_{ba}}{E_a - E_b} = - \sum_a n_F'(E_a) m_a \hat P_a
    \end{align}
\end{enumerate}
In the last case, we have used the expression for orbital magnetic moment
\begin{align}
    m_a = \frac{e}{2 i \hbar } \epsilon_{\mu \nu} \langle \pd_{\mu} u_a | (\hat H - E_a) | \pd_{\nu} u_a \rangle = - \frac{e \hbar}{2 i  } \epsilon_{\mu \nu} \sum_{b \neq a} \frac{ V^{\mu}_{ab} V^{\nu}_{ba} }{E_a - E_b}
\end{align}

Combining the Fermi sea and Fermi surface contributions, we obtain
\begin{align}
    \left. \frac{\rmd \widetilde P}{\rmd B} \right|_{B = 0} & = \frac{e }{2 i \hbar } \epsilon_{\mu\nu} \left\{ -  \big( \bpd_{\mu} \hat P \big) \hat Q \big( \bpd_{\nu} \hat P \big) + \big( \bpd_{\mu} \hat Q \big) \hat P \big( \bpd_{\nu} \hat Q \big) + \hat Q \calL^{-1}\left( V^{\mu} \big( \bpd_{\nu} \hat Q \big) + \big( \bpd_{\mu} \hat P \big) V^{\nu} \right) \hat P \right. \nonumber \\
    & \left. - \hat P \calL^{-1}\left( V^{\mu} \big( \bpd_{\nu} \hat P \big) + \big( \bpd_{\mu} \hat Q \big) V^{\nu} \right) \hat Q \right\} + \frac{e}{2 i \hbar } \epsilon_{\mu\nu}  \sum_a n_F'(E_a) \big( \pd_{\mu} E_a \big) \left[ \big( \pd_{\nu} \hat P_a \big) \hat Q_a - \hat Q_a \big( \pd_{\nu} \hat P_a \big) \right]  - \sum_a n_F'(E_a) m_a  \hat P_a  \label{EqS:dPdBMetal}
\end{align}
Here, $\bpd_\mu$ denotes a momentum derivative that acts only on the eigenstate projectors, while keeping the occupation factors fixed. This distinction is necessary in a metallic system because the full derivative of $\hat P$ or $\hat Q$ also contains derivatives of the Fermi distribution. Those contributions are collected explicitly by the terms proportional to $n_F'(E_a)$ in Eq.~\eqref{EqS:dPdBMetal}.

\section{$\chi_{\rm MCD}$ in insulating states}
For an insulator, Fermi-surface contributions are absent. The projectors $\hat P$ and $\hat Q$ onto the occupied and empty states are therefore smooth functions of momentum. Therefore, $\bpd_{\mu} \hat P = \pd_{\mu} \hat P$. Starting from the expression for $M_{\rm MCD}$, we write
\begin{align}
    M_{\rm MCD} & = \frac{e}{2 \hbar } \epsilon_{\mu\nu} \oIm \mtr\left[ \hat H (\pd_\mu \hat P) (\pd_\nu \hat Q) \right] = \frac{e}{2 \hbar } \epsilon_{\mu\nu} \oIm \mtr\left[ (\pd_\mu \hat H) (\pd_\nu \hat P) \hat Q \right] \  .
\end{align}
In real space, it becomes
\begin{align}
    M_{\rm MCD} = -\frac{e}{2 \hbar } \epsilon_{\mu\nu} \oIm \int \left( \prod_{i=1}^3 \rmd \fvec r_i \right) \mtr\Big[  \big[ \hat r_{\mu}, H_B \big]_{\fvec r_1  \fvec r_2} \big[ \hat r_{\nu}, P_B \big]_{\fvec r_2 \fvec r_3}  Q_B(\fvec r_3, \fvec r_1) \Big]  \   .
\end{align}
where $H_B$ is the finite field Hamiltonian, and $P_B$, and $Q_B$ are the corresponding projectors onto the occupied states and the empty states respectively. Note that this expression remains valid in the presence of a magnetic field. 

We next separate the gauge dependent phase from each real space kernel. This gives
\begin{align}
    M_{\rm MCD} = -\frac{e}{2\hbar} \epsilon_{\mu\nu} \oIm \int \left( \prod_{i = 1}^3 \rmd \fvec r_i \right) \mtr\left\{ e^{i \Phi_B(\fvec r_1 , \fvec r_2, \fvec r_3)} \Big[ \hat r_\mu, \widetilde H \Big]_{\fvec r_1,\fvec r_2} \Big[ \hat r_\nu, \widetilde P \Big]_{\fvec r_2, \fvec r_3} \widetilde Q(\fvec r_3,\fvec r_1) \right\} \ , \label{EqS:MMCDGaugeInv}
\end{align}
where
\begin{align}
    \Phi_B(\fvec r_1, \fvec r_2, \fvec r_3) = \frac{e B}{2 \hbar } \epsilon_{\alpha\beta} (\fvec r_1 - \fvec r_3)_\alpha (\fvec r_2 - \fvec r_1)_\beta \  .
\end{align}
For a tight binding model with the Peierls substitution, the gauge invariant part of the Hamiltonian satisfies
\begin{align}
    \widetilde H(\fvec r_1, \fvec r_2) = H_0(\fvec r_1,\fvec r_2) \ .
\end{align}
i.e.~the gauge invariant tight binding Hamiltonian $\widetilde H$ is identical to zero-field tight binding $H_0$. The MCD susceptibility is therefore
\begin{align}
    \chi_{\rm MCD} & = \left. \frac{\pd M_{\rm MCD}}{\pd B} \right|_{B = 0} = - \frac{e}{2\hbar } \epsilon_{\mu\nu} \oIm \int \left( \prod_{i = 1}^3 \rmd \fvec r_i \right) \Bigg\{ \frac{i e}{2 \hbar} \epsilon_{\alpha\beta} (\fvec r_1 - \fvec r_3)_\alpha (\fvec r_2 -\fvec r_1)_\beta  \mtr\left[ \Big[ \hat r_\mu, H_0 \Big]_{\fvec r_1 \fvec r_2} \Big[\hat r_\nu, P \Big]_{\fvec r_2 \fvec r_3} Q(\fvec r_3, \fvec r_1) \right] \nonumber\\
    &  \qquad + \mtr\left[ \big[ \hat r_\mu, H \big]_{\fvec r_1 \fvec r_2} \Big[ \hat r_{\nu},  \frac{\rmd \widetilde P}{\rmd B} \Big]_{\fvec r_2 \fvec r_3} Q(\fvec r_3, \fvec r_1) + \big[ \hat r_\mu,  H_0 \big]_{\fvec r_1 \fvec r_2} \big[ \hat r_\nu,  P \big]_{\fvec r_2 \fvec r_3} \frac{\rmd \widetilde Q(\fvec r_3, \fvec r_1)}{\rmd B} \right] \Bigg\}   \nonumber \\
    & =  - \frac{e}{2\hbar } \epsilon_{\mu\nu} \oIm \int \left( \prod_{i = 1}^3 \rmd \fvec r_i \right) \Bigg\{ \frac{i e}{2 \hbar } \epsilon_{\alpha\beta}  \mtr\left[ \Big[ \hat r_{\beta}, \Big[ \hat r_\mu, H_0 \Big] \Big]_{\fvec r_1 \fvec r_2} \Big[\hat r_\nu, P \Big]_{\fvec r_2 \fvec r_3} \Big[ \hat r_{\alpha}, Q \Big]_{\fvec r_3 \fvec r_1} \right] \nonumber\\
    &  \qquad + \mtr\left[ \big[ \hat r_\mu, H \big]_{\fvec r_1 \fvec r_2} \Big[ \hat r_{\nu},  \frac{\rmd \widetilde P}{\rmd B} \Big]_{\fvec r_2 \fvec r_3} Q(\fvec r_3, \fvec r_1) + \big[ \hat r_\mu,  H_0 \big]_{\fvec r_1 \fvec r_2} \big[ \hat r_\nu,  P \big]_{\fvec r_2 \fvec r_3} \frac{\rmd \widetilde Q(\fvec r_3, \fvec r_1)}{\rmd B} \right] \Bigg\}  \ .
\end{align}
The spatial arguments in the last expression are suppressed for compactness. This result applies to insulating states. In a metal, additional Fermi surface contributions must be included.

As explained in the main text, all the gauge invariant quantities are translationally invariant, and therefore, the expression can be Fourier transformed back to the $\fvec k$ space. Using the formula 
\begin{align}
    \big[ \hat r_{\mu}, \hat O]  \longrightarrow i \pd_{\mu} \hat O
\end{align}
we rewrite $\chi_{\rm MCD}$ as
\begin{align}
    \chi_{\rm MCD} & = \chi_{\rm MCD}^{(1)} + \chi_{\rm MCD}^{(2)} + \chi_{\rm MCD}^{(3)} \\
    \chi_{\rm MCD}^{(1)} & = -\frac{e^2}{4 \hbar^2} \epsilon_{\mu\nu} \epsilon_{\alpha\beta}  \oRe\mtr\Bigg[ \left( \pd^2_{\beta \mu} \hat H_0 \right) \bigl(  \pd_\nu \hat P \bigr) \bigl(  \pd_\alpha \hat Q \bigr) \Bigg] \\
    \chi_{\rm MCD}^{(2)} & = \frac{e}{2 \hbar } \epsilon_{\mu\nu}  \oIm\mtr\Bigg[ \left( \pd_\mu H_0 \right) \pd_{\nu} \left( \frac{\rmd \widetilde P}{\rmd B} \right)_{B = 0} \hat Q \Bigg] \\
    \chi_{\rm MCD}^{(3)} & = \frac{e}{2 \hbar } \epsilon_{\mu\nu}  \oIm\mtr\Bigg[ \left( \pd_\mu H_0 \right) \left( \pd_{\nu} \hat P \right)  \left( \frac{\rmd \widetilde Q}{\rmd B} \right)_{B = 0} \Bigg] 
    \label{EqS:ChiMCDMomentum}
\end{align}

\subsection{Contribution $\chi_{\mathrm{MCD}}^{(1)}$}

The first contribution $\chi_{\mathrm{MCD}}^{(1)}$ comes from the field derivative of the triangular phase. For notational convenience, we introduce the inverse mass tensor
\begin{align}
    \calM^{-1}_{\alpha \beta} = \frac1{\hbar^2} \frac{\pd^2 \hat H_0}{\pd k_\beta \pd k_\alpha} \  .
\end{align}
In terms of this tensor, the first contribution is
\begin{align}
    \chi_{\mathrm{MCD}}^{(1)} =  -\frac{e^2}{4} \epsilon_{\mu\nu} \epsilon_{\alpha\beta} \oRe \mtr\left[ \calM^{-1}_{\mu\beta} (\pd_\nu \hat P) (\pd_\alpha \hat Q) \right]   \  . \label{Eqn:Chi1TRMCD}
\end{align}
The derivatives of the projectors satisfy
\begin{align}
    \pd_\nu \hat P & = \hat P (\pd_\nu \hat P) \hat Q + \hat Q (\pd_\nu \hat P) \hat P & \pd_\nu \hat Q & = \hat P (\pd_\nu \hat Q) \hat Q + \hat Q (\pd_\nu \hat Q) \hat P \  . \label{Eqn:Identity1}
\end{align}
Substituting these identities into Eq.~\eqref{Eqn:Chi1TRMCD} gives
\begin{align}
    \chi_{\rm MCD}^{(1)} & = -\frac{e^2}{4 } \epsilon_{\mu\nu} \epsilon_{\alpha\beta} \oRe \mtr \Big[ \calM^{-1}_{\mu\beta} \left( \hat P (\pd_\nu \hat P) \hat Q (\pd_\alpha \hat Q) \hat P + \hat Q (\pd_\nu \hat P) \hat P (\pd_\alpha \hat Q) \hat Q \right) \Big]  \  .
\end{align}

Note that the matrix-valued quantum geometric tensors associated with the occupied and empty subspaces are
\begin{align}
 \calQ^{P}_{\nu\alpha} & = \hat P (\pd_\nu \hat P) \hat Q (\pd_\alpha \hat P) \hat P = - \hat P (\pd_\nu \hat P) \hat Q (\pd_\alpha \hat Q) \hat P \\
 \calQ^{Q}_{\nu\alpha} & = \hat Q (\pd_\nu \hat Q) \hat P (\pd_\alpha \hat Q) \hat Q = - \hat Q (\pd_\nu \hat P) \hat P (\pd_\alpha \hat Q) \hat Q  \  . 
\end{align}

The contribution can therefore be written as
\begin{align}
    \chi_{\mathrm{MCD}}^{(1)} =  \frac{e^2}{4 } \epsilon_{\mu\nu} \epsilon_{\alpha\beta} \oRe \mtr\left[ \calM^{-1}_{\mu\beta} \left( \calQ^{P}_{\nu\alpha} + \calQ^{Q}_{\nu\alpha}  \right) \right] = \frac{e^2}{4} \oRe \mtr\left[ \calM^{-1}_{\alpha \beta} \left( \calQ^{P}_{\beta\alpha} + \calQ^{Q}_{\beta\alpha}  \right) - \calM^{-1}_{\alpha \alpha} \left( \calQ^{P}_{\beta \beta} + \calQ^{Q}_{\beta \beta}  \right) \right]
\end{align}
where we have used the two-dimensional identity
\begin{align}
 \epsilon_{\mu\nu} \epsilon_{\alpha\beta} = \delta_{\mu\alpha} \delta_{\nu\beta} - \delta_{\mu\beta} \delta_{\nu\alpha} \ .
\end{align}

The inverse-mass tensor is symmetric under the interchange of its spatial indices. Therefore, only the symmetric parts of the quantum geometric tensors contribute. We define the corresponding non-Abelian quantum metrics as
\begin{align}
    g_{\mu\nu}^{P} & = \frac12 \left( \calQ_{\mu\nu}^{P} + \calQ_{\nu\mu}^{P} \right)  \qquad g_{\mu\nu}^Q  = \frac12 \left( \calQ_{\mu\nu}^Q + \calQ_{\nu\mu}^Q \right) \   . 
\end{align}
The contribution then reduces to
\begin{align}
    \chi_{\mathrm{MCD}}^{(1)} =   \frac{e^2}{4} \oRe \mtr\left[ \calM^{-1}_{\alpha \beta} \left( g^{P}_{\beta\alpha} + g^{Q}_{\beta\alpha}  \right) - \calM^{-1}_{\alpha \alpha} \left( g^{P}_{\beta \beta} + g^{Q}_{\beta \beta}  \right) \right]
\end{align}
Therefore, this contribution can therefore be interpreted as an inverse mass weighted quantum metric contribution.

\subsection{Contribution $\chi_{\rm MCD}^2$}
The second contribution can be rewritten as
\begin{align}
        \chi_{\rm MCD}^{(2)} & = \frac{e}{2 \hbar } \epsilon_{\mu\nu}  \oIm\mtr\Bigg[ \left( \pd_\mu H_0 \right) \pd_{\nu} \left( \frac{\rmd \widetilde P}{\rmd B} \right)_{B = 0} \hat Q \Bigg] = - \frac{e}{2  } \epsilon_{\mu\nu}  \oIm\mtr\Bigg[ \hat V^{\mu}  \left( \frac{\rmd \widetilde P}{\rmd B} \right)_{B = 0} \big( \pd_{\nu} \hat Q \big) \Bigg] \nonumber \\
        & = \frac{e^2}{4 \hbar} \epsilon_{\mu\nu} \epsilon_{\alpha\beta} \oRe\mtr\left[ V^{\mu} \left\{ - \big( \pd_{\alpha} \hat P \big) \hat Q \big( \pd_{\beta} \hat P \big) + \big( \pd_{\alpha} \hat Q \big) \hat P \big( \pd_{\beta} \hat Q \big) + \hat Q \calL^{-1}\left( V^{\alpha} \big( \pd_{\beta} \hat Q \big) + \big( \pd_{\alpha} \hat P \big) V^{\beta} \right) \hat P \right. \right. \nonumber \\
    & \qquad \left. \left. - \hat P \calL^{-1}\left( V^{\alpha} \big( \pd_{\beta} \hat P \big) + \big( \pd_{\alpha} \hat Q \big) V^{\beta} \right) \hat Q \right\} \big( \pd_{\nu} \hat Q \big) \right] \nonumber \\
    & =  \chi_{\rm MCD}^{(2a)} +  \chi_{\rm MCD}^{(2b)} +  \chi_{\rm MCD}^{(2c)} +  \chi_{\rm MCD}^{(2d)}
\end{align}
Now, we simplify it term by term
\begin{align}
    \chi_{\rm MCD}^{(2a)} =  - \frac{e^2}{4 \hbar } \epsilon_{\mu\nu} \epsilon_{\alpha\beta} \oRe\mtr\left[ V^{\mu}  \big( \pd_{\alpha} \hat P \big) \hat Q \big( \pd_{\beta} \hat P \big) \big( \pd_{\nu} \hat Q \big) \right]  =  - \frac{e^2}{4 \hbar} \epsilon_{\mu\nu} \epsilon_{\alpha\beta} \oRe\mtr\left[ V^{\mu}  \big( \pd_{\alpha} \hat P \big) \hat Q \big( \pd_{\beta} \hat P \big) \big( \pd_{\nu} \hat Q \big) \right]  \label{EqS:Chi2aMCD}
\end{align}

For the second term, 
\begin{align}
     \chi_{\rm MCD}^{(2b)}  = \frac{e^2}{4 \hbar} \epsilon_{\mu\nu} \epsilon_{\alpha\beta} \oRe\mtr\left[ V^{\mu} \big( \pd_{\alpha} \hat Q \big) \hat P \big( \pd_{\beta} \hat Q \big) \big( \pd_{\nu} \hat Q \big) \right] = - \frac{e^2}{4 \hbar} \epsilon_{\mu\nu} \epsilon_{\alpha\beta} \oRe\mtr\left[ V^{\mu} \big( \pd_{\alpha} \hat Q \big) \hat P \big( \pd_{\beta} \hat Q \big) \big( \pd_{\nu} \hat P \big) \right] \label{EqS:Chi2bMCD}
\end{align}

For the last two terms, we have 
\begin{align}
    \chi_{\rm MCD}^{(2c)} & =  \frac{e^2}{4 \hbar} \epsilon_{\mu\nu} \epsilon_{\alpha\beta} \oRe\mtr\left[ V^{\mu} \hat Q \calL^{-1}\left( V^{\alpha} \big( \pd_{\beta} \hat Q \big) + \big( \pd_{\alpha} \hat P \big) V^{\beta} \right) \hat P  \big( \pd_{\nu} \hat Q \big) \right] \nonumber \\
    & = - \frac{e^2}{4 \hbar } \epsilon_{\mu\nu} \epsilon_{\alpha\beta} \oRe\mtr\left[  \hat P \calL^{-1}\left( V^{\beta} \big( \pd_{\alpha} \hat P \big) + \big( \pd_{\beta} \hat Q \big) V^{\alpha} \right) \hat Q V^{\mu} \big( \pd_{\nu} \hat Q \big)   \right] \label{EqS:Chi2cMCD} \\
    \chi_{\rm MCD}^{(2d)} & = -\frac{e^2}{4 \hbar } \epsilon_{\mu\nu} \epsilon_{\alpha\beta} \oRe\mtr\left[ V^{\mu} \hat P \calL^{-1}\left( V^{\alpha} \big( \pd_{\beta} \hat P \big) + \big( \pd_{\alpha} \hat Q \big) V^{\beta} \right) \hat Q \big( \pd_{\nu} \hat Q \big) \right] \nonumber \\
    & = \frac{e^2}{4 \hbar } \epsilon_{\mu\nu} \epsilon_{\alpha\beta} \oRe\mtr\left[ V^{\mu} \hat P \calL^{-1}\left( V^{\alpha} \big( \pd_{\beta} \hat P \big) + \big( \pd_{\alpha} \hat Q \big) V^{\beta} \right) \hat Q \big( \pd_{\nu} \hat P \big) \right] \\
    \Longrightarrow \quad \chi_{\rm MCD}^{(2c)} + \chi_{\rm MCD}^{(2d)} & = \frac{e^2}{4 \hbar} \epsilon_{\mu\nu} \epsilon_{\alpha\beta}  \oRe\mtr\left[ \hat P \calL^{-1}\left( V^{\alpha} \big( \pd_{\beta} \hat P \big) + \big( \pd_{\alpha} \hat Q \big) V^{\beta} \right) \hat Q \left(  V^{\mu} \big( \pd_{\nu} \hat Q \big)  - \big( \pd_{\mu} \hat P \big)  V^{\nu} \right)    \right]  \label{EqS:Chi2dMCD}
\end{align}
We define
\begin{align}
    \hat O_1 = \epsilon_{\alpha\beta} \left( V^{\alpha} \big( \pd_{\beta} \hat P \big) + \big( \pd_{\alpha} \hat Q \big) V^{\beta} \right) \quad \mbox{and} \quad \hat O_2 = \epsilon_{\mu\nu} \left( V^{\mu} \big( \pd_{\nu} \hat Q \big) - \big( \pd_{\mu} \hat P \big) V^{\nu} \right) \label{EqS:O1O2}
\end{align}
\begin{align}
    \chi_{\rm MCD}^{(2c)} + \chi_{\rm MCD}^{(2d)} & = \frac{e^2}{4 \hbar} \oRe\mtr\left[ \hat P \calL^{-1}\left( \hat O_1 \right) \hat Q   \hat O_2     \right]
\end{align}

Adding everything together, we obtain
\begin{align}
    \chi_{\rm MCD}^{(2)} & = - \frac{e^2}{4 \hbar} \epsilon_{\mu\nu} \epsilon_{\alpha\beta} \oRe\mtr\left[ V^{\mu} \big( \pd_{\alpha} \hat P \big) \hat Q \big( \pd_{\beta} \hat P \big) \big( \pd_{\nu} \hat Q \big) + V^{\mu} \big( \pd_{\alpha} \hat Q \big) \hat P \big( \pd_{\beta} \hat Q \big) \big( \pd_{\nu} \hat P \big)  \right]  +  \frac{e^2}{4 \hbar} \oRe\mtr\left[ \hat P \calL^{-1}\left( \hat O_1 \right) \hat Q \hat O_2 \right]  
\end{align}
with $\hat O_1$ and $\hat O_2$ defined in Eq.~\eqref{EqS:O1O2}.

\subsection{Contribution $\chi_{\rm MCD}^3$}
\begin{align}
        \chi_{\rm MCD}^{(3)} & = \frac{e}{2 \hbar } \epsilon_{\mu\nu} \oIm\mtr\left( \big( \pd_{\mu} \hat H_0 \big) \big( \pd_{\nu} \hat P \big) \frac{\rmd \widetilde{Q}}{\rmd B}  \right) =  \frac{e}{2 \hbar } \epsilon_{\mu\nu} \oIm\mtr\left( \big( \pd_{\mu} \hat H_0 \big) \big( \pd_{\nu} \hat Q \big) \frac{\rmd \widetilde{P}}{\rmd B}  \right) = - \frac{e}{2 \hbar } \epsilon_{\mu\nu} \oIm\mtr\left( \big( \pd_{\mu} \hat H_0 \big) \big( \pd_{\nu} \hat Q \big) \frac{\rmd \widetilde{P}}{\rmd B}  \right)^\dagger \nonumber \\
        & = - \frac{e}{2 \hbar } \epsilon_{\mu\nu} \oIm\mtr\left( \big( \pd_{\mu} \hat H_0 \big) \frac{\rmd \widetilde{P}}{\rmd B}  \big( \pd_{\nu} \hat Q \big)  \right) = \frac{e}{2 \hbar } \epsilon_{\mu\nu} \oIm\mtr\left( \big( \pd_{\mu} \hat H_0 \big) \pd_{\nu} \left( \frac{\rmd \widetilde{P}}{\rmd B} \right)  \hat Q   \right) = \chi_{\rm MCD}^{(2)} 
\end{align}

\subsection{Total}
Therefore, in total, we have the MCD orbital susceptibility for insulating states:
\begin{align}
    \chi_{\rm MCD}^{\rm ins} & = -\frac{e^2}{4} \epsilon_{\mu\nu} \epsilon_{\alpha\beta} \oRe\mtr\left[ \calM^{-1}_{\mu\beta}  \left(  \hat Q \big( \pd_{\nu} \hat P \big) \hat P \big( \pd_{\alpha} \hat Q \big) + \hat P \big( \pd_{\nu} \hat P \big) \hat Q \big( \pd_{\alpha} \hat Q \big) \hat P \right) \right] \nonumber \\
    & \qquad - \frac{e^2}{2 \hbar} \epsilon_{\mu\nu} \epsilon_{\alpha\beta} \oRe\mtr\left[ V^{\mu} \big( \pd_{\alpha} \hat P \big) \hat Q \big( \pd_{\beta} \hat P \big) \big( \pd_{\nu} \hat Q \big) + V^{\mu} \big( \pd_{\alpha} \hat Q \big) \hat P \big( \pd_{\beta} \hat Q \big) \big( \pd_{\nu} \hat P \big)  \right]  + \frac{e^2}{2 \hbar} \oRe\mtr\left[ \hat P \calL^{-1}\left( \hat O_1 \right) \hat Q \hat O_2 \right]
\end{align}

\section{Systematic Green Function Derivation of $\chi_{\rm MCD}$}

We begin with the optical conductivity for circularly polarized light. In the Matsubara formalism, it can be written as
\begin{align}
    \sigma_{\pm}(\omega) = \left. \frac{i e^2}{\omega} \oint_{\calC}\frac{\rmd z}{2\pi i} n_F(z) \mtr\left[ \hat G(z) \bigl( \hat \varepsilon_{\pm} \cdot \hat{\fvec V} \bigr) \hat G(z + i \omega_n) \bigl( \hat \varepsilon_{\pm}^* \cdot \hat{\fvec V} \bigr) \right] \right|_{i \omega_n \rightarrow \hbar \omega + i 0^+} \ ,
\end{align}
where $\hat \varepsilon_{\pm} = \frac1{\sqrt{2}} \big( \hat x \pm i \hat y \big)$ denotes the polarization vector of right- or left-circularly polarized light. The difference between the corresponding absorptive conductivities is therefore
\begin{align}
    I_{\rm MCD}(\omega) & \equiv \oRe\left[\sigma_+(\omega) - \sigma_-(\omega) \right] = \left. \frac{e^2}{\omega} \oRe\left\{ \epsilon_{\mu\nu} \oint_{\calC} \frac{\rmd z}{2\pi i} n_F(z) \mtr\left[ \hat G(z) \hat V^\mu \hat G(z + i \omega_n) \hat V^\nu \right] \right\} \right|_{i \omega_n \rightarrow \hbar \omega + i 0^+}  \  .
\end{align}
The integrated MCD quantity is defined as
\begin{align}
    M_{\rm MCD} & \equiv \frac{\hbar}{2\pi e} \int_{0^+}^{\infty} \rmd\omega I_{\rm MCD}(\omega) = \frac{e\hbar}{2\pi} \int_{0^+}^{\infty} \frac{\rmd \omega}{\omega} \left. \oRe\left\{ \epsilon_{\mu\nu} \oint_{\calC} \frac{\rmd z}{2\pi i} n_F(z) \mtr\left[ \hat G(z) \hat V^\mu \hat G(z + i \omega_n) \hat V^\nu \right] \right\} \right|_{i \omega_n \rightarrow \hbar \omega + i 0^+} \ . \label{EqS:MMCD}
\end{align}
The lower limit $0^+$ regulates possible zero frequency contributions. Eq.~\eqref{EqS:MMCD} remains well defined for metallic systems at finite temperature and therefore provides a suitable starting point for a systematic derivation of $\chi_{\rm MCD}$. In the zero temperature limit, it reduces to the self-rotation contribution $m_{\rm SR}$.

To evaluate the magnetic response, we first express Eq.~\eqref{EqS:MMCD} in real space. For compactness, we introduce
\begin{align}
    \int_{\fvec r_{1\cdots4}} \equiv \int \prod_{\ell=1}^{4}\rmd\fvec r_\ell \ .
\end{align}
We then obtain
\begin{align}
    M_{\rm MCD} & = \frac{e\hbar}{2\pi} \int_{0^+}^{\infty} \frac{\rmd \omega}{\omega} \left. \oRe\left\{ \epsilon_{\mu\nu} \oint_{\calC} \frac{\rmd z}{2\pi i} n_F(z) \int_{\fvec r_{1\cdots4}} \mtr\left[ G_{12}(z) V^\mu_{23} G_{34}(z + i \omega_n) V^\nu_{41} \right] \right\} \right|_{i \omega_n \rightarrow \hbar \omega + i 0^+} \ , \label{EqS:MMCDReal}
\end{align}
where, for example,
\begin{align}
    G_{12}(z) \equiv G(z; \fvec r_1,\fvec r_2) \ , \qquad V^\mu_{23} \equiv V^\mu(\fvec r_2,\fvec r_3) \ .
\end{align}

We next separate out the gauge dependent phase from each two-point kernel. Introducing the gauge invariant kernel
\begin{align}
    \widetilde O(\fvec r,\fvec r') = e^{-i\phi(\fvec r,\fvec r')} O(\fvec r,\fvec r') \ ,
\end{align}
Eq.~\eqref{EqS:MMCDReal} becomes
\begin{align}
    M_{\rm MCD} & = \frac{e\hbar}{2\pi} \int_{0^+}^{\infty} \frac{\rmd \omega}{\omega} \left. \oRe\left\{ \epsilon_{\mu\nu} \oint_{\calC} \frac{\rmd z}{2\pi i} n_F(z) \int_{\fvec r_{1\cdots4}} \mtr\left[ e^{i\Phi_{1234}} \widetilde G_{12}(z) \widetilde V^\mu_{23} \widetilde G_{34}(z + i \omega_n) \widetilde V^\nu_{41} \right] \right\} \right|_{i \omega_n \rightarrow \hbar \omega + i 0^+} \ . \label{EqS:MMCDGaugeInvReal}
\end{align}
Here,
\begin{align}
    \Phi_{1234} & \equiv \phi_{12} + \phi_{23} + \phi_{34} + \phi_{41}
\end{align}
is the gauge-invariant Aharonov-Bohm phase associated with the magnetic flux through the quadrilateral whose vertices are $\fvec r_1$, $\fvec r_2$, $\fvec r_3$, and $\fvec r_4$. For a uniform perpendicular magnetic field,
\begin{align}
    \Phi_{1234} & = \frac{e B}{2 \hbar} \epsilon_{\alpha\beta} \left[ (\fvec r_2 - \fvec r_1)_\alpha (\fvec r_3 - \fvec r_2)_\beta + (\fvec r_4 -\fvec r_3)_\alpha (\fvec r_1 - \fvec r_4)_\beta \right] \label{EqS:FourPtPhase1} \\
    & = \frac{e B}{4 \hbar} \epsilon_{\alpha\beta} \left[ (\fvec r_2-\fvec r_1)_\alpha (\fvec r_3-\fvec r_2)_\beta + (\fvec r_3-\fvec r_2)_\alpha (\fvec r_4-\fvec r_3)_\beta + (\fvec r_4 - \fvec r_3)_\alpha (\fvec r_1-\fvec r_4)_\beta + (\fvec r_1 - \fvec r_4)_\alpha (\fvec r_2 - \fvec r_1)_\beta \right] \  .  \label{EqS:FourPtPhase2}
\end{align}
The two expressions are equivalent because the four directed edges of the quadrilateral sum to zero.

The gauge invariant velocity kernel is given by
\begin{align}
    \widetilde V^\mu(\fvec r,\fvec r') & = \frac{r_\mu - r'_\mu}{i\hbar} \widetilde H(\fvec r, \fvec r') = \frac1{i\hbar} [\hat r_\mu , \hat{\widetilde H}]_{\fvec r \fvec r'} \ .
\end{align}
For a tight binding Hamiltonian in which the orbital magnetic field enters only through the Peierls substitution, $\widetilde H$ coincides with the zero field Hamiltonian. Consequently, $\widetilde V^\mu$ is also identical to the zero-field velocity operator and has no explicit magnetic field dependence.

We may now differentiate Eq.~\eqref{EqS:MMCDGaugeInvReal} with respect to $B$. At fixed temperature and chemical potential $\mu$, the magnetic response contains three contributions. They arise from the field derivatives of the loop phase, $\widetilde G(z)$, and $\widetilde G(z + i\omega_n)$, respectively. We therefore write
\begin{align}
    \chi_{\rm MCD} \equiv \left. \frac{\pd M_{\rm MCD}}{\pd B} \right|_{B=0} = \chi_1 + \chi_2 + \chi_3  \    .
\end{align}
Here, $\chi_1$ originates from the explicit magnetic field dependence of $\Phi_{1234}$, whereas $\chi_2$ and $\chi_3$ arise from the field dependence of the two gauge invariant Green functions.

\subsection{$\chi_1$}
\begin{figure}[htbp]
	\centering
	\includegraphics[width=0.9\textwidth]{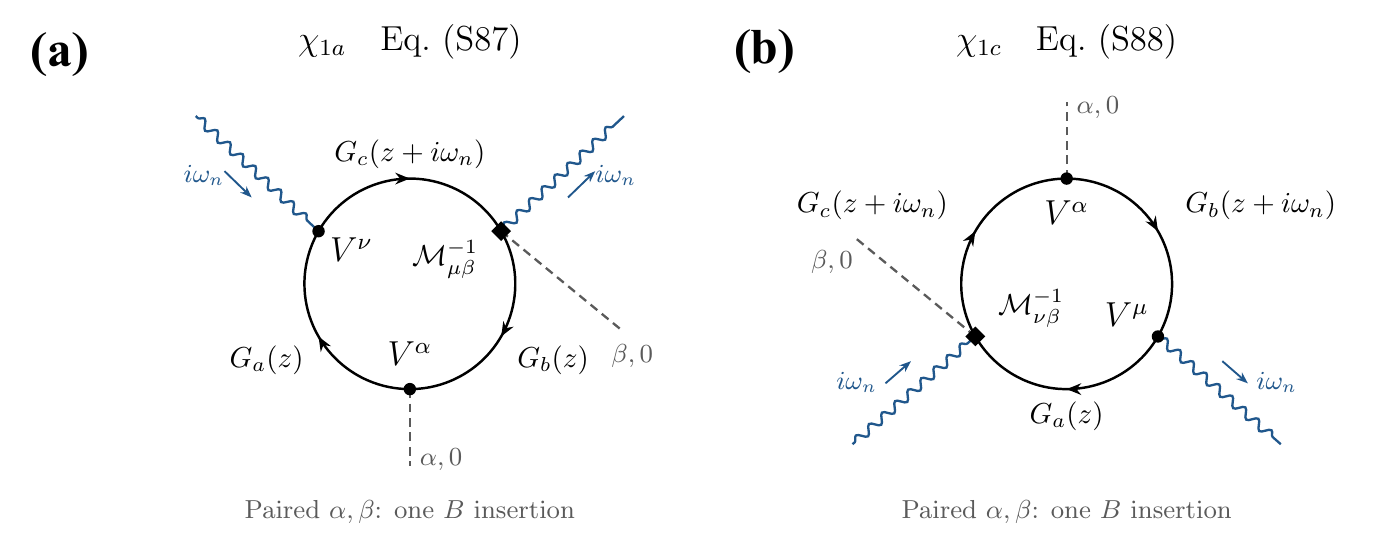}
	\caption{Loop-phase contributions to the MCD susceptibility: (a) $\chi_{1a}$ in Eq.~\eqref{EqS:Chi1aMCD} and (b) $\chi_{1c}$ in Eq.~\eqref{EqS:Chi1cMCD}. Both three-vertex kernels arise from differentiating the AB loop phase. Solid directed lines denote $G_j(\zeta)=(\zeta-E_j)^{-1}$; all propagators and vertices within a loop are evaluated at the same zero-field crystal momentum $\fvec k$, with the Brillouin-zone sum implicit. Wavy external legs denote optical probes carrying energy $i\omega_n$, continued to $\hbar\omega+i0^+$. Filled circles are velocity vertices, and diamonds are inverse-mass contact vertices carrying one optical index and one magnetic-gradient index. The zero-frequency factors marked by the dashed $\alpha,\beta$ stubs are contracted with $\epsilon_{\alpha\beta}$ and together represent one linear magnetic-field insertion, not two independent magnetic photons. Band sums, the $n_F(z)$ contour integral, the optical-frequency integral, the $\epsilon_{\mu\nu}$ contraction, and prefactors are supplied by the corresponding equations.}
	\label{FigS:MCD_chi1_kernels}
\end{figure}

We first consider the contribution arising from the magnetic field derivative of the loop phase. Although two equivalent expressions for this phase were given in Eqs.~\eqref{EqS:FourPtPhase1} and \eqref{EqS:FourPtPhase2}, it is convenient to use Eq.~\eqref{EqS:FourPtPhase1}. Note that
\begin{align}
    \pd_\alpha G_0(z) = G_0 \big( \pd_\alpha \hat H_0 \big) G_0 = \hbar G_0 \hat V^\alpha  G_0 \ .
\end{align}
This gives
\begin{align}
    \chi_1 & = - \frac{e^2}{4 \pi} \int_{0^+}^{\infty} \frac{\rmd \omega}{\omega} \oIm\left\{ \epsilon_{\mu\nu} \epsilon_{\alpha\beta} \oint_{\mcC'} \frac{\rmd z}{2 \pi i} n_F(z) \mtr\left[ [\hat r_{\alpha}, G_0(z)] [\hat r_{\beta}, V^{\mu}] G_0(z + i \omega_n) V^{\nu} \right. \right. \nonumber \\
    & \qquad \left. \left. + G_0(z) V^{\mu} [\hat r_{\alpha}, G_0(z + i \omega_n) ] [\hat r_{\beta}, V^{\nu}] \right]  \right\}_{i\omega_n \rightarrow \hbar \omega + i 0^+} \nonumber \\
    & = \frac{e^2 \hbar}{4 \pi} \int_{0^+}^{\infty} \frac{\rmd \omega}{\omega}  \oIm\left\{ \epsilon_{\mu\nu} \epsilon_{\alpha\beta} \oint_{\mcC'} \frac{\rmd z}{2 \pi i} n_F(z) \mtr\left[ \big( \pd_{\alpha} G_0(z) \big) \calM^{-1}_{\mu\beta} G_0(z + i \omega_n) V^{\nu} \right. \right. \nonumber \\
    & \qquad \left. \left. + G_0(z) V^{\mu} \big( \pd_{\alpha} G_0(z + i \omega_n) \big)  \calM^{-1}_{\nu\beta}  \right] \right\}_{i\omega_n \rightarrow \hbar \omega + i 0^+}  \nonumber \\
    & = \frac{e^2 \hbar^2}{4 \pi} \int_{0^+}^{\infty} \frac{\rmd \omega}{\omega} \oIm\left\{ \epsilon_{\mu\nu} \epsilon_{\alpha\beta} \oint_{\mcC'} \frac{\rmd z}{2 \pi i} n_F(z) \mtr\left[ G_0(z) V^{\alpha} G_0(z)  \calM^{-1}_{\mu\beta} G_0(z + i \omega_n) V^{\nu} \right. \right. \nonumber \\
    & \qquad \left. \left. + G_0(z) V^{\mu} G_0(z + i \omega_n) V^{\alpha} G_0(z + i \omega_n) \calM^{-1}_{\nu\beta}  \right] \right\}_{i\omega_n \rightarrow \hbar \omega + i 0^+}  \nonumber \\
    & = \chi_{1a} + \chi_{1c} \ .
\end{align}

The two terms correspond to Fig.~\ref{FigS:MCD_chi1_kernels}(a,b). In each, a velocity insertion from $\pd_\alpha G_0=\hbar G_0V^\alpha G_0$ is paired with an inverse-mass contact vertex from the momentum derivative of an optical velocity vertex. Expanding the Green functions in the eigenstate basis gives
\begin{align}
    \chi_{1a} & = \frac{e^2 \hbar^2}{4 \pi} \int_{0^+}^{\infty} \frac{\rmd \omega}{\omega} \oIm\left\{ \epsilon_{\mu\nu} \epsilon_{\alpha\beta} V^{\alpha}_{ab} \left( \calM^{-1}_{\mu\beta} \right)_{bc} V^{\nu}_{ca} \oint \frac{\rmd z}{2 \pi i} \frac{n_F(z)}{(z - E_a) (z - E_b) (i \omega_n + z -E_c)}  \right\}_{i\omega_n \rightarrow \hbar \omega + i 0^+}   \label{EqS:Chi1aMCD} \\
    \chi_{1c} & = \frac{e^2 \hbar^2}{4 \pi} \int_{0^+}^{\infty} \frac{\rmd \omega}{\omega} \oIm\left\{ \epsilon_{\mu\nu} \epsilon_{\alpha\beta} V^{\mu}_{ab} V^{\alpha}_{bc} \left( \calM^{-1}_{\nu\beta} \right)_{ca} \oint \frac{\rmd z}{2 \pi i} \frac{n_F(z)}{(z - E_a) (i \omega_n + z - E_b) (i \omega_n + z - E_c)}  \right\}_{i\omega_n \rightarrow \hbar \omega + i 0^+}   \label{EqS:Chi1cMCD}
\end{align}

After analytic continuation, each energy denominator can be decomposed according to
\begin{align}
    \frac{1}{x + i 0^+} = \calP \frac1x -i \pi \delta(x) \  ,
\end{align}
where $\calP$ denotes the Cauchy principal value. Accordingly, both $\chi_{1a}$ and $\chi_{1c}$ contain a principal-value contribution and an on-shell contribution proportional to an energy conserving delta function. As shown below, the principal value terms cancel against the corresponding contributions from $\chi_2$ and $\chi_3$. We therefore evaluate the delta function contributions first.

\subsubsection{$\chi_{1a}$}
We now evaluate the on-shell contribution to $\chi_{1a}$. The energies below are measured relative to the chemical potential. We distinguish the following occupation configurations.
\begin{itemize}
    \item $E_a < 0$ and $E_b, E_c > 0$:  In this case, the contribution is
    \begin{align}
        \chi_{1a}^{(1)} & = - \frac{e^2 \hbar^2}4 \epsilon_{\mu\nu} \epsilon_{\alpha\beta} \oRe\left[ \frac{V^{\alpha}_{ab} \left( \calM^{-1}_{\mu\beta} \right)_{bc} V^{\nu}_{ca}}{(E_a - E_b) (E_c - E_a)}   \right] = \frac{e^2}4 \epsilon_{\mu\nu} \epsilon_{\alpha\beta} \oRe\mtr\left[ \hat P \big( \bpd_{\alpha} \hat Q \big) \calM^{-1}_{\mu\beta} \big( \bpd_{\nu} \hat Q \big) \right] \label{EqS:Chi1aCase1}
    \end{align}
    \item $E_b < 0$ and $E_a, E_c > 0$: The corresponding contribution is
    \begin{align}
        \chi_{1a}^2 & = - \frac{e^2 \hbar^2}4 \epsilon_{\mu\nu} \epsilon_{\alpha\beta} \oRe\left[ \frac{V^{\alpha}_{ab} \left( \calM^{-1}_{\mu\beta} \right)_{bc} V^{\nu}_{ca}}{(E_b - E_a) (E_c - E_b)}   \right] = - \frac{e^2}4 \epsilon_{\mu\nu} \epsilon_{\alpha\beta} \oRe\mtr\left[  \big( \bpd_{\alpha} \hat Q \big) \hat P \calL^{-1} \left( \calM^{-1}_{\mu\beta} \right) \hat Q V^{\nu} \right] \label{EqS:Chi1aCase2}
    \end{align}
    This term will cancel against a corresponding contribution obtained below.
    
    \item $E_a,  E_b < 0$ and  $E_c > 0$: For this occupation configuration, the relevant frequency dependent factor can be simplified as
    \begin{align}
        & - \frac1{(E_c - E_a - i \omega_n ) (E_c - E_b - i \omega_n)} = - \frac1{E_a - E_b} \left( \frac1{E_c - E_a - i \omega_n} - \frac1{E_c - E_b - i \omega_n} \right) \nonumber \\
        \longrightarrow &  - i \pi \hbar \frac1{E_a - E_b} \left( \frac1{E_c - E_a} - \frac1{E_c - E_b} \right) = - i \pi \hbar \frac1{(E_c - E_a) (E_c - E_b)} \label{Eqn:Formula1}
    \end{align}
    where in the second line, we keep the on-shell part after the analytic continuation $i\omega_n \rightarrow \hbar\omega + i 0^+$ and perform the frequency integral. We therefore obtain
    \begin{align}
        \chi_{1a}^3 & =  - \frac{e^2 \hbar^2}4 \epsilon_{\mu\nu} \epsilon_{\alpha\beta} \oRe\left[ \frac{V^{\alpha}_{ab} \left( \calM^{-1}_{\mu\beta} \right)_{bc} V^{\nu}_{ca}}{(E_c - E_b) (E_c - E_a)}   \right] =  - \frac{e^2}4 \epsilon_{\mu\nu} \epsilon_{\alpha\beta} \oRe\mtr\left[ \big( \bpd_{\nu} \hat P \big) V^{\alpha} \hat P \calL^{-1} \left( \calM^{-1}_{\beta\mu} \right) \hat Q \right]   \label{EqS:Chi1aCase3}
    \end{align}
    This contribution will also cancel against terms derived below.
    
    \item $a = b$ and $E_c > 0$: The coincident poles at $E_a$ generate a Fermi surface contribution proportional to $n_F'(E_a)$.
    \begin{align}
        \chi_{1a}^{\rm FS} & = - \frac{e^2 \hbar }4 \epsilon_{\mu\nu} \epsilon_{\alpha\beta} \sum_{a, c\neq a} \frac{n_F'(E_a)}{E_c - E_a} \big( \pd_{\alpha} E_a \big) \oRe\mtr\left( \left( \calM^{-1}_{\mu\beta} \right)_{ac} V^{\nu}_{ca} \right)  \nonumber \\
        & =  - \frac{e^2}4 \epsilon_{\mu\nu} \epsilon_{\alpha\beta} \sum_a n_F'(E_a) \big( \pd_{\alpha} E_a \big) \oRe\mtr\left( \hat P_a \calM^{-1}_{\beta \mu} \big( \pd_{\nu} \hat Q_{\sa} \big) \right)  \label{EqS:Chi1aFS}
    \end{align}
\end{itemize}

All remaining occupation configurations have no poles on the positive frequency integration contour and therefore do not produce on-shell delta function contributions.

\subsubsection{$\chi_{1c}$}
We next evaluate $\chi_{1c}$, defined in Eq.~\eqref{EqS:Chi1cMCD}. As in the preceding subsection, we keep only the on-shell contributions generated after analytic continuation.
\begin{itemize}
    \item $E_c > 0$ and $E_a, E_b < 0$:  After performing the frequency integral, we obtain
    \begin{align}
        \chi_{1c}^1 & = - \frac{e^2}4 \epsilon_{\mu\nu} \epsilon_{\alpha\beta} \oRe\mtr\left( \hat Q \calL^{-1}\big( \calM^{-1}_{\beta\mu} \big) \hat P V^{\alpha} \big( \bpd_{\nu} \hat P \big) \right) \label{EqS:Chi1cCase1}
    \end{align}
    Upon relabeling the contracted spatial indices and using the cyclicity of the trace, this contribution cancels Eq.~\eqref{EqS:Chi1aCase3}: $\chi_{1c}^1 + \chi_{1a}^3 = 0$.

    \item  $E_b > 0$ and $E_a, E_c < 0$:  In this case,
    \begin{align}
        \chi_{1c}^2 & = - \frac{e^2}4 \epsilon_{\mu\nu} \epsilon_{\alpha\beta} \Re\mtr\left( \hat Q \big( \bpd_{\alpha} \hat P \big)  \calM^{-1}_{\beta \nu} \big( \bpd_{\mu} \hat P \big) \right) \label{EqS:Chi1cCase2}
    \end{align}

    \item $E_a < 0$ and $E_b, E_c > 0$: Using Eq.~\eqref{Eqn:Formula1}, the on-shell contribution becomes 
    \begin{align}
        \chi_{1c}^3 & = \frac{e^2}4 \epsilon_{\mu\nu} \epsilon_{\alpha\beta}\Re\mtr\left[ \hat Q \calL^{-1}\left( \calM^{-1}_{\beta\nu} \right) \hat P \big( \bpd_{\mu} \hat Q \big) V^{\alpha} \right] \label{EqS:Chi1cCase3}
    \end{align}
    This term cancels Eq.~\eqref{EqS:Chi1aCase2}: $\chi_{1c}^3 + \chi_{1a}^2 = 0$.
    
    \item $b = c$ and $E_a < 0$:  The coincident poles at $E_b$ generate an additional Fermi-surface term
     \begin{align}
        \chi_{1c}^{\rm FS} & =  \frac{e^2}4 \epsilon_{\mu\nu} \epsilon_{\alpha\beta} \sum_b n_F'(E_b) \big( \pd_{\alpha} E_b \big) \oRe\mtr\left( \big( \pd_{\mu} \hat P_b \big) \calM^{-1}_{\beta\nu} P_{\slashed{b}}  \right)  \label{EqS:Chi1cFS}
    \end{align}
\end{itemize}

The remaining occupation configurations do not generate on-shell delta function contributions for positive frequencies. After the cancellations listed above, the Fermi-sea part of $\chi_1$ is determined by $\chi_{1a}^{(1)}+\chi_{1c}^{(2)}$, while its Fermi surface part is given by
$\chi_1^{\rm FS} = \chi_{1a}^{\rm FS} + \chi_{1c}^{\rm FS}$.

\subsubsection{Cancellation of the Principal-Value Contributions}
In the preceding subsections, we kept only the on-shell delta-function contributions to $\chi_1$. We now show that the remaining principal value terms cancel identically. For this purpose, we use the symmetrized decomposition
\begin{align}
    \chi_1 = \frac12\left( \chi_{1a} + \chi_{1b} + \chi_{1c} + \chi_{1d} \right)
\end{align}
The contributions $\chi_{1a}$ and $\chi_{1c}$ were defined in Eqs.~\eqref{EqS:Chi1aMCD} and \eqref{EqS:Chi1cMCD}, respectively. The remaining two terms are
\begin{align}
    \chi_{1b} & = \frac{e^2 \hbar^2}{4 \pi} \int_{0^+}^{\infty} \frac{\rmd \omega}{\omega} \oIm\left\{ \epsilon_{\mu\nu} \epsilon_{\alpha\beta} \oint_{\mcC'} \frac{\rmd z}{2 \pi i} n_F(z) \mtr\left[ G_0(z) \calM^{-1}_{\mu\alpha}  G_0(z + i \omega_n) V^{\beta} G_0(z + i \omega_n) V^{\nu}   \right] \right\}_{i\omega_n \rightarrow \hbar \omega + i 0^+} \label{EqS:Chi1bMCD} \\
    \chi_{1d} & = \frac{e^2 \hbar^2}{4 \pi} \int_{0^+}^{\infty} \frac{\rmd \omega}{\omega} \oIm\left\{ \epsilon_{\mu\nu} \epsilon_{\alpha\beta} \oint_{\mcC'} \frac{\rmd z}{2 \pi i} n_F(z) \mtr\left[ G_0(z) V^{\beta} G_0(z) V^{\mu} G_0(z + i \omega_n)  \calM^{-1}_{\nu\alpha}   \right] \right\}_{i\omega_n \rightarrow \hbar \omega + i 0^+} \label{EqS:Chi1dMCD}
\end{align}
We first compare $\chi_{1d}$ with $\chi_{1a}$. Expanding Eq.~\eqref{EqS:Chi1dMCD} in the eigenstate basis gives
%
\begin{align}
    \chi_{1d} & = \frac{e^2 \hbar^2}{4 \pi} \int_{0^+}^{\infty} \frac{\rmd \omega}{\omega} \oIm\left\{ \epsilon_{\mu\nu} \epsilon_{\alpha\beta} V^{\beta}_{ab} V^{\mu}_{bc} \left( \calM^{-1}_{\nu\alpha} \right)_{ca}  \oint_{\calC'} \frac{\rmd z}{2\pi i} \frac{n_F(z)}{(z - E_a) (z - E_b) (i \omega_n +z - E_c)} \right\}_{i\omega_n \rightarrow \hbar \omega + i 0^+} \nonumber \\
    & = \frac{e^2 \hbar^2}{4 \pi} \int_{0^+}^{\infty} \frac{\rmd \omega}{\omega} \oIm\left\{ \epsilon_{\mu\nu} \epsilon_{\alpha\beta}  \left( V^{\alpha}_{ab} V^{\nu}_{ca} \big( \calM^{-1}_{\mu\beta} \big)_{bc} \right)^*  \oint_{\calC'} \frac{\rmd z}{2\pi i} \frac{n_F(z)}{(z - E_a) (z - E_b) (i \omega_n +z - E_c)} \right\}_{i\omega_n \rightarrow \hbar \omega + i 0^+}
\end{align}
Thus, the matrix element factors in $\chi_{1a}$ and $\chi_{1d}$ are complex conjugates of one another.  Similarly,  expansion of $\chi_{1b}$ in the eigenstate basis gives
\begin{align}
    \chi_{1b} & = \frac{e^2 \hbar^2}{4 \pi} \int_{0^+}^{\infty} \frac{\rmd \omega}{\omega} \oIm\left\{ \epsilon_{\mu\nu} \epsilon_{\alpha\beta}  \left( \calM^{-1}_{\mu\alpha} \right)_{ac} V^{\beta}_{c b} V^{\nu}_{ba}  \oint_{\calC'} \frac{\rmd z}{2\pi i} \frac{n_F(z)}{(z - E_a) (i \omega_n + z - E_b) (i \omega_n +z - E_c)} \right\}_{i\omega_n \rightarrow \hbar \omega + i 0^+} \nonumber \\
    & = \frac{e^2 \hbar^2}{4 \pi} \int_{0^+}^{\infty} \frac{\rmd \omega}{\omega} \oIm\left\{ \epsilon_{\mu\nu} \epsilon_{\alpha\beta}  \left( V^{\mu}_{ab} V^{\alpha}_{bc} \big( \calM^{-1}_{\nu\beta} \big)_{ca} \right)^*  \oint_{\calC'} \frac{\rmd z}{2\pi i} \frac{n_F(z)}{(z - E_a) (i \omega_n + z - E_b) (i \omega_n +z - E_c)} \right\}_{i\omega_n \rightarrow \hbar \omega + i 0^+}
\end{align}
The matrix-element factors in $\chi_{1b}$ and $\chi_{1c}$ are therefore also complex conjugates.

\subsubsection{Final Result for $\chi_1$}
Combining the results derived above, we decompose $\chi_1$ into a regular Fermi-sea contribution and a Fermi-surface contribution:
\begin{align}
    \chi_1    & =  \chi_1^{\rm TR}  +  \chi_1^{\rm FS} \ .    \label{Eqn:Chi1Overall}
\end{align}
The Fermi-sea contribution is
\begin{align}
    \chi_1^{\rm TR} & =  \chi_{1a}^{(1)}  + \chi_{1c}^{(2)} = - \frac{e^2}4 \epsilon_{\mu\nu}\epsilon_{\alpha\beta} \oRe\mtr\left\{ \calM^{-1}_{\nu\beta}   \left[ \bigl(\bpd_\mu\hat P\bigr) \hat Q \bigl(\bpd_\alpha\hat P\bigr) + \bigl(\bpd_\mu\hat Q\bigr) \hat P \bigl( \bpd_\alpha \hat Q\bigr)  \right] \right\}   \  . \label{EqS:Chi1TRFinal}
\end{align}
The Fermi surface contribution is obtained by combining Eqs.~\eqref{EqS:Chi1aFS} and \eqref{EqS:Chi1cFS}:
\begin{align}
    \chi_1^{\rm FS} & = \chi_{1a}^{\rm FS}  + \chi_{1c}^{\rm FS} = - \frac{e^2}4  \epsilon_{\mu\nu} \epsilon_{\alpha\beta} \sum_a n_F'(E_a) \bigl( \pd_\alpha E_a \bigr) \oRe\mtr\left\{ \calM^{-1}_{\mu\beta} \left[ \bigl( \pd_\nu \hat Q_{\sa} \bigr) \hat P_a + \hat P_{\sa} \bigl( \pd_\nu \hat P_a \bigr) \right] \right\} \ . \label{EqS:Chi1FSFirst}
\end{align}
Since $\hat Q_{\sa} \hat P_a = 0$, differentiation with respect to $k_\nu$ gives
\begin{align}
    \bigl(\pd_\nu\hat Q_{\sa}\bigr)\hat P_a = - \hat Q_{\sa} \bigl(\pd_\nu \hat P_a \bigr)  \  .
\end{align}
Eq.~\eqref{EqS:Chi1FSFirst} can therefore be rewritten as
\begin{align}
    \chi_1^{\rm FS} & = \frac{e^2}4  \epsilon_{\mu\nu} \epsilon_{\alpha\beta} \sum_a n_F'(E_a)  \bigl( \pd_\alpha E_a \bigr) \oRe\mtr\left[ \calM^{-1}_{\mu\beta} \bigl( \hat Q_{\sa} - \hat P_{\sa} \bigr) \bigl( \pd_\nu \hat P_a \bigr)  \right]  \   .  \label{Eqn:Chi1FSFinal}
\end{align}
Here, the band index $a$ runs over the bands intersecting the Fermi surface.

Using $\hat P + \hat Q=\hat I$, one can verify that $\chi_1^{\rm TR}$ in Eq.~\eqref{EqS:Chi1TRFinal} is identical to $\chi_{\rm MCD}^{(1,\rm TR)}$ given in Eq.~\eqref{Eqn:Chi1TRMCD}. The Green function derivation therefore reproduces the result obtained directly from the projector formulation.

\subsection{$\chi_2$}
\begin{figure}[htbp]
	\centering
	\includegraphics[width=0.48\textwidth]{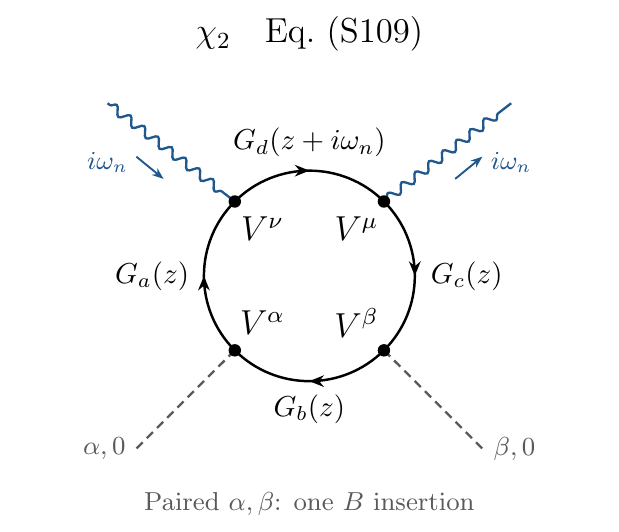}
	\caption{Diagrammatic representation of $\chi_2$ in Eq.~\eqref{EqS:Chi2DiagramKernel}, arising from the magnetic-field derivative of $\widetilde G(z)$. The paired velocity insertions $V^\alpha$ and $V^\beta$ lie on the unshifted-frequency branch: $G_a$, $G_b$, and $G_c$ carry energy $z$, whereas $G_d$ carries $z+i\omega_n$. The optical vertices are $V^\nu$ and $V^\mu$. Solid lines, wavy optical legs, and dashed static markers follow the conventions of Fig.~\ref{FigS:MCD_chi1_kernels}; the $\alpha,\beta$ pair represents one linear magnetic-field insertion contracted with $\epsilon_{\alpha\beta}$. All factors share the same zero-field momentum $\fvec k$, which is summed over the Brillouin zone. Band sums, contour and frequency integrals, and prefactors are as in Eq.~\eqref{EqS:Chi2DiagramKernel}.}
	\label{FigS:MCD_chi2_kernel}
\end{figure}

The contribution $\chi_2$ originates from the magnetic-field derivative of the first Green function in Eq.~\eqref{EqS:MMCDGaugeInvReal}, as illustrated in Fig.~\ref{FigS:MCD_chi2_kernel}. The replacement $\pd_B\widetilde G(z)|_{B=0}=(e\hbar/2i)\epsilon_{\alpha\beta}G(z)V^\alpha G(z)V^\beta G(z)$ inserts two velocity factors on the unshifted-frequency branch. Consequently, the propagators labeled $a,b,c$ carry energy $z$, while the remaining propagator $d$ carries $z+i\omega_n$:
\begin{align}
    \chi_2 & = \frac{e \hbar}{2\pi} \int_{0^+}^{\infty} \frac{\rmd \omega}{\omega} \epsilon_{\mu\nu} \oRe\left\{ \oint_{\calC'} \frac{\rmd z}{2\pi i} n_F(z) \mtr\left[ \frac{\rmd \widetilde{G}(z)}{\rmd B} V^{\mu} G(z + i \omega_n) V^{\nu}  \right] \right\}_{i \omega_n \rightarrow \hbar \omega + i 0^+} \nonumber \\
    & = \frac{e^2 \hbar^2}{4\pi} \epsilon_{\mu\nu} \epsilon_{\alpha \beta}  \int_{0^+}^{\infty} \frac{\rmd \omega}{\omega} \oIm\left\{ \oint_{\calC'} \frac{\rmd z}{2\pi i}n_F(z) \mtr\left[ G(z) V^{\alpha} G(z) V^{\beta} G(z) V^{\mu} G(z+ i \omega_n) V^{\nu} \right] \right\}_{i \omega_n \rightarrow \hbar \omega + i 0^+} \nonumber \\
    & = \frac{e^2 \hbar^2}{4\pi} \epsilon_{\mu\nu} \epsilon_{\alpha \beta} \int_{0^+}^{\infty} \frac{\rmd \omega}{\omega} \oIm\left( V^{\alpha}_{ab} V^{\beta}_{bc} V^{\mu}_{cd} V^{\nu}_{da} \oint_{\calC'} \frac{\rmd z}{2 \pi i} \frac{n_F(z)}{(z - E_a) (z - E_b) (z - E_c) (z + i \omega_n - E_d)}  \right)_{i \omega_n \rightarrow \hbar \omega + i 0^+}
    \label{EqS:Chi2DiagramKernel}
\end{align}

Note that hermiticity, together with the interchange $a \leftrightarrow c$, shows that the principal value contribution vanishes after summation. We therefore keep only the on-shell delta-function terms.

\subsubsection{Band Transition contributions}
The energies below are measured relative to the chemical potential. We distinguish the following occupation configurations.
\begin{itemize}
    \item $E_a < 0$ and $E_b, E_c, E_d > 0$:
    \begin{align}
        \chi_{2a} & = - \frac{e^2 \hbar^2}4 \epsilon_{\mu\nu} \epsilon_{\alpha \beta} \oRe \frac{ V^{\alpha}_{ab} V^{\beta}_{bc} V^{\mu}_{cd} V^{\nu}_{da} }{ (E_a - E_b) (E_a - E_c) (E_d - E_a) }  = \frac{e^2}{4 \hbar} \epsilon_{\mu\nu} \epsilon_{\alpha \beta}  \oRe\mtr\left( \hat P \calL^{-1} \left( \big( \bpd_{\alpha} \hat Q \big) V^{\beta} \right) \hat Q V^{\mu} \big( \bpd_{\nu} \hat Q \big) \right) \nonumber \\
        & = - \frac{e^2}{4 \hbar} \epsilon_{\mu\nu} \epsilon_{\alpha \beta}  \oRe\mtr\left( \hat P \calL^{-1} \left( \big( \bpd_{\beta} \hat Q \big) V^{\alpha} \right) \hat Q V^{\mu} \big( \bpd_{\nu} \hat Q \big) \right) 
    \end{align}
    This is one of the terms entering $\chi_{\rm MCD}^{(2c)}$ in Eq.~\eqref{EqS:Chi2cMCD}. 
    
    \item $E_b < 0$ and $E_a, E_c, E_d > 0$:
    \begin{align}
        \chi_{2b} & =  - \frac{e^2 \hbar^2}4 \epsilon_{\mu\nu} \epsilon_{\alpha \beta} \oRe \frac{ V^{\alpha}_{ab} V^{\beta}_{bc} V^{\mu}_{cd} V^{\nu}_{da} }{ (E_b - E_a) (E_b - E_c) (E_d - E_b) }  = \frac{e^2}{4 \hbar} \epsilon_{\mu\nu} \epsilon_{\alpha \beta}  \oRe\mtr\left( \hat P \calL^{-1} \left( \big( \bpd_{\beta} \hat Q \big) V^{\mu} \right) \hat Q V^{\nu} \big( \bpd_{\alpha} \hat Q \big) \right) \label{EqS:Chi2b}
    \end{align}
    
    \item $E_c < 0$ and $E_a, E_b, E_d > 0$:
    \begin{align}
        \chi_{2c} & =  - \frac{e^2 \hbar^2}4 \epsilon_{\mu\nu} \epsilon_{\alpha \beta} \oRe \frac{ V^{\alpha}_{ab} V^{\beta}_{bc} V^{\mu}_{cd} V^{\nu}_{da} }{ (E_c - E_a) (E_c - E_b) (E_d - E_c) }  = - \frac{e^2}{4 \hbar} \epsilon_{\mu\nu} \epsilon_{\alpha \beta}  \oRe\mtr\left( \hat Q \calL^{-1} \left( V^{\alpha} \big( \bpd_{\beta} \hat Q \big)  \right) \hat P  \big( \bpd_{\mu} \hat Q \big) V^{\nu} \right) \nonumber \\
        & =  \frac{e^2}{4 \hbar} \epsilon_{\mu\nu} \epsilon_{\alpha \beta}  \oRe\mtr\left( \hat P \calL^{-1} \left(  \big( \bpd_{\beta} \hat Q \big) V^{\alpha} \right) \hat Q  V^{\nu} \big( \bpd_{\mu} \hat Q \big)  \right)
    \end{align}
    This contribution provides a term in $\chi_{\rm MCD}^{(2c)}$ in Eq.~\ref{EqS:Chi2cMCD}. 
    
    \item $E_a, E_b < 0$ and $E_c, E_d > 0$: The two occupied-state residues can be combined using
    \begin{align}
        & \quad \frac1{(E_a - E_b) (E_a - E_c) (E_d - E_a) } + \frac1{(E_b - E_a) (E_b - E_c) (E_d - E_b) }  \nonumber \\
        & = \frac{E_a + E_b - (E_c + E_d)}{ (E_a - E_c) (E_d - E_a) (E_b - E_c) (E_d - E_b)} \nonumber \\
        & = \frac1{(E_d - E_a) (E_d - E_b) (E_b - E_c)} - \frac1{ (E_a - E_c) (E_d - E_a) (E_b - E_c)}
    \end{align}
    Consequently,
    \begin{align}
        \chi_{2d} & = \chi_{2d}^1 + \chi_{2d}^2 \\
        \chi_{2d}^1 & = - \frac{e^2 \hbar^2}4 \epsilon_{\mu\nu} \epsilon_{\alpha \beta} \oRe\frac{ V^{\alpha}_{ab} V^{\beta}_{bc} V^{\mu}_{cd} V^{\nu}_{da} }{ (E_d - E_a) (E_d - E_b) (E_b - E_c) } = \frac{e^2}{4 \hbar} \epsilon_{\mu\nu} \epsilon_{\alpha \beta} \oRe\mtr\left( \hat P \calL^{-1} \left( \big( \bpd_{\beta} \hat Q \big) V^{\mu} \right) \hat Q \big( \bpd_{\nu} \hat P \big) V^{\alpha} \right) \\
        \chi_{2d}^2 & = \frac{e^2 \hbar^2}4 \epsilon_{\mu\nu} \epsilon_{\alpha \beta} \oRe\frac{ V^{\alpha}_{ab} V^{\beta}_{bc} V^{\mu}_{cd} V^{\nu}_{da} }{ (E_a - E_c) (E_d - E_a) (E_b - E_c) } = \frac{e^2}{4 \hbar} \epsilon_{\mu\nu} \epsilon_{\alpha \beta} \oRe\mtr\left( \hat P \calL^{-1} \left(  V^{\alpha} \big( \bpd_{\beta} \hat P \big)  \right) \hat Q V^{\mu} \big( \bpd_{\nu} \hat Q \big) \right)
    \end{align}
    The second term  $\chi_{2d}^2$ contributes to $\chi_{\rm MCD}^{(2c)}$ in Eq.~\ref{EqS:Chi2cMCD}.
    
    \item $E_c, E_b < 0$ and $E_a, E_d > 0$: In this case, we use
    \begin{align}
        & \quad \frac1{(E_b - E_a) (E_b - E_c) (E_d - E_b) } + \frac1{(E_c - E_a) (E_c - E_b) (E_d - E_c) }  \nonumber \\
        & = \frac{E_c + E_b - (E_a + E_d)}{ (E_b - E_a) (E_d - E_b) (E_c - E_a) (E_d - E_c)} \nonumber \\
        & = \frac1{(E_b - E_a) (E_d - E_b) (E_d - E_c)} - \frac1{ (E_b - E_a) (E_c - E_a) (E_d - E_c)}
    \end{align}
    This gives
    \begin{align}
        \chi_{2e} & = \chi_{2e}^1 + \chi_{2e}^2 \\
        \chi_{2e}^1 & = - \frac{e^2 \hbar^2}4 \epsilon_{\mu\nu} \epsilon_{\alpha \beta} \oRe\frac{ V^{\alpha}_{ab} V^{\beta}_{bc} V^{\mu}_{cd} V^{\nu}_{da} }{ (E_b - E_a) (E_d - E_b) (E_d - E_c) } = \frac{e^2}{4 \hbar} \epsilon_{\mu\nu} \epsilon_{\alpha \beta} \oRe\mtr\left( \hat P \calL^{-1} \left( V^{\beta} \big( \bpd_{\mu} \hat P \big)  \right) \hat Q V^{\nu} \big( \bpd_{\alpha} \hat Q \big)  \right) \\
        \chi_{2e}^2 & = \frac{e^2 \hbar^2}4 \epsilon_{\mu\nu} \epsilon_{\alpha \beta} \oRe\frac{ V^{\alpha}_{ab} V^{\beta}_{bc} V^{\mu}_{cd} V^{\nu}_{da} }{ (E_b - E_a) (E_c - E_a) (E_d - E_c) } =  - \frac{e^2}{4 \hbar} \epsilon_{\mu\nu} \epsilon_{\alpha \beta} \oRe\mtr\left( \hat Q \calL^{-1} \left(   \big( \bpd_{\alpha} \hat P \big) V^{\beta} \right) \hat P  \big( \bpd_{\mu} \hat Q \big) V^{\nu} \right)
    \end{align}
    The second term  $\chi_{2e}^2$ also contributes to $\chi_{\rm MCD}^{(2 c)}$ in Eq.~\ref{EqS:Chi2cMCD}.
    
    \item $E_c, E_a < 0$ and $E_b, E_d > 0$: The relevant identity is
    \begin{align}
        & \quad \frac1{(E_a - E_b) (E_a - E_c) (E_d - E_a) } + \frac1{(E_c - E_a) (E_c - E_b) (E_d - E_c) }  \nonumber \\
        & = \frac1{(E_a - E_b) (E_d - E_a) (E_d - E_c)} - \frac1{ (E_a - E_b) (E_c - E_b) (E_d - E_c)}
    \end{align}
    Accordingly,
    \begin{align}
        \chi_{2f} & = \chi_{2f}^1 + \chi_{2f}^2 \\
        \chi_{2f}^1 & = - \frac{e^2 \hbar^2}4 \epsilon_{\mu\nu} \epsilon_{\alpha \beta} \oRe\frac{ V^{\alpha}_{ab} V^{\beta}_{bc} V^{\mu}_{cd} V^{\nu}_{da} }{(E_a - E_b) (E_d - E_a) (E_d - E_c)} = - \frac{e^2}{4 \hbar} \epsilon_{\mu\nu} \epsilon_{\alpha \beta} \oRe\mtr\left( \hat P  \big( \bpd_{\alpha} \hat Q \big)  V^{\beta} \big( \bpd_{\mu} \hat P \big) \big( \bpd_{\nu} \hat Q \big)  \right) \\
        \chi_{2f}^2 & = \frac{e^2 \hbar^2}4 \epsilon_{\mu\nu} \epsilon_{\alpha \beta} \oRe\frac{ V^{\alpha}_{ab} V^{\beta}_{bc} V^{\mu}_{cd} V^{\nu}_{da} }{ (E_a - E_b) (E_c - E_b) (E_d - E_c) } =  \frac{e^2}{4 \hbar} \epsilon_{\mu\nu} \epsilon_{\alpha \beta} \oRe\mtr\left(  \big( \bpd_{\alpha} \hat P \big) \hat Q \big( \bpd_{\beta} \hat P \big)  \big( \bpd_{\mu} \hat Q \big) V^{\nu}  \right)
    \end{align}
    Relabeling the contracted indices, we may rewrite
    \begin{align}
        \chi_{2f}^1 & = - \frac{e^2}{4 \hbar} \epsilon_{\mu\nu} \epsilon_{\alpha \beta} \oRe\mtr\left( \hat P \big( \bpd_{\nu} \hat Q \big)  V^{\mu} \big( \bpd_{\beta} \hat P \big) \big( \bpd_{\alpha} \hat Q \big)  \right) = \frac{e^2}{4 \hbar} \epsilon_{\mu\nu} \epsilon_{\alpha \beta} \oRe\mtr\left( \hat P \big( \bpd_{\nu} \hat Q \big)  V^{\mu} \big( \bpd_{\alpha} \hat P \big) \big( \bpd_{\beta} \hat Q \big)  \right) \nonumber \\
        & = - \frac{e^2}{4 \hbar} \epsilon_{\mu\nu} \epsilon_{\alpha \beta} \oRe\mtr\left( \big( \bpd_{\beta} \hat P \big) \big( \bpd_{\nu} \hat Q \big)  V^{\mu} \big( \bpd_{\alpha} \hat P \big)  \hat Q  \right) 
    \end{align}
    This is precisely $\chi_{\rm MCD}^{(2a)}$ in Eq.~\eqref{EqS:Chi2aMCD}. The same manipulations show that $\chi_{2f}^2 = \chi_{2f}^1$, so both terms give the same contribution.
    
    \item $E_c, E_a, E_b < 0$ and $E_d > 0$: The three occupied-state residues satisfy
    \begin{align}
        & \quad \frac1{(E_a - E_b) (E_a - E_c) (E_d - E_a) } + \frac1{(E_b - E_a) (E_b - E_c) (E_d - E_b)} + \frac1{(E_c - E_a) (E_c - E_b) (E_d - E_c) }  \nonumber \\
        & = \frac1{(E_d - E_a) (E_d - E_b) (E_d - E_c)} 
    \end{align}
    It follows that
    \begin{align}
        \chi_{2g} & = - \frac{e^2 \hbar^2}4 \epsilon_{\mu\nu} \epsilon_{\alpha \beta} \oRe\frac{ V^{\alpha}_{ab} V^{\beta}_{bc} V^{\mu}_{cd} V^{\nu}_{da} }{(E_d - E_a) (E_d - E_b) (E_d - E_c)} =  \frac{e^2}{4 \hbar} \epsilon_{\mu\nu} \epsilon_{\alpha \beta} \oRe\mtr\left( \hat P \calL^{-1} \left(  V^{\beta} \big( \bpd_{\mu} \hat P \big) \right) \hat Q  \big( \bpd_{\nu} \hat P \big)  \hat V^{\alpha}  \right) 
    \end{align}
\end{itemize}

The terms
\begin{align} 
 \chi_{2a},\quad \chi_{2c},\quad \chi_{2d}^{(2)},\quad \chi_{2e}^{(2)},\quad \chi_{2f}^{(1)},\quad \chi_{2f}^{(2)}
\end{align}
combine into the regular contribution $\chi_{\rm MCD}^{(2)}$. The four remaining terms,
\begin{align}
    \chi_{2b},\qquad \chi_{2d}^{(1)},\qquad \chi_{2e}^{(1)},\qquad \chi_{2g} \ ,
\end{align}
are not part of this expression. As shown in the next section, they cancel against the corresponding terms generated by $\chi_3$.

\subsubsection{Fermi-Surface Contributions}

We now consider the terms proportional to $n_F'$, which arise from coincident poles and are constrained at the Fermi surface. There are three distinct cases.
\begin{itemize}
    \item $a= b \neq c$, $E_d > 0$: The coincident poles at $E_a$ give
    \begin{align}
        \chi_{2a}^{\rm FS} & = - \frac{e^2 \hbar}4 \epsilon_{\mu\nu} \epsilon_{\alpha \beta} \oRe \sum_{\substack{a, c\neq a \\ E_d > 0}} \big(\pd_{\alpha} E_a \big) \frac{n_F'(E_a)}{ (E_a - E_c) (E_d - E_a) } V^{\beta}_{ac} V^{\mu}_{cd} V^{\nu}_{da} \nonumber \\
        & = \frac{e^2}{4 \hbar} \epsilon_{\mu\nu} \epsilon_{\alpha \beta} \oRe\sum_a  n_F'(E_a) \big(\pd_{\alpha} E_a \big) \mtr\left( \hat P_a \big( \bpd_{\beta} \hat Q_a \big) V^{\mu} \big( \bpd_{\nu} \hat Q_{\sa} \big) \right) \nonumber \\
        & = \frac{e^2}{4 \hbar} \epsilon_{\mu\nu} \epsilon_{\alpha \beta} \oRe\sum_a  n_F'(E_a) \big(\pd_{\alpha} E_a \big) \mtr\left( \big( \pd_{\beta} \hat P_a \big)  \hat Q_a \hat V^{\mu}  \hat Q_{\sa} \big( \pd_{\nu} \hat P_a \big) \right)  \label{EqS:Chi2aFS}
    \end{align}
    
    \item $a \neq b = c$, $E_d > 0$: The repeated pole at $E_b$ yields
    \begin{align}
        \chi_{2b}^{\rm FS} & = - \frac{e^2 \hbar}4 \epsilon_{\mu\nu} \epsilon_{\alpha \beta} \oRe\sum_{\substack{b, a \neq b \\ E_d > 0}} \big(\pd_{\beta} E_b \big) \frac{n_F'(E_b)}{ (E_b - E_a) (E_d - E_b) } V^{\alpha}_{ab} V^{\mu}_{bd} V^{\nu}_{da} \nonumber \\
        & = \frac{e^2}{4 \hbar} \epsilon_{\mu\nu} \epsilon_{\alpha \beta} \oRe\sum_a  n_F'(E_a) \big(\pd_{\alpha} E_a \big) \mtr\left( \big( \pd_{\beta}  \hat P_a \big) \big( \pd_{\nu}  \hat P_a \big) \hat Q_{\sa}  V^{\mu} \hat Q_a \right) = \chi_{2a}^{\rm FS}
    \end{align}
    where we have relabeled the band and spatial indices, and used the cyclicity of the trace.
    
    \item $b \neq a = c$, $E_d > 0$: In this case, the coincident poles at $E_a$ give
    \begin{align}
        \chi_{2c}^{\rm FS} & = - \frac{e^2 \hbar^2}4 \epsilon_{\mu\nu} \epsilon_{\alpha \beta} \oRe\sum_{\substack{a, b\neq a \\ E_d > 0}} \frac{n_F'(E_a)}{ (E_a - E_b) (E_d - E_a) } V^{\alpha}_{ab} V^{\beta}_{ba} V^{\mu}_{ad} V^{\nu}_{da} \nonumber \\
        & = - \frac{e^2}4 \epsilon_{\mu\nu} \epsilon_{\alpha \beta} \sum_a  n_F'(E_a) \oRe\mtr\left( \big( \pd_{\alpha}  \hat P_a \big) \hat Q_a V^{\beta} \hat P_a \big( \pd_{\mu}  \hat Q_{\sa} \big) V^{\nu} \right) \nonumber \\
        & = - \frac{e}2 \epsilon_{\mu\nu} \sum_a n_F'(E_a) m_a \oIm\mtr\left( \hat P_a \big( \pd_{\mu}  \hat Q_{\sa} \big) V^{\nu} \right) =  - \frac{e}2 \epsilon_{\mu\nu} \sum_a n_F'(E_a) m_a \oIm\mtr\left( \big( \pd_{\nu}  \hat P_a \big) \hat Q_{\sa} V^{\mu} \right) 
    \end{align}
    where we have used the formula
    \[  \epsilon_{\alpha\beta}  \big( \pd_{\alpha}  \hat P_a \big) \hat Q_a V^{\beta} \hat P_a  = -\frac{2 i}e m_a \hat P_a  \]
\end{itemize}

No other coincident-pole configurations generate terms proportional to $n_F'$ in $\chi_2$. Combining the three contributions, and using
$\chi_{2a}^{\rm FS} = \chi_{2b}^{\rm FS}$, we obtain
\begin{align}
    \chi_2^{\rm FS} & = \frac{e^2}{2\hbar} \epsilon_{\mu\nu} \epsilon_{\alpha\beta} \sum_a n_F'(E_a) \bigl( \pd_\alpha E_a \bigr) \oRe\mtr\left[ \bigl( \pd_\beta \hat P_a \bigr) \hat Q_aV^\mu \hat Q_{\sa} \bigl(\pd_\nu \hat P_a \bigr) \right] - \frac{e}2 \epsilon_{\mu\nu} \sum_a n_F'(E_a) m_a \oIm\mtr\left[ \bigl( \pd_\nu \hat P_a \bigr) \hat Q_{\sa}V^\mu \right] \ . \label{EqS:Chi2FS}
\end{align}
Thus, $\chi_2^{\rm FS}$ contains two qualitatively distinct terms. The first originates from the field-induced mixing of the Fermi-surface band with the complementary subspace. The second is weighted directly by the orbital magnetic moment $m_a$ of the states at the Fermi surface.

\subsection{$\chi_3$}
\begin{figure}[htbp]
	\centering
	\includegraphics[width=0.48\textwidth]{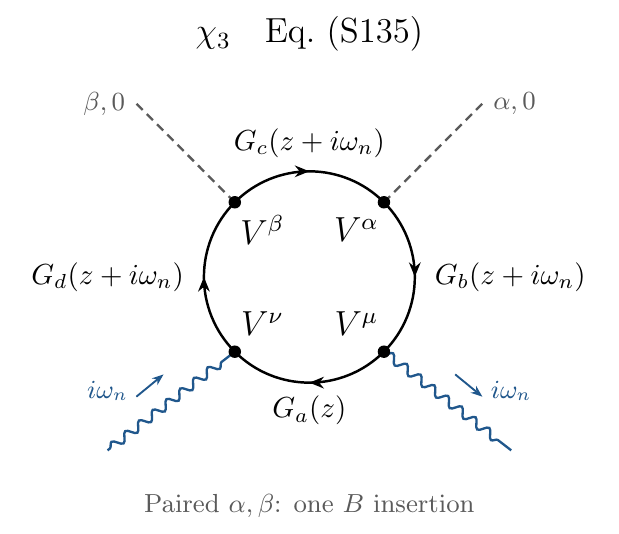}
    \caption{Diagrammatic representation of $\chi_3$ in Eq.~\eqref{EqS:Chi3DiagramKernel}, arising from the magnetic-field derivative of $\widetilde G(z+i\omega_n)$. The paired velocity insertions $V^\alpha$ and $V^\beta$ lie on the shifted-frequency branch: $G_b$, $G_c$, and $G_d$ carry energy $z+i\omega_n$, whereas $G_a$ carries $z$. The optical vertices are $V^\nu$ and $V^\mu$. Line and vertex conventions are those of Fig.~\ref{FigS:MCD_chi1_kernels}; the dashed $\alpha,\beta$ markers together represent one linear magnetic-field insertion contracted with $\epsilon_{\alpha\beta}$. All factors share the same zero-field momentum $\fvec k$, which is summed over the Brillouin zone. Band sums, contour and frequency integrals, and prefactors are as in Eq.~\eqref{EqS:Chi3DiagramKernel}.}
\label{FigS:MCD_chi3_kernel}
\end{figure}

The contribution $\chi_3$ arises from the magnetic-field derivative of the second Green function in Eq.~\eqref{EqS:MMCDGaugeInvReal}, as shown in Fig.~\ref{FigS:MCD_chi3_kernel}. Here the paired velocity factors are inserted on the shifted-frequency branch: the propagators $b,c,d$ carry $z+i\omega_n$, while $a$ carries $z$. The corresponding response is
\begin{align}
    \chi_3 & = \frac{e \hbar}{2\pi} \int_{0^+}^{\infty} \frac{\rmd \omega}{\omega} \epsilon_{\mu\nu} \oRe\left\{ \oint_{\calC'} \frac{\rmd z}{2\pi i} n_F(z) \mtr\left[ G(z) V^{\mu} \frac{\rmd G(z + i \omega_n)}{\rmd B} V^{\nu}  \right] \right\}_{i \omega_n \rightarrow \hbar \omega + i 0^+}
\end{align}
Substituting the magnetic-field derivative of the gauge-invariant Green function gives
\begin{align}
    \chi_3 & = \frac{e^2 \hbar^2}{4\pi} \epsilon_{\mu\nu} \epsilon_{\alpha \beta}  \int_{0^+}^{\infty} \frac{\rmd \omega}{\omega} \oIm \left\{ \oint_{\calC'} \frac{\rmd z}{2\pi i} n_F(z)\mtr\left[  G(z) V^{\mu} G(z + i \omega_n) V^{\alpha} G(z + i \omega_n) V^{\beta} G(z+ i \omega_n) V^{\nu} \right] \right\}_{i \omega_n \rightarrow \hbar \omega + i 0^+} \nonumber \\
    & = \frac{e^2 \hbar^2}{4\pi} \epsilon_{\mu\nu} \epsilon_{\alpha \beta} \int_{0^+}^{\infty} \frac{\rmd \omega}{\omega} \oIm\left( V^{\mu}_{ab} V^{\alpha}_{bc} V^{\beta}_{cd} V^{\nu}_{da} \oint_{\calC'} \frac{\rmd z}{2 \pi i} \frac{n_F(z)}{(z - E_a) (z + i \omega_n - E_b) (z + i\omega_n - E_c) (z + i \omega_n - E_d)}  \right)_{i \omega_n \rightarrow \hbar \omega + i 0^+} \ . 
    \label{EqS:Chi3DiagramKernel}
\end{align}

As in the derivation of $\chi_2$, hermiticity and relabeling of the intermediate state indices lead to the cancellation of the principal value terms. We may therefore restrict the following analysis to the on-shell contributions generated by the delta functions.

\subsubsection{Band-Transition Contributions}
We omit the intermediate residue calculations and summarize the non-vanishing band transition contributions below. As before, all energies are measured relative to the chemical potential.

For compactness, we define
\begin{align}
    \calW_{abcd}^{\mu\alpha\beta\nu} \equiv V^\mu_{ab} V^\alpha_{bc} V^\beta_{cd} V^\nu_{da} \ .
\end{align}

\begin{itemize}
    \item $E_a < 0$ and $E_b, E_c, E_d > 0$:  We first use the formula
    \begin{align}
        & \quad \frac1{(i \omega_n + E_a - E_b) (i \omega_n + E_a - E_c) (i \omega_n + E_a - E_d)} \nonumber \\
        & = \frac1{(E_b - E_c) (E_b - E_d) (i\omega_n + E_a - E_b)} + \frac1{(E_c - E_b) (E_c - E_d) (i\omega_n + E_a - E_c)} + \frac1{(E_d - E_b) (E_d - E_c) (i\omega_n + E_a - E_d)}   \nonumber \\
        & \rightarrow - i \pi  \left( \frac1{ (E_b - E_c) (E_b - E_d) (E_b - E_a)} + \frac1{ (E_c - E_b) (E_c - E_d) (E_c - E_a)} + \frac1{ (E_d - E_b) (E_d - E_c) (E_d - E_a)} \right)  \nonumber \\
        & = \frac{- i \pi }{ (E_b - E_a) (E_c - E_a) (E_d - E_a) } 
    \end{align}
    Consequently,
    \begin{align}
        \chi_{3a} & = - \frac{e^2 \hbar^2}4 \epsilon_{\mu\nu} \epsilon_{\alpha \beta} \oRe \frac{ \calW_{abcd}^{\mu\alpha\beta\nu} }{ (E_b - E_a) (E_c - E_a) (E_d - E_a) }  = \frac{e^2}{4\hbar} \epsilon_{\mu\nu} \epsilon_{\alpha \beta}  \Re\mtr\left( \hat P \calL^{-1} \left( \big( \bpd_{\mu} \hat Q \big) V^{\alpha} \right) \hat Q V^{\beta} \big( \bpd_{\nu} \hat Q \big) \right) \label{EqS:Chi3a}
    \end{align}
    Relabeling the contracted indices shows that $\chi_{3a} + \chi_{2b} = 0$.
    
    \item $E_a, E_b < 0$ and $E_c, E_d > 0$: The relevant frequency dependent terms satisfy
    \begin{align}
        & \quad - \left( \frac1{(E_c - E_d) (E_c - E_b) (E_c - i \omega_n - E_a)} + \frac1{(E_d - E_c) (E_d - E_b) (E_d - i \omega_n - E_a)} \right) \nonumber \\
        & \rightarrow - i \pi \left( \frac1{(E_c - E_d) (E_c - E_b) (E_c - E_a)} +  \frac1{(E_d - E_c) (E_d - E_b) (E_d - E_a)} \right) \nonumber \\
        & = i \pi \left( \frac1{ (E_c - E_a) (E_c - E_b) (E_d - E_a) } + \frac1{ (E_c - E_b) (E_d - E_b) (E_d - E_a) }  \right)
    \end{align}
    leading to
    \begin{align}
        \chi_{3b} & = \chi_{3b}^{(1)} + \chi_{3b}^{(2)}   \\ 
        \chi_{3b}^{(1)} & = \frac{e^2 \hbar^2}4 \epsilon_{\mu\nu} \epsilon_{\alpha \beta} \oRe\left( \frac{ \calW_{abcd}^{\mu\alpha\beta\nu} }{ (E_c - E_a) (E_c - E_b) (E_d - E_a) } \right) =  \frac{e^2}{4\hbar} \epsilon_{\mu\nu} \epsilon_{\alpha \beta} \oRe\mtr\left( \hat P \calL^{-1}\left( V^{\mu} \big( \bpd_{\alpha} \hat P \big)  \right)  \hat Q \hat V^{\beta} \big( \bpd_{\nu} \hat Q \big) \right) = -\chi_{2e}^{(1)}  \\
        \chi_{3b}^{(2)} & = \frac{e^2 \hbar^2}4 \epsilon_{\mu\nu} \epsilon_{\alpha \beta} \oRe\left( \frac{\calW_{abcd}^{\mu\alpha\beta\nu} }{ (E_c - E_b) (E_d - E_b) (E_d - E_a) }  \right) = \frac{e^2}{4\hbar} \epsilon_{\mu\nu} \epsilon_{\alpha \beta} \Re\mtr\left( \hat P \calL^{-1}\left( \big( \bpd_{\alpha} \hat Q \big) V^{\beta} \right)  \hat Q  \big( \bpd_{\nu} \hat P \big) \hat V^{\mu} \right)
    \end{align}
    Thus, $\chi_{3b}^{(1)}$ cancels $\chi_{2e}^{(1)}$, while $\chi_{3b}^{(2)}$ contributes to the second structure in $\chi_{\rm MCD}^{(2d)}$ in Eq.~\eqref{EqS:Chi2dMCD}.
    
    \item $E_a, E_d < 0$ and $E_b, E_c > 0$: Proceeding in the same manner, we find
    \begin{align}
        \chi_{3c} & = \frac{e^2 \hbar^2}4 \epsilon_{\mu\nu} \epsilon_{\alpha\beta} \oRe\left\{ \calW_{abcd}^{\mu\alpha\beta\nu} \left[ \frac1{ (E_c - E_a)(E_c - E_d)(E_b - E_a)} + \frac1{(E_c - E_d)(E_b - E_d)(E_b - E_a)} \right] \right\} \ . \label{EqS:Chi3c}
    \end{align}
    This expression separates into two terms. After relabeling the dummy indices, one term cancels $\chi_{2d}^{(1)}$. The other is, up to Hermitian conjugation, identical to $\chi_{3b}^{(2)}$. Together, the surviving terms reproduce the second structure in $\chi_{\rm MCD}^{(2d)}$ in Eq.~\eqref{EqS:Chi2dMCD}.
    
    \item $E_a, E_c < 0$ and $E_b, E_d > 0$: In this case,
    \begin{align}
        \chi_{3d} & = \chi_{3d}^{(1)} + \chi_{3d}^{(2)} \\
        \chi_{3d}^{(1)} & = \frac{e^2 \hbar^2}4 \epsilon_{\mu\nu} \epsilon_{\alpha \beta} \oRe\frac{ \calW_{abcd}^{\mu\alpha\beta\nu} }{ (E_b - E_a) (E_d - E_a) (E_b - E_c) } = \frac{e^2}{4\hbar} \epsilon_{\mu\nu} \epsilon_{\alpha \beta} \oRe\mtr\left( \big( \bpd_{\mu} \hat P \big) \hat Q \big( \bpd_{\alpha} \hat P \big) V^{\beta}  \big( \bpd_{\nu} \hat Q \big) \right) \\
        \chi_{3d}^2 & = \frac{e^2 \hbar^2}4 \epsilon_{\mu\nu} \epsilon_{\alpha \beta} \oRe\frac{ \calW_{abcd}^{\mu\alpha\beta\nu} }{ (E_b - E_c) (E_d - E_c) (E_d - E_a) } =  \frac{e^2}{4\hbar} \epsilon_{\mu\nu} \epsilon_{\alpha \beta} \oRe\mtr\left( \hat P   V^{\mu} \big( \bpd_{\alpha} \hat Q \big)   \hat P  \big( \bpd_{\beta} \hat Q \big) \big( \bpd_{\nu} \hat Q \big) \right)
    \end{align}
    The two terms are related by relabeling the contracted indices and using the cyclicity of the trace. Together, they reproduce the contribution $\chi_{\rm MCD}^{(2b)}$ in Eq.~\eqref{EqS:Chi2bMCD}.
    
    \item $E_c, E_b, E_a < 0$ and $E_d > 0$: Combining the three occupied-state residues gives
    \begin{align}
        \chi_{3e} & =  - \frac{e^2 \hbar^2}4 \epsilon_{\mu\nu} \epsilon_{\alpha \beta} \oRe\frac{ \calW_{abcd}^{\mu\alpha\beta\nu} }{ (E_d - E_a) (E_d - E_b) (E_d - E_c) } = \frac{e^2}{4\hbar} \epsilon_{\mu\nu} \epsilon_{\alpha \beta} \oRe\mtr\left( \hat P \calL^{-1} \left( V^{\alpha} \big( \bpd_{\beta} \hat P \big)  \right) \hat Q \big( \bpd_{\nu} \hat P \big) V^{\mu}  \right) \label{EqS:Chi3e}
    \end{align}
    
    \item $E_c, E_a, E_d < 0$ and $E_b > 0$: The corresponding result is
    \begin{align}
        \chi_{3f} & = - \frac{e^2 \hbar^2}4 \epsilon_{\mu\nu} \epsilon_{\alpha \beta} \oRe\frac{ \calW_{abcd}^{\mu\alpha\beta\nu} }{(E_b - E_a) (E_b - E_c) (E_b - E_d)} = - \frac{e^2}{4\hbar} \epsilon_{\mu\nu} \epsilon_{\alpha \beta} \oRe\mtr\left( \big( \bpd_{\mu} \hat P \big) \hat Q \calL^{-1}\left( \big( \bpd_{\alpha} \hat P \big)  V^{\beta} \right) \hat P V^{\nu}   \right) 
    \end{align}
    After relabeling indices and taking the Hermitian conjugate where necessary, $\chi_{3e}$ and $\chi_{3f}$ combine to reproduce the first structure in $\chi_{\rm MCD}^{(2d)}$.
    
    \item $E_d, E_a, E_b < 0$ and $E_c > 0$: Finally,
    \begin{align}
        \chi_{3g} & = - \frac{e^2 \hbar^2}4 \epsilon_{\mu\nu} \epsilon_{\alpha \beta} \oRe\frac{ \calW_{abcd}^{\mu\alpha\beta\nu} }{(E_c - E_a) (E_c - E_b) (E_c - E_d)} =  \frac{e^2}{4\hbar} \epsilon_{\mu\nu} \epsilon_{\alpha \beta} \oRe\mtr\left( \hat P \calL^{-1} \left(  V^{\mu} \big( \bpd_{\alpha} \hat P \big) \right) \hat Q  \big( \bpd_{\beta} \hat P \big)  \hat V^{\nu}  \right) 
    \end{align}
    This contribution cancels $\chi_{2g}$.
\end{itemize}

We have therefore established the cancellations
\begin{align}
    \chi_{3a} + \chi_{2b} & = 0 \ , & \chi_{3b}^{(1)} + \chi_{2e}^{(1)} & = 0 \ , & \chi_{3c}^{(1)} + \chi_{2d}^{(1)} & = 0 \ , & \chi_{3g} + \chi_{2g} & = 0 \ .
\end{align}
The remaining terms combine precisely into the regular band transition contribution obtained from the projector formulation. This completes the consistency check between the Green function and projector derivations of the Fermi sea part of $\chi_{\rm MCD}$.

\subsubsection{Fermi-Surface Contributions}
We next evaluate the terms proportional to $n_F'$. These contributions arise from coincident poles associated with states at the Fermi surface. Three distinct configurations must be considered.
\begin{itemize}
    \item $b = c \neq d$ and $E_a < 0$: Since $V^\alpha_{bb}=\hbar^{-1}\pd_\alpha E_b$, the repeated pole at $E_b$ gives
    \begin{align}
        \chi_{3a}^{\rm FS} & = \frac{e^2 \hbar}4 \epsilon_{\mu\nu} \epsilon_{\alpha \beta} \oRe\sum_{\substack{b, d \neq b \\ E_a < 0}} \big( \pd_{\alpha} E_b \big) \frac{n_F'(E_b)}{(E_b - E_d) (E_b - E_a) }  V^{\mu}_{ab} V^{\beta}_{bd} V^{\nu}_{da} \nonumber \\
        & = - \frac{e^2}{4 \hbar} \epsilon_{\mu\nu} \epsilon_{\alpha \beta} \oRe\sum_a n_F'(E_a) \big( \pd_{\alpha} E_a \big) \Re\mtr\left( \big( \pd_{\beta} \hat P_a \big) \hat Q_a \hat V^{\mu} \hat P_{\sa} \big( \pd_{\nu} \hat P_a \big)  \right) \label{EqS:Chi3aFS}
    \end{align}
    
    \item $d = c \neq b$ and $E_a < 0$: In this case, $V^\beta_{cc}=\hbar^{-1}\pd_\beta E_c$, and we obtain
    \begin{align}
        \chi_{3b}^{\rm FS} & = \frac{e^2 \hbar}4 \epsilon_{\mu\nu} \epsilon_{\alpha \beta} \oRe\sum_{\substack{c, b \neq c \\ E_a < 0}} \big( \pd_{\beta} E_c \big) \frac{n_F'(E_c)}{(E_c - E_b) (E_c - E_a) }  V^{\mu}_{ab} V^{\alpha}_{bc} V^{\nu}_{ca} \nonumber \\
        & = - \frac{e^2}{4 \hbar} \epsilon_{\mu\nu} \epsilon_{\alpha \beta}  \sum_a n_F'(E_a) \big( \pd_{\alpha} E_a \big) \oRe\mtr\left( \big( \pd_{\nu} \hat P_a \big) \hat P_{\sa} \hat V^{\mu} \hat Q_a \big( \pd_{\beta} \hat P_a \big)  \right) \label{EqS:Chi3bFS}
    \end{align}
    
    \item $d = b \neq c$ and $E_a < 0$: The two Green functions associated with band $b$ produce a repeated pole. Unlike the preceding cases, this contribution does not involve a diagonal velocity matrix element. 
    \begin{align}
        \chi_{3c}^{\rm FS} & = \frac{e^2 \hbar^2}4 \epsilon_{\mu\nu} \epsilon_{\alpha \beta} \oRe\sum_{\substack{b, c \neq b \\ E_a < 0}}  \frac{n_F'(E_b)}{(E_b - E_c) (E_b - E_a) }  V^{\mu}_{ab} V^{\alpha}_{bc} V^{\beta}_{cb} V^{\nu}_{ba} \nonumber \\
        & = \frac{e}2 \epsilon_{\mu\nu} \sum_a n_F'(E_a) m_a \oIm\left( \big( \bpd_{\nu} \hat P_a \big) \hat P_{\sa} \hat V^{\mu}  \right) \label{Eqn:Chi3cFS}
    \end{align}
    where we have used the orbital moment identity
    \begin{align}
        m_b = - \frac{e \hbar}{2i}  \epsilon_{\alpha\beta} \sum_{c\neq b} \frac{ V^\alpha_{bc} V^\beta_{cb} }{E_b-E_c} \ .
    \end{align}
\end{itemize}

No other coincident-pole configurations generate terms proportional to $n_F'$ in $\chi_3$. Combining all the equations, we obtain
\begin{align}
    \chi_3^{\rm FS} & = - \frac{e^2}{2\hbar} \epsilon_{\mu\nu} \epsilon_{\alpha\beta} \sum_a n_F'(E_a) \bigl( \pd_\alpha E_a \bigr) \oRe\mtr\left[ \bigl( \pd_\beta \hat P_a \bigr) \hat Q_aV^\mu\hat P_{\sa} \bigl( \pd_\nu \hat P_a \bigr) \right] + \frac{e}2 \epsilon_{\mu\nu} \sum_a n_F'(E_a) m_a \oIm\mtr\left[ \bigl( \pd_\nu \hat P_a \bigr) \hat P_{\sa} V^\mu \right] \  . \label{EqS:Chi3FS}
\end{align}

\subsection{Final Expression for $\chi_{\rm MCD}$}

Combining the results for $\chi_1$, $\chi_2$, and $\chi_3$, we obtain
\begin{align}
    \chi_{\rm MCD} & = \chi_{\rm MCD}^{\rm TR} + \chi_{\rm MCD}^{\rm FS} \label{EqS:ChiMCD} \\
        \chi_{\rm MCD}^{\rm TR} & = -\frac{e^2}4 \epsilon_{\mu\nu} \epsilon_{\alpha\beta} \oRe\mtr\left[ \calM^{-1}_{\mu\beta}  \big( \bpd_{\nu} \hat P \big) \big( \bpd_{\alpha} \hat Q \big) + \frac{2}{\hbar} V^{\mu} \left[ \big( \bpd_{\alpha} \hat P \big) \hat Q \big( \bpd_{\beta} \hat P \big) \big( \bpd_{\nu} \hat Q \big) + \big( \bpd_{\alpha} \hat Q \big)   \hat P  \big( \bpd_{\beta} \hat Q \big) \big( \bpd_{\nu} \hat P \big) \right]  \right]  \nonumber \\
    & \qquad + \frac{e^2}{2 \hbar} \oRe\mtr\left[ \hat P \mcL^{-1}\left( \hat O_1 \right) \hat Q \hat O_2 \right]  \label{EqS:ChiTRMCD}
\end{align}
where
\begin{align}
    \hat O_1 & = \epsilon_{\alpha\beta} \left( V^{\alpha} \big( \bpd_{\beta} \hat P \big) + \big( \bpd_{\alpha} \hat Q \big) V^{\beta} \right) \qquad  \hat O_2 = \epsilon_{\mu\nu} \left( V^{\mu} \big( \bpd_{\nu} \hat Q \big) - \big( \bpd_{\mu} \hat P \big) V^{\nu} \right)   
\end{align}
The remaining contribution is from the Fermi surface:
\begin{align}
    \chi_{\rm MCD}^{\rm FS}  & = \frac{e^2}4 \epsilon_{\mu\nu} \epsilon_{\alpha\beta} \sum_a n_F'(E_a) \big( \pd_{\alpha} E_a \big) \oRe\mtr\left[ \calM^{-1}_{\beta \mu}  \big( \hat Q_{\sa} -  \hat P_{\sa} \big)  \big( \pd_{\nu} \hat P_a \big)  + \frac2{\hbar} \big( \pd_{\beta} \hat P_a \big)  \hat Q_a \hat V^{\mu}  \big( \hat Q_{\sa} - \hat P_{\sa} \big) \big( \pd_{\nu} \hat P_a \big) \right] \nonumber \\
    & \quad - \frac{e}2 \epsilon_{\mu\nu} \sum_a n_F'(E_a) m_a \oIm\mtr\left( \big( \bpd_{\nu}  \hat P_a \big) \big( \hat Q_{\sa} - \hat P_{\sa} \big) V^{\mu} \right) \label{EqS:ChiFSMCD}
\end{align}
Here, the sum over $a$ is restricted to bands intersecting the Fermi surface. Because each term in Eq.~\eqref{EqS:ChiFSMCD} is proportional to $n_F'(E_a)$, this contribution is supported entirely at the Fermi surface in the zero temperature limit.

Eq.~\eqref{EqS:ChiTRMCD} reproduces the results obtained directly from the projector formulation. The Green function calculation therefore provides an independent derivation of both the regular Fermi sea response and the additional Fermi surface contribution.

\section{Two Band model}
To obtain further physical insight into $\chi_{\rm MCD}$, it is useful to express the general result in the band basis. For simplicity, we consider a two band model and label the lower and upper bands by $a$ and $b$, respectively. We assume that the upper band $b$ remains empty, whereas the lower band $a$ may be occupied or empty depending on the chemical potential $\mu$. At a given momentum $\fvec k$, the Fermi sea contribution is therefore present only when $E_a(\fvec k)<\mu<E_b(\fvec k)$.

\subsection{Fermi Sea Contribution}
We first consider $\chi_{\rm TR}^{(1)}$. Using $\hat P = |u_a \rangle \langle u_a|$ and $\hat Q = | u_b \rangle \langle u_b |$, we obtain
\begin{align}
    \chi^{(1)}_{\rm TR} & =  -\frac{e^2}4 \epsilon_{\mu\nu} \epsilon_{\alpha\beta} \sum_{E_a(\fvec k) < \mu < E_b(\fvec k)} \oRe\mtr\left[ \calM^{-1}_{\mu\beta}  \big( \pd_{\nu} \hat P \big) \big( \pd_{\alpha} \hat Q \big) \right] \nonumber \\
    & =  -\frac{e^2}4 \epsilon_{\mu\nu} \epsilon_{\alpha\beta} \sum_{E_a(\fvec k) < \mu < E_b(\fvec k)} \oRe\left( \big( \calM^{-1}_{\beta\mu} \big)_{aa} \bra \wpd_{\nu} u_a | u_b \ket \bra \wpd_{\alpha} u_b | u_a \ket + \big( \calM^{-1}_{\beta\mu} \big)_{bb} \bra u_b | \wpd_{\nu} u_a \ket \bra u_a |  \wpd_{\alpha} u_b \ket   \right)  \nonumber \\
    & = \frac{e^2}4 \epsilon_{\mu\nu} \epsilon_{\alpha\beta} \sum_{E_a(\fvec k) < \mu < E_b(\fvec k)}  \oRe\left(  \big( \calM^{-1}_{\beta\mu} \big)_{aa} \calQ_a^{\nu\alpha} + \big( \calM^{-1}_{\beta\mu} \big)_{bb} \calQ^{\alpha\nu}_a  \right) \nonumber \\
    & = \frac{e^2}4 \epsilon_{\mu\nu} \epsilon_{\alpha\beta} \sum_{E_a(\fvec k) < \mu < E_b(\fvec k)}  \left(  \big( \calM^{-1}_{\beta\mu} \big)_{aa}  + \big( \calM^{-1}_{\beta\mu} \big)_{bb} \right) g^{(a)}_{\nu\alpha} 
\end{align}
where $\calQ_a^{\mu\nu}$ is the quantum geometric tensor of band $a$, and $g_{\mu\nu}^{(a)} = \oRe\calQ_a^{\mu\nu}$ is the corresponding quantum metric. Further, 
\begin{equation}  
    \big( \calM^{-1}_{\beta\mu} \big)_{aa} = \hbar^{-2} \pd^2_{\beta \mu} E_a - \sum_{b \neq a} \frac{V^{\beta}_{ab} V^{\mu}_{b a} + c.c.}{E_a - E_b} \ ,    \label{EqS:InvM}
\end{equation}
we conclude that for two band system
\begin{align}
     \chi^{(1)}_{\rm TR} & =  \frac{e^2}{4 \hbar^2} \epsilon_{\mu\nu} \epsilon_{\alpha\beta} \sum_{E_a(\fvec k) < \mu < E_b(\fvec k)} g^{(a)}_{\alpha \nu} \pd^2_{\beta \mu} (E_a + E_b) \label{EqS:ChiTR1TwoBand}
\end{align}
Thus, $\chi_{\rm TR}^{(1)}$ is controlled by the quantum metric together with the curvature of the total band energy $E_a + E_b$. In particular, it vanishes when $E_a(\fvec k) + E_b(\fvec k) = 0$ at every momentum.

We next consider $\chi_{\rm TR}^{(2)}$. For a two band system,
\begin{align}
	\epsilon_{\alpha\beta} (\pd_\alpha\hat P)\hat Q(\pd_\beta\hat P) & = -i \calF^{(a)} \hat P \ , \qquad	\epsilon_{\alpha\beta} (\pd_\alpha\hat Q)\hat P(\pd_\beta\hat Q)  = -i \calF^{(b)} \hat Q,
\end{align}
where $\calF^{(a)}$ and $\calF^{(b)}$ are the Berry curvatures of the two bands. Hence,
\begin{align}
	\chi_{\rm TR}^{(2)} = -\frac{e^2}{2\hbar} \epsilon_{\mu\nu} \sum_{E_a < \mu < E_b} \oRe\mtr\left[ -i \calF^{(a)} V^\mu\hat P(\pd_\nu\hat Q)	-i \calF^{(b)}	V^\mu \hat Q (\pd_\nu\hat P) \right] \ .
\end{align}
The two traces satisfy
\begin{align}
	\epsilon_{\mu\nu} \mtr\left[ V^\mu \hat P (\pd_\nu \hat Q) \right]	& =	\frac{2m^{(a)}}{i e} \ ,  \qquad
	\epsilon_{\mu\nu} \mtr\left[ V^\mu \hat Q (\pd_\nu \hat P) \right] = \frac{2m^{(b)}}{ie} \ ,
\end{align}
where
\begin{align}
	m^{(n)}	= \frac{e}{2 i \hbar} \epsilon_{\mu\nu}	\langle	\pd_\mu u_n	| (H - E_n)	| \pd_\nu u_n \rangle
\end{align}
is the orbital magnetic moment of band $n$. For a two band model, $m^{(a)} = m^{(b)}$, while the Berry curvatures satisfy $\calF^{(a)} + \calF^{(b)}=0$. The two contributions therefore cancel exactly, giving
\begin{align}
	\chi_{\rm TR}^{(2)} = 0	\  . \label{EqS:ChiTR2TwoBand}
\end{align}

Finally, we consider $\chi_{\rm TR}^{(3)}$. The relevant matrix elements of $\hat O_1$ and $\hat O_2$ are
\begin{align}
	\langle u_a | \hat O_1 | u_b \rangle & = \frac1{\hbar} \epsilon_{\alpha\beta} \left[	( \pd_\alpha E_a )	\langle\pd_\beta u_a | u_b \rangle	+ (\pd_\beta E_b) \langle u_a | \pd_\alpha u_b \rangle \right] = \frac1{\hbar} \epsilon_{\alpha\beta} \pd_\alpha( E_a + E_b ) \langle \pd_\beta u_a | u_b \rangle \ ,
\end{align}
and
\begin{align}
	\langle u_b | \hat O_2 | u_a \rangle & = \frac1{\hbar} \epsilon_{\alpha\beta} \left[ ( \pd_\alpha E_b ) \langle \pd_\beta u_b | u_a \rangle	- ( \pd_\beta E_a )	\langle u_b | \pd_\alpha u_a \rangle \right] = \frac1{\hbar} \epsilon_{\alpha\beta}	\pd_\alpha(E_b - E_a) \langle \pd_\beta u_b | u_a \rangle \ .
\end{align}
Using
\begin{align}
	\langle u_a | \calL^{-1}(\hat O_1) | u_b \rangle = \hbar \frac{	\langle u_a | \hat O_1 | u_b \rangle }{	E_a - E_b } \ ,
\end{align}
we obtain
\begin{align}
	\chi_{\rm TR}^{(3)}	& = \frac{e^2}{2 \hbar^2} \epsilon_{\mu\nu} \epsilon_{\alpha\beta} \sum_{E_a < \mu < E_b} \pd_\alpha(E_b - E_a)	\pd_\mu(E_a + E_b) \oRe	\frac{	\langle \pd_\nu u_a | u_b \rangle \langle \pd_\beta u_b | u_a \rangle }{ E_a - E_b } \ . \label{EqS:ChiTR3TwoBandIntermediate}
\end{align}
Since
\begin{align}
	\oRe\left[ \langle \pd_\nu u_a | u_b \rangle \langle \pd_\beta u_b | u_a \rangle \right] = -g_{\nu\beta}^{(a)} \ ,
\end{align}
it is convenient to introduce
\begin{align}
	\Sigma(\fvec k)	& =	E_a(\fvec k) + E_b(\fvec k) \ ,	& \Delta(\fvec k) & = E_b(\fvec k) - E_a(\fvec k) > 0 \ .
\end{align}
Equation~\eqref{EqS:ChiTR3TwoBandIntermediate} then becomes
\begin{align}
	\chi_{\rm TR}^{(3)}	= \frac{e^2}{2\hbar^2} \epsilon_{\mu\nu} \epsilon_{\alpha\beta}	\sum_{E_a < \mu < E_b} \frac{ (\pd_\alpha \Delta)(\pd_\mu \Sigma) }{ \Delta } g_{\nu\beta}^{(a)} \ .
	\label{EqS:ChiTR3TwoBand}
\end{align}
Eq.~\eqref{EqS:ChiTR1TwoBand}, \eqref{EqS:ChiTR2TwoBand}, and \eqref{EqS:ChiTR3TwoBand} show that, for a two band system, the entire Fermi sea contribution is controlled by the particle-hole asymmetric part of the dispersion through $\Sigma = E_a + E_b$. In particular, when the two bands satisfy $E_b(\fvec k) = - E_a(\fvec k)$ at every momentum, both $\chi_{\rm TR}^{(1)}$ and $\chi_{\rm TR}^{(3)}$ vanish, while $\chi_{\rm TR}^{(2)}$ vanishes identically. Therefore,
\begin{align}
	\chi_{\rm MCD}^{\rm TR} = 0
\end{align}
for a particle-hole symmetric two band model.

\subsection{Fermi surface contribution}
The first Fermi surface contribution then becomes
\begin{align}
    \chi_{\rm FS}^{(1)} & = \frac{e^2}4 \epsilon_{\mu\nu} \epsilon_{\alpha\beta} \sum_{\fvec k} n_F'(E_a) (\pd_\alpha E_a) \oRe\mtr\left[ \calM^{-1}_{\beta\mu} \hat Q (\pd_\nu\hat P_a) \right] \nonumber \\
    & = \frac{e^2}4 \epsilon_{\mu\nu} \epsilon_{\alpha\beta} \sum_{\fvec k} n_F'(E_a) (\pd_\alpha E_a) \oRe\left[ \left( \calM^{-1}_{\beta\mu} \right)_{ab} \langle u_b | \pd_\nu u_a \rangle \right] \ .
\end{align}
Note that this can be further simplified by noting that
\begin{align}
   &  \epsilon_{\mu\nu} \oRe\mtr\left[ \calM^{-1}_{\beta\mu} \hat Q (\pd_\nu\hat P) \right] = - \epsilon_{\mu\nu} \oRe\mtr\left[ \calM^{-1}_{\beta\mu} (\pd_\nu \hat Q) \hat P \right] = - \epsilon_{\mu\nu} \oRe\mtr\left[ \calM^{-1}_{\beta\mu}  \hat P (\pd_\nu \hat Q) \right] = \epsilon_{\mu\nu} \oRe\mtr\left[ \calM^{-1}_{\beta\mu}  \hat P (\pd_\nu \hat P) \right]  \nonumber \\
   & \Longrightarrow \quad \epsilon_{\mu\nu} \oRe\mtr\left[ \calM^{-1}_{\beta\mu} \hat Q (\pd_\nu\hat P) \right] = \half \epsilon_{\mu\nu} \oRe\mtr\left[ \calM^{-1}_{\beta\mu}  (\pd_\nu\hat P) \right] = \half \epsilon_{\mu\nu} \oRe\mtr\left\{  \pd_\nu \left[ \calM^{-1}_{\beta\mu} \hat P  \right] - \big( \pd_\nu \calM^{-1}_{\beta\mu} \big) \hat P\right\}
\end{align}
By definition of $\calM^{-1}$, the last term vanishes when contracting with the tensor $\epsilon_{\mu\nu}$. Using Eq.~\eqref{EqS:InvM}, we have
\[  \epsilon_{\mu\nu} \pd_\nu \big( \calM^{-1}_{\beta\mu} \big)_{aa}  = \epsilon_{\mu\nu} \pd_\nu \left( \frac{V^{\beta}_{ab} V^{\mu}_{ba} + c.c.}{E_b - E_a} \right) = \epsilon_{\mu\nu} \pd_\nu \left( \frac{E_b - E_a}{\hbar^2} 2 g^{(a)}_{\beta \mu} \right) \]

Consequently,
\begin{align}
    \chi_{\rm FS}^{(1)} & = \frac{e^2}{4\hbar^2} \epsilon_{\mu\nu} \epsilon_{\alpha\beta} \sum_{\fvec k} n_F'(E_a) (\pd_\alpha E_a)  \pd_\nu \Big( \Delta g^{(a)}_{\beta\mu} \Big)  \   .
\end{align}

The second contribution can be expressed in terms of the quantum metric of band $a$:
\begin{align}
    \chi_{\rm FS}^{(2)} & = \frac{e^2}{2\hbar} \epsilon_{\mu\nu} \epsilon_{\alpha\beta} \sum_{\fvec k} n_F'(E_a) ( \pd_\alpha E_a ) \oRe\mtr\left[ ( \pd_\beta\hat P_a ) \hat Q \hat V^\mu \hat Q ( \pd_\nu\hat P_a ) \right] \nonumber \\
    & = \frac{e^2}{2\hbar^2} \epsilon_{\mu\nu} \epsilon_{\alpha\beta} \sum_{\fvec k} n_F'(E_a) ( \pd_\alpha E_a ) ( \pd_\mu E_b ) g^{(a)}_{\beta\nu} \nonumber \\
    & = \frac{e^2}{2\hbar^2} \sum_{\fvec k} n_F'(E_a) \left[ ( \pd_\alpha E_a )( \pd_\alpha E_b ) g^{(a)}_{\beta\beta} - (\pd_\alpha E_a)(\pd_\beta E_b) g^{(a)}_{\alpha\beta} \right]    \  .
\end{align}

Finally, the third contribution is determined by the orbital magnetic moment:
\begin{align}
    \chi_{\rm FS}^{(3)} & = -\frac{e}2 \epsilon_{\mu\nu} \sum_{\fvec k} n_F'(E_a) m_a \oIm\mtr\left[ (\pd_\nu \hat P_a) \hat Q \hat V^\mu \right] = - \sum_{\fvec k} n_F'(E_a) m_a^2   \   .
\end{align}
This term often dominates $\chi_{\rm MCD}$ near a small band gap, where the orbital magnetic moment is strongly enhanced.

\section{Correction from the Zeeman coupling}
In the main text, we neglect the Zeeman coupling and focus on the orbital effect of the magnetic field, which enters through the vector potential $\fvec A$. Here, we discuss the additional correction arising from the Zeeman coupling. We neglect the direct magnetic dipole coupling between light and spin. Nevertheless, the Zeeman perturbation can modify both the band energies and, in the presence of spin-orbit coupling, the Bloch eigenstates. To linear order in the magnetic field, the MCD susceptibility can be decomposed as
\begin{align}
    \chi_{\rm MCD} = \chi_{\rm MCD}^{\rm gauge} + \chi_{\rm MCD}^{\rm Z} \   .
\end{align}
The orbital contribution $\chi_{\rm MCD}^{\rm gauge}$ is from the minimal coupling with gauge field, and is well discussed in the main text. We focus below on the Zeeman contribution $\chi_{\rm MCD}^{\rm Z}$.

We here derive the response to a more general parameter $\lambda$, representing a static and uniform perturbation:
\begin{align}
    \chi_{\rm MCD}^{\lambda} = \left. \pd_\lambda M_{\rm MCD} \right|_{\lambda=0} \ .
\end{align}
For later use, we define
\begin{align}
    H_\mu & \equiv \pd_\mu H \ ,  & H_\lambda & \equiv \pd_\lambda H \ ,  &  H_{\lambda\mu} &\equiv \pd_\lambda \pd_\mu H \ , & P_\lambda &\equiv \pd_\lambda P \   .
\end{align}
For Zeeman coupling with spins, $H_\lambda$ is given by the electron spin operator.

The first order perturbation theory gives
\begin{align}
    P_\lambda = \frac{1}{\hbar}\left[ P \calL^{-1}(H_\lambda) Q - Q \calL^{-1}(H_\lambda) P \right] \ ,  \qquad  Q_\lambda = - P_\lambda \ .
\end{align}

\subsection{Fermi sea contribution}
At zero temperature, the integrated MCD is
\begin{align}
    M_{\rm MCD}  = \frac{e}{2 \hbar} \epsilon_{\mu\nu} \sum_i \sum_{\fvec k\in S_i} \oIm\mtr\left[ H_\mu P_\nu Q \right] \  ,
\end{align}
where $S_i$ denotes a region of the Brillouin zone in which the rank of the occupied subspace is constant. Differentiating with respect to $\lambda$ gives
\begin{align}
    \chi_{\rm MCD}^{\lambda} = \frac{e}{2\hbar} \epsilon_{\mu\nu} \sum_i\sum_{\fvec k \in S_i} \oIm\mtr\left[ H_{\lambda\mu} P_\nu Q + H_\mu P_{\nu\lambda} Q + H_\mu P_\nu Q_\lambda \right] \ .
\end{align}
Using
\begin{align}
    H_\mu P_{\nu\lambda} Q & = \pd_\nu(H_\mu P_\lambda Q) - H_{\mu\nu} P_\lambda Q - H_\mu P_\lambda Q_\nu
\end{align}
and $\epsilon_{\mu\nu} H_{\mu\nu} = 0$, we obtain
\begin{align}
    \chi_{\rm MCD}^{\lambda} = \frac{e}{2\hbar} \epsilon_{\mu\nu} \sum_i \sum_{\fvec k \in S_i} \oIm\mtr\left[ \pd_\nu(H_\mu P_\lambda Q) + H_{\lambda\mu} P_\nu Q + H_\mu (P_\nu Q_\lambda - P_\lambda Q_\nu) \right].
\end{align}
The total derivative produces a boundary contribution. The remaining terms give the Fermi-sea response:
\begin{align}
    \chi_{\rm MCD}^{\lambda,{\rm TR}} = \frac{e}{2\hbar}\epsilon_{\mu\nu} \sum_{\fvec k} \oIm\mtr\left[ H_{\lambda\mu} P_\nu Q + H_\mu(P_\nu Q_\lambda - P_\lambda Q_\nu) \right] \ .
\end{align}
A similar approach was employed in Ref.~\cite{VenderbosArXiv2512} to study the response of the orbital magnetization $M_{\rm orb}$ to a general perturbation $\lambda$ in insulating systems. For metallic systems, the derivative of $\hat P$ and $\hat Q$ should be understood as the Fermi sea derivative.

\subsection{Fermi surface contribution}
For an insulator, the Brillouin zone contains no internal boundaries, and the total derivative obtained above integrates to zero. The situation is different in a metal. In this case, the Brillouin zone is divided into regions $S_i$ with different occupied subspaces. The boundaries between these regions are the Fermi surfaces. Therefore, the total derivative generates a boundary contribution that must be retained.

Stokes' theorem gives
\begin{align}
    \epsilon_{\mu\nu} \int_{S_i} \rmd^2k\, \pd_\nu F_\mu = - \oint_{\partial S_i} F_\mu\, \rmd k_\mu \  .
\end{align}
Accordingly, the total derivative term contributes
\begin{align}
    \chi_{\rm MCD}^{\lambda,{\rm FS}, (1)} = -\frac{e}{2\hbar} \sum_i \oint_{\pd S_i} \rmd k_\mu\, \oIm\mtr\left[ H_\mu P_\lambda Q \right] \ . \label{EqS:LambdaFSBoundary}
\end{align}
To evaluate this contribution, consider a band $a$ that crosses the Fermi level. On the occupied side of the Fermi surface,
\begin{align}
    P^{(-)} = P_{\sa} + P_a \ ,  \qquad  Q^{(-)} = Q_{\sa},
\end{align}
whereas on the empty side,
\begin{align}
    P^{(+)} = P_{\sa} \  , \qquad Q^{(+)} = Q_{\sa} + P_a  \  .
\end{align}
The discontinuity of the integrand across the Fermi surface is therefore
\begin{align}
    & \oIm\mtr\left[ H_\mu P_\lambda^{(-)} Q^{(-)} \right] - \oIm\mtr\left[ H_\mu P_\lambda^{(+)} Q^{(+)} \right] = \oIm\mtr\left[ H_\mu (\pd_\lambda P_a) \left( Q_{\sa} - P_{\sa} \right) \right]  \   . \label{EqS:LambdaFSJump}
\end{align}
Here, we have used the orthogonality of $P_a$, $P_{\sa}$, and $Q_{\sa}$, together with the derivatives of the corresponding orthogonality relations.

At zero temperature, a Fermi surface integral can be converted into a BZ integral using
\begin{align}
    n_F'(E_a) = - \delta(E_a - \mu) \  .
\end{align}
The boundary contribution in Eq.~\eqref{EqS:LambdaFSBoundary} then becomes
\begin{align}
    \chi_{\rm MCD}^{\lambda,{\rm FS}, (1)} = -\frac{e}{2 \hbar} \epsilon_{\mu\nu} \sum_{a, \fvec k} n_F'(E_a) (\pd_\nu E_a) \oIm\mtr\left[ H_\mu (\pd_\lambda P_a) \left( Q_{\sa} - P_{\sa} \right) \right]  \  . \label{EqS:LambdaFS1}
\end{align}
The factor $n_F'(E_a - \mu)$ restricts this term to the Fermi surface. Meanwhile, $\pd_\nu E_a$ is normal to the Fermi
surface and arises when the boundary integral is converted into a momentum space integral. This term describes the change of the Fermi surface MCD weight caused by the deformation of the Bloch eigenstates under the perturbation $\lambda$. 

A second Fermi surface contribution arises from the shift of the band energy. Under a variation of $\lambda$, the occupation of band $a$ changes according to
\begin{align}
    \pd_\lambda n_F(E_a) = n_F'(E_a) \left( \pd_\lambda E_a \right) \  .
\end{align}
The resulting contribution is
\begin{align}
    \chi_{\rm MCD}^{\lambda,{\rm FS}, (2)} = \frac{e}{2\hbar} \epsilon_{\mu\nu} \sum_{a,\fvec k} n_F'(E_a) (\pd_\lambda E_a) \oIm\mtr\left[ H_\mu (\pd_\nu P_a) \left( Q_{\sa} - P_{\sa} \right) \right]   \  . \label{EqS:LambdaFS2}
\end{align}
Unlike Eq.~\eqref{EqS:LambdaFS1}, this term does not require a change of the Bloch eigenstates. It originates from the
displacement of the Fermi surface caused by the energy shift $\pd_\lambda E_a$.

Combining Eqs.~\eqref{EqS:LambdaFS1} and \eqref{EqS:LambdaFS2}, we obtain
\begin{align}
    \chi_{\rm MCD}^{\lambda, \rm FS} = \frac{e}{2\hbar} \epsilon_{\mu\nu} \sum_{a,\fvec k} n_F'(E_a) \Big\{ (\pd_\lambda E_a) \oIm\mtr\left[ H_\mu (\pd_\nu P_a) (Q_{\sa} - P_{\sa}) \right] -  (\pd_\nu E_a) \oIm\mtr\left[ H_\mu (\pd_\lambda P_a) (Q_{\sa} - P_{\sa}) \right] \Big \}   \   .  \label{EqS:LambdaFSTotal}
\end{align}

The two terms in Eq.~\eqref{EqS:LambdaFSTotal} have distinct physical origins. The first describes the change in occupation caused by the shift of the band energy. The second describes the change in the MCD weight caused by the deformation of the Bloch wave functions. 

The operator $Q_{\sa} - P_{\sa}$ assigns opposite signs to intermediate states above and below the Fermi level. This
structure reflects the fact that a Fermi surface state can appear on either side of an optical transition: it can serve as an occupied initial state or as an empty final state. These two processes enter the variation of the MCD spectrum with opposite signs.

\section{Projector Formalism for $\chi_{\rm orb}$}
In this section, we present a projector based expression for the thermodynamic orbital magnetic susceptibility $\chi_{\rm orb}$. We focus on non-interacting systems, or equivalently on the band contribution $\chi_{\rm band}$ within the Hartree-Fock approximation. The orbital susceptibility can be written in two equivalent Green-function forms~\cite{StauberPRL2011, RaouxPRB2015, KangPRL2026},
\begin{align}
    \chi_{\rm orb} & = \frac{(e\hbar)^2}2 \oint_{\calC} \frac{\rmd z}{2\pi i}\, n_F(z) \left\{ \mtr\left[ (G_0 V^x G_0 V^y)^2 \right] + \half \mtr\left[ G_0 \calM^{-1}_{xy} \left( G_0 V^x G_0V^y + G_0 V^y G_0 V^x \right) \right] \right\} \ , \label{EqS:ChiOrb1} \\
    & = \frac{\hbar^2e^2}{12} \oint_{\calC} \frac{\rmd z}{2\pi i}\, n_F(z) \mtr\left[ G_0 \calM^{-1}_{xx} G_0 \calM^{-1}_{yy} - ( G_0 \calM^{-1}_{xy} )^2 - 4\left( (G_0 V^x)^2 (G_0 V^y)^2 - (G_0V^x G_0V^y)^2 \right) \right] \ . \label{EqS:ChiOrb2}
\end{align}
Here, $G_0 = (z - H_0)^{-1}$ is the zero field Green function, and the contour $\calC$ encloses the energy spectrum with the corresponding Fermi-function weight. We use
\begin{align}
    V^\mu = \frac1{\hbar} \pd_\mu H \ , \qquad \calM^{-1}_{\mu\nu} = \frac1{\hbar^2} \pd_{\mu\nu}^2 H_0 
\end{align}
We evaluate the contour integral at $T = 0$ and express the final result in terms of the occupied and empty state projectors $\hat P$ and $\hat Q$. Since the derivation is lengthy, we present only the final expressions here. A detailed derivation and further discussion will be given in future manuscript.

For notational convenience, we introduce
\begin{align}
    h_a^{\mu\nu}(\fvec k) & = 2\oRe\mtr\left[ (\pd_\mu \hat P_a) \hat Q_aV^\nu \right] = -2\oRe\mtr\left[ \hat P_a(\pd_\mu\hat Q_a)V^\nu \right] = 2\oRe\mtr\left[ \hat P_a (\pd_\mu \hat P_a) V^\nu \right] \ ,  \\
    m_a(\fvec k) & = \frac{e}{2i \hbar} \epsilon_{\mu\nu} \langle \pd_\mu u_a | (H_0 - E_a) | \pd_\nu u_a \rangle \  , \\
    g_a^{\mu\nu}(\fvec k) & = \oRe\mtr\left[ (\pd_\mu \hat P_a) \hat Q_a (\pd_\nu \hat P_a) \right] = \oRe\mtr\left[ \hat P_a (\pd_\mu \hat P_a) (\pd_\nu \hat P_a) \right] \ ,
\end{align}
where $a$ is the band index.

We focus on Eq.~\eqref{EqS:ChiOrb1}. The orbital susceptibility can be decomposed into Fermi-surface and interband contributions,
\begin{align}
    \chi_{\rm orb} = \chi_{\rm orb}^{\rm FS} + \chi_{\rm orb}^{\rm TR}  \   .
\end{align}
The Fermi surface contribution can be further separated as
\begin{align}
    \chi_{\rm orb}^{\rm FS} = \chi_{\rm orb}^{\rm LP} + \chi_{\rm orb}^{\rm GEOM} + \chi_{\rm orb}^{\rm AB}   \   .
\end{align}

The contribution $\chi_{\rm orb}^{\rm AB}$ is
\begin{align}
    \chi_{\rm orb}^{\rm AB} & = \frac{(e\hbar)^2}2 \sum_{a,\fvec k} n_F'(E_a) \left\{ - \frac14 \left[ (\pd_x E_a) (\pd_y h_a^{xy}) + (x \leftrightarrow y) \right]  + 2\left[ (\pd_x E_a) \mtr\left[ (\pd_y \hat P_a) \hat Q_a V^x \hat Q_a (\pd_y \hat P_a) \right] + (x \leftrightarrow y) \right] \right. \nonumber \\
    & \qquad \left. - \half \left[ (\pd_x E_a) \mtr\left[ \hat P_a \calM^{-1}_{xy} (\pd_y \hat Q_a) + {\rm c.c.} \right] + (x \leftrightarrow y) \right] \right\}   \   .
\end{align}

The Landau-Peierls contribution is
\begin{align}
    \chi_{\rm orb}^{\rm LP} = \frac{e^2}{12 \hbar^2} \sum_{a, \fvec k} n_F'(E_a) \left[ (\pd_x^2 E_a) (\pd_y^2 E_a) - (\pd_{xy}^2 E_a)^2 \right] \ ,
\end{align}
while the geometric contribution is
\begin{align}
    \chi_{\rm orb}^{\rm GEOM} = -  \sum_{a,\fvec k} n_F'(E_a) \left[ \frac14 m_a^2 + 2 \frac{e^2}{\hbar^2} (\pd_x E_a) (\pd_y E_a) g_a^{xy} \right] \ .
\end{align}

We next consider the contribution associated with interband transitions,
\begin{align}
    \chi_{\rm orb}^{\rm TR} = \chi_{\rm orb}^{\rm TR,(1)} + \chi_{\rm orb}^{\rm TR,(2)} \ .
\end{align}
The first term originates from the velocity operators and is given by
\begin{align}
    \chi_{\rm orb}^{\rm TR,(1)} & = 2 \hbar^2 e^2 \sum_{\fvec k} \oRe\mtr\left\{ \hat P \calL^{-1}(\hat W_1) \hat Q \hat W_2 + \hat Q V^x (\pd_y \hat P) (\pd_x \hat Q) (\pd_y \hat P) \right\} \ ,
\end{align}
where
\begin{align}
    \hat W_1 & = (\pd_x \hat Q) V^y + V^x (\pd_y \hat P) \ , & \hat W_2 & = V^x (\pd_y \hat Q) + (\pd_x \hat P) V^y \ .
\end{align}

The second contribution originates from the inverse mass operator $\calM^{-1}_{xy}$ and is given by
\begin{align}
    \chi_{\rm orb}^{\rm TR,(2)} & = \frac{\hbar^2e^2}2 \sum_{\fvec k} \oRe\mtr\left\{ (\pd_y \hat P) \calM^{-1}_{xy} (\pd_x \hat P) \hat Q - (\pd_y \hat Q) \calM^{-1}_{xy} (\pd_x \hat Q) \hat P \right. \nonumber \\
    & \qquad \left. - \hat P \calL^{-1}(\calM^{-1}_{xy}) \hat Q \left[ (\pd_x \hat P) V^y + (\pd_y \hat P) V^x \right] + \hat Q \calL^{-1}( \calM^{-1}_{xy} ) \hat P \left[ (\pd_x \hat Q) V^y + (\pd_y \hat Q) V^x \right] \right\} \ .
\end{align}


\section{Gauge Choice in Tight Binding Models}
An important subtlety in applying the above formulas to tight-binding models is the choice of Bloch basis, or equivalently, the gauge convention for the Bloch Hamiltonian~\cite{OkumaJPSJ2024}~\footnote{Part of the discussion in this section overlaps with Ref.~\cite{OkumaJPSJ2024}.}. In real space, the velocity operator is unambiguously defined as
\begin{equation}
    \hat V^\mu=\frac{1}{i\hbar}[\hat r_\mu,\hat H].
\end{equation}
Its commonly used momentum space representation,
\begin{equation}
    V^\mu(\fvec k)=\frac{1}{\hbar}\partial_\mu \calH(\fvec k)  \  , \qquad  \calH(\fvec k) = e^{-i\fvec k \cdot \hat{\fvec r}} \hat H e^{i\fvec k \cdot \hat{\fvec r}} \  ,  \label{eq:velocity_nonperiodic_gauge}
\end{equation}
is valid when the Bloch Hamiltonian is defined using the full position operator $\hat{\fvec r}$. Here and below, $\partial_\mu \equiv \pd/\pd k_\mu$.

For a tight binding basis, the position of orbital $a$ can be written as $\fvec r = \fvec R + \fvec \tau_a$, where $\fvec R$ labels the unit cell and $\fvec \tau_a$ denotes the position of the orbital within the unit cell. The Hamiltonian $\calH(\fvec k)$ defined in Eq.~\eqref{eq:velocity_nonperiodic_gauge} is generally not periodic in the Brillouin zone. In practical tight binding calculations, it is often more convenient to use the periodic gauge Hamiltonian
\begin{equation}
    \overline{\calH}(\fvec k) = e^{-i \fvec k \cdot \hat{\fvec R}} \hat H e^{i\fvec k \cdot \hat{\fvec R}} \ ,  \qquad \overline{\calH}(\fvec k + \fvec G) = \overline{\calH}(\fvec k)   \  .
\end{equation}
The two representations are related by
\begin{equation}
    \calH = \calD \, \overline{\calH} \, \calD^\dagger \ ,    \qquad  \calD_{ab}(\fvec k) = e^{-i \fvec k \cdot \fvec \tau_a}\delta_{ab} \ . \label{eq:gauge_transformation}
\end{equation}

To express momentum derivatives in the periodic gauge, we introduce the diagonal matrices
\begin{equation}
    (\sfT^\mu)_{ab} = \tau_a^\mu \delta_{ab} \ ,  \qquad \pd_\mu \calD = -i \calD \sfT^\mu \ ,
\end{equation}
and define the covariant derivative
\begin{equation}
    \nabla_\mu X\equiv \pd_\mu X - i [ \sfT^\mu , X ]  \ .   \label{eq:covariant_derivative_tb}
\end{equation}
Equation~\eqref{eq:gauge_transformation} then immediately gives
\begin{equation}
    \pd_\mu (\calD X \calD^\dagger)  = \calD (\nabla_\mu X) \calD^\dagger \ .
\end{equation}
Therefore, the physical velocity operator represented in the periodic gauge is
\begin{equation}
    V^\mu_{\rm per} =\frac{1}{\hbar} \nabla_\mu \overline{\calH} = \overline V^\mu - \frac{i}{\hbar} [ \sfT^\mu , \overline{\calH} ] \ , \qquad
    \overline V^\mu \equiv \frac1{\hbar} \pd_\mu \overline{\calH} \  .  \label{eq:velocity_periodic_gauge}
\end{equation}
Thus, simply replacing the velocity operator by $\hbar^{-1} \pd_\mu \overline{\calH}$ in the periodic gauge generally misses the contribution arising from the intra unit cell orbital positions.

The same consideration applies to the inverse-mass operator. In real space,
\begin{equation}
    \hat \calM^{-1}_{\mu\nu} = -\frac1{\hbar^2} [\hat r_\mu, [\hat r_\nu, \hat H]] = \frac1{\hbar^2} \pd^2_{\mu\nu} \calH \ .
\end{equation}
Its representation in the periodic gauge is therefore
\begin{align}
    \calM^{-1}_{\rm per, \mu\nu} = \frac{1}{\hbar^2} \nabla_\mu \nabla_\nu \overline{\calH} = \overline{\calM}^{-1}_{\mu\nu} - \frac{i}{\hbar} \left( [ \sfT^\mu, \overline V^\nu] + [\sfT^\nu, \overline V^\mu] \right) - \frac1{\hbar^2} [\sfT^\mu \ ,  [\sfT^\nu, \overline{\calH} ] ] \ , \label{eq:inverse_mass_periodic_gauge}
\end{align}
where
\begin{equation}
    \overline{\calM}^{-1}_{\mu\nu} \equiv \frac1{\hbar^2} \pd_\mu \pd_\nu \overline{\calH} \  .
\end{equation}
Hence, when the periodic gauge is employed in a tight binding calculation, all momentum derivatives entering the magnetic response formulas should be understood as the covariant derivatives defined in Eq.~\eqref{eq:covariant_derivative_tb}.

In the previous sections and in the main text, we have repeatedly used the identity
\begin{align}
    \sum_{\fvec k} \pd_\mu \mtr\!\left[ \hat O(\fvec k) \right] = 0   \ .
\end{align}
One may wonder whether this identity remains valid when $\hat O(\fvec k)$ is not explicitly periodic over the Brillouin zone. To see this more generally, it is useful to express the left hand side in real space. Using the correspondence between a momentum derivative and the commutator with the position operator, we obtain
\begin{align}
    \sum_{\fvec k} \pd_\mu \mtr\!\left[ \hat O(\fvec k) \right] = \int \rmd \fvec r\, \frac{1}{i} \mtr\!\left( [\hat r_\mu, \hat O]_{\fvec r\fvec r} \right)  \ .
\end{align}
The integral on the right hand side is simply the trace of a commutator over the full real space Hilbert space and therefore vanishes. Hence,
\begin{align}
    \sum_{\fvec k} \pd_\mu \mtr\!\left[ \hat O(\fvec k) \right] = 0
\end{align}
does not rely on the explicit periodicity of $\hat O(\fvec k)$ in a particular Bloch basis, but instead follows from the basis independent real space representation of the corresponding operator, provided that the trace and the thermodynamic limit are well defined.

\section{Toy Model}
\subsection{Formula Verification: Two-band model}
In this subsection, we start from a toy model, analytically solve the magnetic spectrum of a toy model, obtain a closed-form expression for $M_{\rm MCD}$ as a function of $B$ in the low temperature, and compare the slope of $M_{\rm MCD}(B)$ with our expression for zero -temperature $\chi_{\rm MCD}$.

\subsubsection{Magnetic Spectrum}
We consider the two-dimensional Hamiltonian
\begin{equation}
H(\fvec k) = \begin{pmatrix}
    \alpha k^2 + \Delta  & \beta k_-   \\  \beta k_+ & \alpha k^2 - \Delta
\end{pmatrix}  \qquad \mbox{with} \quad  \alpha > 0,   \label{Eqn:ToyModel}
\end{equation}
where $k_{\pm} = k_x \pm i k_y$. Note that the energy is bounded below. For simplicity, we set the chemical potential $\mu=0$. For notational convenience, we also introduce  a series of dimensionless variables as
\[ \alpha' = \alpha  \frac{\Delta}{\beta^2} \quad \mbox{and} \quad  k' = k \frac{\beta}{\Delta}  \]
such that the Hamiltonian can be rewritten as
\begin{align}
    \frac{H}{\Delta} = \begin{pmatrix}
        \alpha' k'^2 + 1 &  k_-' \\  k_+' & \alpha' k'^2 - 1
    \end{pmatrix}
\end{align}
such that $E_{\pm} / \Delta = \alpha' k'^2 \pm \sqrt{1 +  k'^2}$.

In the presence of the magnetic field, we also introduce the corresponding dimensionless magnetic field
\begin{align}
    B' = B \frac{e \beta^2}{\hbar \Delta^2} 
\end{align}

The Hamiltonian becomes 
\begin{align}
    H(\fvec k) = \begin{pmatrix}
    \alpha (\pi / \hbar )^2 + \Delta  & \beta \pi_-/ \hbar   \\  \beta \pi_+/\hbar  & \alpha (\pi / \hbar)^2 - \Delta
\end{pmatrix} 
\end{align}
where $\fvec \pi = \fvec p + e \fvec A$. Since $[\pi_x, \pi_y] = -i \hbar e B = -i B' ( \hbar \Delta / \beta )^2$,  
\begin{align}
    a = \frac{\beta/ \hbar \Delta}{\sqrt{2 B'}} \big( \pi_x - i \pi_y \big) \quad a^{\dagger} = \frac{\beta/ \hbar \Delta}{\sqrt{2 B'}}  \big( \pi_x + i \pi_y \big) \  .
\end{align}
where $a$ and $a^{\dagger}$ is introduced as the ladder operators. In addition,
\[  \pi^2 = \half \big( \pi_+ \pi_- + \pi_- \pi_+ \big) = 2 B ' \left( \frac{\hbar \Delta}{\beta} \right)^2 \left( a^{\dagger} a + \half \right) \]

Now, we rewrite the Hamiltonian $H$ as 
\begin{align}
    \frac{H}{\Delta} = \begin{pmatrix}
        2 \alpha' B' \big(  a^{\dagger} a + \half) + 1 & \sqrt{2 B'}  a  \\  \sqrt{2 B'}  a^{\dagger} &  2 \alpha' B' \big(  a^{\dagger} a + \half) - 1
    \end{pmatrix}
\end{align}
We try the eigenstate as  $\psi = \big( x_n | n - 1 \ket , \ y_n | n \ket \big)^T$, thus the energy and the associated eigenstate is
\begin{align}
    E_n^{\pm}/\Delta & = 2\alpha' B' n \pm R_n \quad \mbox{with} \quad R_n = \sqrt{ \big( 1 - \alpha'B' \big)^2 + 2 B' n} \qquad E_0^-/\Delta = \alpha' B' - 1
\end{align}
For notational convenience, we define
\begin{align}
    \cos\theta_n & = \frac{1 - \alpha' B'}{R_n} \quad \ , \quad \sin\theta_n = \frac{\sqrt{2 n B'}}{R_n} \qquad \psi_0^- = \big( 0 ,  |0 \ket \big)^T \\
    \psi_{n \geq 1}^+ & = \left( \cos\frac{\theta_n}2 | n - 1 \ket \ , \  \sin\frac{\theta_n}2 | n \ket \right)^T \qquad \psi_{n \geq 1}^-  = \left( -\sin\frac{\theta_n}2 | n - 1 \ket \ , \  \cos\frac{\theta_n}2 | n \ket \right)^T  
\end{align}

\subsubsection{Velocity Operator}
In the presence of the magnetic field, we have 
\begin{align}
    \hat v_x & = \frac1{i \hbar} [\hat x, \hat H] =  \hbar^{-1} \begin{pmatrix}
        2 \alpha \hat \pi_x/\hbar & \beta \\ \beta & 2\alpha \hat \pi_x/\hbar
    \end{pmatrix} = \frac{\beta}{\hbar} \begin{pmatrix}
        \alpha' \sqrt{2 B'} \left( a + a^{\dagger} \right) & 1 \\ 1 & \alpha' \sqrt{2 B'} \left( a + a^{\dagger} \right)
    \end{pmatrix}  \\
    \hat v_y & = \frac1{i \hbar} [\hat y, \hat H] = \hbar^{-1} \begin{pmatrix}
        2 \alpha \hat \pi_y /\hbar & - i \beta \\ i \beta & 2\alpha \hat \pi_y /\hbar
    \end{pmatrix} = i \frac{\beta}{\hbar} \begin{pmatrix}
        \alpha' \sqrt{2 B'}  \left( a - a^{\dagger} \right) & - 1  \\ 1  & \alpha' \sqrt{2 B'}  \left( a - a^{\dagger} \right)
    \end{pmatrix} 
\end{align}

Now, we consider the matrix elements of the velocity operator. For simplicity, we consider only the inter-band one. For $\hat v_x$,
\begin{align}
    \bra \psi_m^+ | \hat v_x | \psi_n^- \ket & = \frac{\beta}{\hbar} \begin{pmatrix} \cos\frac{\theta_m}2 \bra m - 1 | \\ \sin\frac{\theta_m}2 \bra m | \end{pmatrix} 
    \begin{pmatrix}  \alpha' \sqrt{2 B'} \left( a + a^{\dagger} \right) & 1 \\ 1 & \alpha' \sqrt{2 B'} \left( a + a^{\dagger} \right)  \end{pmatrix} 
    \begin{pmatrix} - \sin\frac{\theta_n}2 | n - 1 \ket \\ \cos\frac{\theta_n}2  | n \ket  \end{pmatrix} \nonumber \\
    & = \frac{\beta}\hbar \left\{ \begin{array}{ll}
        \alpha' \sqrt{2 B'}  \left( - \sqrt{n} \cos\frac{\theta_{n + 1}}2 \sin\frac{\theta_n}2 + \sqrt{n + 1} \sin\frac{\theta_{n + 1}}2 \cos\frac{\theta_n}2 \right) + \cos\frac{\theta_{n + 1}}2 \cos\frac{\theta_n}2  \ , &  m =  n + 1\\
         \alpha' \sqrt{2 B'} \left( - \sqrt{n - 1} \cos\frac{\theta_{n - 1}}2 \sin\frac{\theta_n}2 + \sqrt{n} \sin\frac{\theta_{n - 1}}2 \cos\frac{\theta_n}2 \right) - \sin\frac{\theta_{n - 1}}2 \sin\frac{\theta_n}2  \ ,  &   m =  n - 1
    \end{array}    \right.
\end{align}
and also
\begin{align}
    \bra \psi_1^+ | \hat v_x | \psi_0^- \ket & = \frac{\beta}{\hbar} \left( \alpha' \sqrt{2 B'}  \sin\frac{\theta_1}2 + \cos\frac{\theta_1}2 \right) \  .
\end{align}

Similarly, for $\hat v_y$, we have
\begin{align}
    i \bra \psi_m^+ | \hat v_y | \psi_n^- \ket & = \frac{\beta}{\hbar} \left\{ \begin{array}{ll}
        \alpha' \sqrt{2 B'} \left( - \sqrt{n} \cos\frac{\theta_{n + 1}}2 \sin\frac{\theta_n}2 + \sqrt{n + 1} \sin\frac{\theta_{n + 1}}2 \cos\frac{\theta_n}2 \right) + \cos\frac{\theta_{n + 1}}2 \cos\frac{\theta_n}2  \ , &  m =  n + 1\\
         \alpha' \sqrt{2 B'} \left( \sqrt{n - 1} \cos\frac{\theta_{n - 1}}2 \sin\frac{\theta_n}2 - \sqrt{n} \sin\frac{\theta_{n - 1}}2 \cos\frac{\theta_n}2 \right) + \sin\frac{\theta_{n - 1}}2 \sin\frac{\theta_n}2  \ ,  &   m =  n - 1
    \end{array}    \right.
\end{align}
and also
\begin{align}
    i\bra \psi_1^+ | \hat v_y | \psi_0^- \ket & = \frac{\beta}{\hbar} \left( \alpha' \sqrt{2 B'} \sin\frac{\theta_1}2 +  \cos\frac{\theta_1}2 \right) \  .
\end{align}

It is obvious that only $\bra \psi_{n - 1}^+ | \big( \hat v_x - i \hat v_y \big) | \psi_n^- \ket$ and $\bra \psi_{n+1}^+ | \big( \hat v_x + i \hat v_y \big) | \psi_n^- \ket$ are non-zero, consistent with the fact that $\hat v_+ = \hat v_x + i \hat v_y$  and $\hat v_- = \hat v_x - i \hat v_y$ are raising and lowering operators respectively.

\subsubsection{Magnetization}
For notational convenience, we introduce the dimensionless variables
\begin{align}
     A_{n \geq 0} & = \alpha' \sqrt{2 B'}  \left( - \sqrt{n} \cos\frac{\theta_{n + 1}}2 \sin\frac{\theta_n}2 + \sqrt{n + 1} \sin\frac{\theta_{n + 1}}2 \cos\frac{\theta_n}2 \right) +  \cos\frac{\theta_{n + 1}}2 \cos\frac{\theta_n}2  \\
     B_{n \geq 1} & = \alpha' \sqrt{2 B'}  \left( - \sqrt{n} \cos\frac{\theta_n}2 \sin\frac{\theta_{n+1}}2 + \sqrt{n + 1} \sin\frac{\theta_n}2 \cos\frac{\theta_{n+1}}2 \right) -  \sin\frac{\theta_n}2 \sin\frac{\theta_{n+1}}2  \\
     \Longrightarrow \quad &  \bra \psi^+_{n + 1} | \hat v_x | \psi^-_n \ket = i \bra \psi^+_{n + 1} | \hat v_y | \psi^-_n \ket =  A_n \beta/\hbar \quad \bra \psi^+_{n + 1} | \hat v^+ | \psi^-_n \ket = 2 A_n \beta/\hbar \\
     & \bra \psi^+_n | \hat v_x | \psi^-_{n+1} \ket = - i \bra \psi^+_n | \hat v_y | \psi^-_{n+1} \ket = B_n \beta/\hbar  \quad \bra \psi^+_n | \hat v^- | \psi^-_{n+1} \ket = 2 B_n \beta / \hbar
\end{align}
where $\theta_0 = 0$. We also notice that when $n = 0$ and $\theta_0 = 0$, $B_0 = 0$. Thus, the above formula for $B_n$ holds also at $n = 0$. 

It is interesting to notice that
\begin{align}
    A_n^2 - B_n^2 & = 2(\alpha')^2 B' \left[ n \left(  \cos^2\frac{\theta_{n + 1}}2 \sin^2\frac{\theta_n}2 - \cos^2\frac{\theta_n}2 \sin^2\frac{\theta_{n+1}}2 \right) + (n + 1) \left( \sin^2\frac{\theta_{n + 1}}2 \cos^2\frac{\theta_n}2 - \sin^2\frac{\theta_n}2 \cos^2\frac{\theta_{n+1}}2 \right)  \right] \nonumber \\
    & \quad + \left( \cos^2\frac{\theta_{n + 1}}2 \cos^2\frac{\theta_n}2 - \sin^2\frac{\theta_n}2 \sin^2\frac{\theta_{n+1}}2  \right) \nonumber \\
    & \quad + 2\alpha' \sqrt{2B'} \left[ \sqrt{n}  \left( - \cos^2\frac{\theta_{n + 1}}2 \cos\frac{\theta_n}2 \sin\frac{\theta_n}2 - \sin^2\frac{\theta_{n+1}}2 \cos\frac{\theta_n}2  \sin\frac{\theta_n}2  \right) \right. \nonumber \\
    & \qquad \left. + \sqrt{n + 1} \left(  \cos^2\frac{\theta_n}2 \cos\frac{\theta_{n + 1}}2  \sin\frac{\theta_{n+1}}2 + \sin^2\frac{\theta_n}2 \sin\frac{\theta_{n+1}}2 \cos\frac{\theta_{n+1}}2 \right) \right] \nonumber \\
    & = 2(\alpha')^2 B' \left[ \frac{n}2 \left( \cos\theta_{n + 1} - \cos\theta_n \right) + \frac{n + 1}2 \left( - \cos\theta_{n + 1} + \cos\theta_n  \right)   \right] + \half \left( \cos\theta_n + \cos\theta_{n + 1} \right) \nonumber \\
    & \qquad + \alpha' \sqrt{2B'}\left( -\sqrt{n} \sin\theta_n + \sqrt{n + 1} \sin\theta_{n + 1} \right) \nonumber \\
    & = (\alpha')^2 B' \left( - \cos\theta_{n + 1} + \cos\theta_n \right)  + \half \left( \cos\theta_n + \cos\theta_{n + 1} \right) + \alpha' \sqrt{2B'}\left( -\sqrt{n} \sin\theta_n + \sqrt{n + 1} \sin\theta_{n + 1} \right)
\end{align}

Now, we are ready to calculate $M_{\rm MCD}$ under finite temperature with $B \neq 0$.  The degeneracy of each Landau level is the number of quantum flux through the system, i.e.~$S B/\Phi_0 = S e B/h = S \frac{B' \Delta^2}{2\pi \beta^2}$, or equivalently $\frac{B' \Delta^2}{2\pi \beta^2}$ per unit area. Assuming that $\mu = 0$, we start from the formula
\begin{align}
    \frac{M_{\rm MCD}}S & = \hbar e \frac{B' \Delta^2}{2\pi \beta^2}  \sum_{m, n} \Theta(E_m - E_n) \frac{n_F(E_n^-) - n_F(E_m^+)}{E_m^+ - E_n^-} \Im\left( v^x_{n-, m+} v^y_{m+, n-} \right) \nonumber \\
    & = \frac{e \Delta}{2\pi \hbar} B' \left( - \sum_{n = 0}^{\infty} \frac{n_F(E_n^-)}{2\alpha' B' + R_n + R_{n+1}} \big( A_n \big)^2 + \sum_{n = 0}^{\infty} \frac{n_F(E^-_{n+1})}{-2\alpha' B' + R_n + R_{n+1}} \big( B_n \big)^2    \right)   \label{Eqn:MMCDLL}
\end{align}


\subsubsection{Zero field limit}
The MCD magnetization can be expressed as
\begin{align}
    M_{\rm MCD}(B) & = S \frac{e \Delta}{2\pi \hbar} B'  \sum_{n = 0}^{\infty} f(n) 
\end{align}
where
\begin{align}
    f(n) & = - \frac{n_F(E_n^-)}{2\alpha' B' + R_n + R_{n+1}} \big( A_n \big)^2 + \frac{n_F(E_{n+1}^-)}{-2\alpha' B' + R_n + R_{n+1}} \big( B_n \big)^2
\end{align}
with $E_n^- = \Delta (2\alpha' B' n - R_n)$ and $R_n = \sqrt{(1- \alpha' B')^2 + 2 B' n}$, and thus becomes infinite as $n \rightarrow \infty$. This implies that $f(n \rightarrow \infty) = 0$.

\subsection{Absence of the Sum Rule}
In this subsection, we consider the two band Qi-Wu-Zhang (QWZ) Hamiltonian $\hat H(\fvec k) = d_0(\fvec k) \sigma_0 + \fvec d(\fvec k) \cdot \fvec \sigma$, where $d_0(\fvec k) = \alpha \big( \cos k_x + \beta \cos k_y)$, and $\fvec d(\fvec k) = (\sin k_x, \sin k_y, m + \cos k_x + \cos k_y)$ is a three-component vector. Note that in this subsection, for notational convenience, we set the lattice spacing and the coefficient of the Pauli matrix terms to unity and use $e = \hbar = 1$. The energies and the corresponding spectral projectors are
\begin{align}
    E_\pm(\fvec k) & = d_0(\fvec k) \pm \left| \fvec d(\fvec k) \right| & \hat P_\pm(\fvec k) & = \frac{\sigma_0 \pm \hat{\fvec d} \cdot \fvec \sigma}2   \  .
\end{align}
We further set $m = -1$ such that
\[ |\fvec d|^2 = 1 + 2(1 - \cos k_x) (1 - \cos k_y) \geq 1    \]
so that the gap is always at every momentum. Obviously, the spectrum is bounded above and below. We choose $\alpha = 0.6$ and $\beta = 0.4$. The corresponding scalar term $d_0$ leaves the projectors, Berry curvature, and quantum metric unchanged, but changes the band energies and their derivatives and thus breaks the particle-hole symmetry.

We calculate the integral
\begin{equation}
    \calI =\int_{-\infty}^{\infty}\rmd \mu\, \chi_{\rm MCD}(\mu) 
\end{equation}
where $\chi_{\rm MCD}$ is given by Eq.~\eqref{EqS:ChiMCD}. Here $\rm TR$ denotes the Fermi sea contribution and $\rm FS$ the Fermi surface contribution. We use the Brillouin zone convention $\int_{\rm BZ}\rmd^2 k/(2\pi)^2$.

First, consider the Fermi sea contribution. At zero temperature, the occupied projector at fixed momentum is
\begin{equation}
    \hat P(\fvec k,\mu)= \begin{cases} 0, & \mu < E_- \ , \\ \hat P_-, & E_- < \mu < E_+  \ ,    \\ \hat I , & \mu > E_+ \  \end{cases} \qquad \hat Q = \hat I - \hat P.
\end{equation}
The integrand of the stated projector formula vanishes when both bands are empty or both are occupied. When $\mu$ is inside the gap, $\hat P = \hat P_-$ and $\hat Q = \hat P_+$. Its value is independent of $\mu$. Denoting this value by $\chi_{\rm TR}(\fvec k)$, we have
\begin{equation}
    \chi_{\rm TR}(\fvec k,\mu) = \Theta(\mu - E_-) \Theta(E_+ - \mu) \chi_{\rm TR}(\fvec k) \quad \Longrightarrow \quad \int_{-\infty}^{\infty}\rmd \mu\, \chi_{\rm TR}(\fvec k, \mu) = ( E_+ - E_- ) \chi_{\rm TR}(\fvec k) = 2| \fvec d| \chi_{\rm TR}(\fvec k) \ .
\end{equation}
Thus
\begin{equation}
    \calI_{\rm TR} = \int_{\rm BZ} \frac{\rmd^2 k}{(2\pi)^2} \, 2 |\fvec d(\fvec k)| \chi_{\rm TR}(\fvec k). \label{EqS:ITR}
\end{equation}
Although $E_+ - E_- = 2 |\fvec d|$ is independent of $d_0$, the TR response need not be independent of $d_0$ since it depends on the velocity operator $V_{\mu}$.

Now, we consider the Fermi-surface contribution. Note that $n_F(E;\mu,T) = \left[ \exp((E-\mu)/(k_B T)) + 1 \right]^{-1}$, then
\begin{equation}
    n_F'(E_a; \mu, T) \xrightarrow{T \rightarrow 0} -\delta(E_a - \mu) \ , \qquad \int_{-\infty}^{\infty} \rmd \mu\ n_F'(E_a; \mu, T) = -1 \ .
\end{equation}
In particular, if the FS integrand is written as
\begin{equation}
    \chi_{\rm FS}(\fvec k, \mu) = \sum_{a=\pm} n_F'(E_a; \mu) \calF_a(\fvec k, \mu) \quad \Longrightarrow \quad 
    \calI_{\rm FS} = \int \rmd \mu\, \int_{\rm BZ} \frac{\rmd^2 k}{(2\pi)^2}\, \chi_{\rm FS}(\fvec k, \mu) =  -\int_{\rm BZ} \frac{\rmd^2 k}{(2\pi)^2} \sum_{a = \pm} \calF_a(\fvec k, E_a) \   . \label{EqS:IFS}
\end{equation}
Here $\calF_a$ represents the coefficient supplied by the formula for $\chi_{\rm FS}$, evaluated on the corresponding Fermi surface. If $\calF_a$ is independent of $\mu$, the integration simply replaces each $n_F'(E_a)$ by $-1$. Eq.~\eqref{EqS:ITR} and \eqref{EqS:IFS} establish the chemical potential integration identities, rather than a closed form expression of the remaining BZ integrals.

The obtained numerical results for the integral are listed below:
\begin{align}
    \calI_{\rm TR} & = 0.0179 \ ,  & \calI_{\rm FS} & = 0.0367 \ , \quad \Longrightarrow \quad \calI = \calI_{\rm TR} + \calI_{\rm FS} = 0.0546 \neq 0 \ . 
\end{align}

\section{Continuum model and numerical evaluation for twisted bilayer graphene}
\label{SecS:TBG}
\subsection{Hamiltonian and parameters}
We specify the continuum model used for Fig.~\ref{Fig:TBG_responses}. The moir\'e lattice constant is $L_M=a_0/[2\sin(\theta/2)]$, with $a_0=0.246$~nm. We choose
\begin{align}
\bG_1&=\frac{1}{L_M}\left(-\frac{2\pi}{\sqrt{3}},-2\pi\right),&
\bG_2&=\frac{1}{L_M}\left(\frac{4\pi}{\sqrt{3}},0\right),&
\bK_1&=\frac{\bG_1+2\bG_2}{3},&
\bK_2&=\frac{-\bG_1+\bG_2}{3}.
\end{align}
In a layer-dependent gauge in which the tunneling is moir\'e periodic, the valley-$\tau$ Hamiltonian is
\begin{align}
H_{\rm BM}^{\tau}&=
\begin{pmatrix}
h_1^\tau(-i\boldsymbol{\nabla}) & U_\tau(\br)\\
U_\tau^\dagger(\br) & h_2^\tau(-i\boldsymbol{\nabla})
\end{pmatrix},\label{EqS:TBGHamiltonian}\\
h_\ell^\tau(\bk)&=-\hbar v_F(\bk-\bK_\ell^\tau)\cdot(\tau\sigma_x,\sigma_y)+\Delta_\ell\sigma_z,
\qquad \bK_\ell^\tau=-\tau\bK_\ell,\\
U_\tau(\br)&=T_0^\tau+T_1^\tau e^{-i\tau\bG_1\cdot\br}
+T_2^\tau e^{-i\tau(\bG_1+\bG_2)\cdot\br},\\
T_j^\tau&=\begin{pmatrix}u_0&u_1e^{-i\tau\phi_j}\\u_1e^{i\tau\phi_j}&u_0\end{pmatrix},
\qquad(\phi_0,\phi_1,\phi_2)=(0,2\pi/3,-2\pi/3).
\end{align}
Here $\sigma_i$ act on the sublattice degrees of freedom, $\ell=1,2$ labels the layers, and $\tau=\pm1$ labels the valleys. We neglect the layer-dependent rotations of the Dirac matrices and strain. The parameters are $\theta=1.4^\circ$, $u_0=80~{\rm meV}$, $u_1=110~{\rm meV}$, 
$\Delta_1=\Delta_2=10~{\rm meV}$,
$\frac{\hbar v_F}{a_0}=2.365~{\rm eV}$.
The calculation shown uses $\tau=-1$. In particular, its Dirac blocks have $\partial_{k_x}h_\ell=\hbar v_F\sigma_x$ and $\partial_{k_y}h_\ell=-\hbar v_F\sigma_y$. The full Bloch Hamiltonian has zero second momentum derivatives, which explains the vanishing first TR and FS terms in the main text.

\subsection{Numerical conventions and scope of the comparison}
The Hamiltonian is represented in a plane-wave basis with $\bG=n_1\bG_1+n_2\bG_2$, $n_1,n_2=-5,\ldots,5$, and four layer-sublattice components, giving $N_b=484$ bands. The susceptibilities in the main text are evaluated on a uniform $120\times120$ moir\'e Brillouin-zone mesh; below we also compare $n_k\times n_k$ meshes with $n_k=60,90,120$. Brillouin-zone sums are normalized per moir\'e cell, and no spin or valley degeneracy factor is included. The band plot follows $K_1$--$\Gamma$--$M$--$K_2$. The energy origin and all model parameters are shared by the MCD and thermodynamic calculations.

The TR contribution uses sharp occupied and empty subspaces with the Fermi-sea derivatives defined in the main text. In the FS contribution, the zero-temperature occupation derivative is represented by
\begin{align}
n_F'(E,\mu)\ \longrightarrow\ -\delta_\eta(E-\mu)
=-\frac{1}{\sqrt{\pi}\eta}\exp\left[-\frac{(E-\mu)^2}{\eta^2}\right],
\qquad \eta=0.5~{\rm meV}.
\end{align}
The plotted chemical potentials are separated by $0.5$~meV. The MCD FS curves use the band-summation evaluation of the projector derivatives and orbital moments. 

\subsection{Band-summation cutoffs and momentum-mesh dependence}
The plane-wave basis size, the band-summation cutoffs, and the momentum mesh are independent numerical parameters. We denote the TR cutoff by $n_{\rm cut}\equiv n_{\rm cut,TR}$ which was first introduced in Ref.~\cite{ZhangArXiv2026} and the FS cutoff by $n_{\rm cut,FS}$. At each momentum, the $N_b$ eigenvalues are ordered increasingly. A cutoff $c$ selects the central band window
\begin{align}
\mathcal W(c)=\left\{\frac{N_b}{2}-c,\ldots,\frac{N_b}{2}+c+1\right\},
\qquad |\mathcal W(c)|=2(c+1).
\end{align}
Thus $c=0$ retains the two central bands, $c=100$ retains $202$ bands, and $c=241$ includes all $484$ bands of the fixed plane-wave Hamiltonian. These windows are centered on the central band pair, not on the chemical potential. 

\begin{figure}[htbp]
\centering
\includegraphics[width=0.78\textwidth]{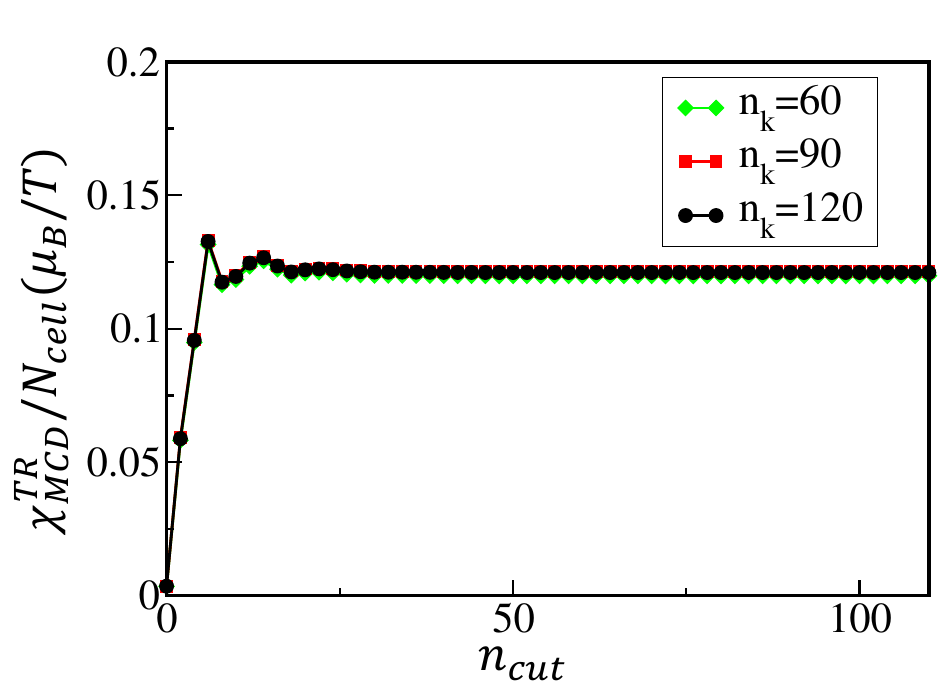}
\caption{Band-cutoff and momentum-mesh dependence of $\chi_{\rm MCD}^{\rm TR}$ in TBG at $\mu=15$~meV. The three curves correspond to $n_k\times n_k$ meshes with $n_k=60,90,120$; $n_{\rm cut,FS}=100$ is fixed. }
\label{FigS:TBG_TR_convergence}
\end{figure}

For the FS calculation, $n_{\rm cut,FS}$ controls the intermediate-band sums independently of $n_{\rm cut}$. For a nondegenerate band $a$, the component of the eigenstate derivative orthogonal to $|u_a\rangle$ is evaluated as
\begin{align}
|D_\alpha u_a\rangle_{\rm FS}
&=\sum_{\substack{b\in\mathcal W(n_{\rm cut,FS})\\ b\ne a}}
|u_b\rangle\frac{\langle u_b|\partial_{k_\alpha}H|u_a\rangle}{E_a-E_b},\\
(\partial_{k_\alpha}P_a)_{\rm FS}
&=|D_\alpha u_a\rangle_{\rm FS}\langle u_a|
+|u_a\rangle\,{}_{\rm FS}\langle D_\alpha u_a|.
\end{align}
All quantities in these expressions are evaluated at the same momentum. In the full-band limit this is the usual spectral representation of the projector derivative; a finite window truncates its intermediate states. The same window is used for the orbital moment,
\begin{align}
m_a^{\rm FS}=\frac{e}{\hbar}\operatorname{Im}
\sum_{\substack{b\in\mathcal W(n_{\rm cut,FS})\\b\ne a}}
\frac{\langle u_a|\partial_{k_x}H|u_b\rangle
\langle u_b|\partial_{k_y}H|u_a\rangle}{E_b-E_a},
\end{align}
and for the occupied/empty intermediate-state contractions in the FS kernel. Hence $n_{\rm cut,FS}$ is not only a cutoff for $m_a$: it controls the band-summation evaluation of all FS contributions. We use a fixed FS cutoff $n_{\rm cut,FS}=100$, retaining $202$ bands, for all reported FS responses, which is enough to get accurate Fermi surface derivatives. The main-text TR curves use $n_{\rm cut}=100$. Figure~\ref{FigS:TBG_TR_convergence} shows that the TR response is nearly independent of the band cutoff beyond this value at $\mu=15$~meV. At $n_{\rm cut}=100$, the results on the $90\times90$ and $120\times120$ momentum meshes differ by approximately $0.26\%$.

\end{document}